\documentclass[twocolumn]{aastex7}

\begin{document}

\title{MIRIS P\MakeLowercase{a}$\alpha$ Galactic Plane Survey. II. The Catalog of P\MakeLowercase{a}$\alpha$ Emission-line Sources}

\author[orcid=0000-0002-9541-3988]{Il-Joong Kim}
\affiliation{Korea Astronomy and Space Science Institute, Daejeon, 34055, Republic of Korea}
\email[show]{ijkim@kasi.re.kr}

\author[orcid=0000-0001-9937-8270]{Jeonghyun Pyo}
\affiliation{Korea Astronomy and Space Science Institute, Daejeon, 34055, Republic of Korea}
\email{jhpyo@kasi.re.kr}

\author[orcid=0000-0001-9561-8134]{Kwang-Il Seon}
\affiliation{Korea Astronomy and Space Science Institute, Daejeon, 34055, Republic of Korea}
\affiliation{Astronomy and Space Science Major, University of Science and Technology, Daejeon, 34113, Republic of Korea}
\email{kiseon@kasi.re.kr}

\author[orcid=0000-0002-2770-808X]{Woong-Seob Jeong}
\affiliation{Korea Astronomy and Space Science Institute, Daejeon, 34055, Republic of Korea}
\affiliation{Astronomy and Space Science Major, University of Science and Technology, Daejeon, 34113, Republic of Korea}
\email{jeongws@kasi.re.kr}

\author[orcid=0000-0002-6660-9375]{Takao Nakagawa}
\affiliation{Institute of Space and Astronautical Science, Japan Aerospace Exploration Agency, Kanagawa, 252-5210, Japan}
\affiliation{Advanced Research Laboratories, Tokyo City University, Tokyo, 158-8557, Japan}
\email{nakagawa@ir.isas.jaxa.jp}

\author{Toshio Matsumoto}
\altaffiliation{Deceased on 2022 July}
\affiliation{Institute of Space and Astronautical Science, Japan Aerospace Exploration Agency, Kanagawa, 252-5210, Japan}
\email{matsumoto@ir.isas.jaxa.jp}

\begin{abstract}
Using data from the MIRIS Pa$\alpha$ Galactic Plane Survey (MIPAPS), we present a Pa$\alpha$ 1.87 $\mu$m line image of the entire Galactic plane within the latitude range of $-3\arcdeg \la b \la +3\arcdeg$, revealing numerous Pa$\alpha$ features. Based on the MIPAPS Pa$\alpha$ image and the {\it WISE} \ion{H}{2} region catalog, we compile a catalog of 1489 Pa$\alpha$ emission-line sources in the Galactic plane within $90\arcdeg \leq \ell \leq 330\arcdeg$. By comparing our Pa$\alpha$ images with H$\alpha$ images constructed from the IPHAS and VPHAS+ survey data, we demonstrate the advantages of Pa$\alpha$ line observations. We identify 902 Pa$\alpha$ sources associated with \ion{H}{2} regions, and newly confirm 619 \ion{H}{2} region candidates as definitive \ion{H}{2} regions through Pa$\alpha$ or H$\alpha$ detections. We also identify 261 extended and 326 point-like Pa$\alpha$ sources not included in the {\it WISE} catalog, most of which have H$\alpha$ counterparts in the IPHAS or VPHAS+ images. A search of the SIMBAD database indicates that these sources originate from diverse object types. By measuring Pa$\alpha$ and H$\alpha$ fluxes, we estimate the $E(\bv)$ color excesses derived from extended emissions for 138 Pa$\alpha$ sources, showing good agreement with values obtained from spectrophotometry of ionizing stars in previous studies. Futhermore, we calculate total Lyman continuum luminosities for 42 Pa$\alpha$ sources, providing constraints on the distances to \ion{H}{2} regions and the spectral types of their ionizing stars. These results highlight the scientific potential of Pa$\alpha$ line observations and the benefits of combining multiple hydrogen recombination lines in exploring ionized regions.
\end{abstract}


\section{Introduction} \label{sec:intro}

Using the Multipurpose InfraRed Imaging System (MIRIS), a compact near-infrared space telescope designed for wide-field surveys \citep{2014PASP..126..853H}, the MIRIS Pa$\alpha$ Galactic Plane Survey (MIPAPS) was performed in order to study the nature and distribution of ionized hydrogen gas in the Milky Way. Over approximately ten months of observations, the first Pa$\alpha$ 1.87 $\mu$m line map covering the entire Galactic plane was completed, and then preliminary results for the longitude range of $\ell = 96\arcdeg.5$--$116\arcdeg.3$ ($\sim$5\% of the entire MIPAPS survey) were presented in \citet{2018ApJS..238...28K}, hereafter referred to as Paper I. By comparing the MIPAPS data with the INT/WFC Photometric H$\alpha$ Survey (IPHAS) data \citep{2005MNRAS.362..753D} and the {\it WISE} \ion{H}{2} region catalog \citep{2014ApJS..212....1A}, we demonstrated the scientific potential of the MIPAPS data despite its moderate spatial resolution ($\sim$52$\arcsec$). In addition to the detection of Pa$\alpha$ lines from 27 \ion{H}{2} regions listed in the {\it WISE} \ion{H}{2} region catalog, 53 \ion{H}{2} region candidates were confirmed as definitive \ion{H}{2} regions through the detection of Pa$\alpha$ hydrogen recombination lines. Furthermore, 47 Pa$\alpha$ sources not included in the {\it WISE} \ion{H}{2} region catalog were newly reported. The $E(\bv)$ color excesses and total Lyman continuum luminosities for some \ion{H}{2} regions were also estimated by combining the MIPAPS Pa$\alpha$ and IPHAS H$\alpha$ photometry data.

To investigate the origins of the observed Pa$\alpha$ sources, we conducted high-resolution near-infrared spectroscopic follow-up observations of three Pa$\alpha$ sources. Among these, \citet{2021AJ....162...24K} reported the results for MWC 1080, one of the brightest Pa$\alpha$ sources detected in the Cepheus region ($\ell = 96\arcdeg.5$--$116\arcdeg.3$). MWC 1080, identified as a Herbig Ae/Be star in the SIMBAD database, was detected as a point-like source in the MIPAPS Pa$\alpha$ image. However, the IPHAS H$\alpha$ data, with its higher spatial resolution, revealed extended H$\alpha$ features surrounding a central H$\alpha$ point source (Paper I). By detecting various emission lines (six hydrogen Brackett lines, seven H$_{2}$ lines, and an {[}\ion{Fe}{2}{]} line) from the extended features, we found that these features primarily consist of stellar emission from MWC 1080A scattered by dust, along with additional components associated with stellar outflows and nearby young stars. The follow-up spectroscopy demonstrated that Pa$\alpha$ sources observed by MIPAPS can have diverse origins, not limited to photo-ionized hydrogen gas.

Paper I primarily focused on \ion{H}{2} regions; however, hydrogen recombination lines can originate from various types of objects. These lines from ionized hydrogen gas have been observed mostly via radio recombination lines, the near-infrared Br$\gamma$ line, and the optical H$\alpha$ line, all of which are detectable with ground-based telescopes. In particular, large-scale surveys have predominantly been conducted using the optical H$\alpha$ line. Recent H$\alpha$ surveys with the highest spatial resolutions and sensitivities include IPHAS, which covers the entire northern Galactic plane, and the VST Photometric H$\alpha$ Survey of the Southern Galactic Plane and Bulge \citep[VPHAS+;][]{2014MNRAS.440.2036D}, which covers the entire southern Galactic plane. Studies utilizing the IPHAS and/or VPHAS+ H$\alpha$ data have been published on various types of objects, including supernova remnants \citep{2013MNRAS.431..279S}, planetary nebulae \citep{2009A&A...504..291V,2009A&A...502..113V,2010PASA...27..166S,2014MNRAS.443.3388S,2021MNRAS.501.6156D,2024A&A...692A.103L}, classical Be stars \citep{2013MNRAS.430.2169R,2015MNRAS.446..274R,2016A&A...591A.140G,2020A&A...638A..21V}, Herbig Ae/Be stars \citep{2020A&A...638A..21V}, young stellar objects \citep{2008MNRAS.387..308V,2011MNRAS.415..103B,2015MNRAS.453.1026K,2019MNRAS.484.5102K}, symbiotic stars \citep{2008A&A...480..409C,2010A&A...509A..41C,2014A&A...567A..49R,2021MNRAS.502.2513A}, cataclysmic variables \citep{2007MNRAS.382.1158W,2015MNRAS.451.2863S}, and open clusters \citep{2017MNRAS.465.1505D}.

Although the H$\alpha$ line is the most commonly used hydrogen recombination line for observations, its detection is strongly limited by attenuation due to interstellar dust. As a result, detecting the H$\alpha$ line can be challenging not only for distant sources with significant interstellar dust along the line of sight but also for sources embedded within dense molecular clouds. In contrast, the Pa$\alpha$ 1.87 $\mu$m line experiences significantly less dust extinction than the H$\alpha$ line due to its longer wavelength. According to the case B hydrogen recombination spectrum, the intrinsic H$\alpha$ to Pa$\alpha$ line ratio is approximately 8.5 at a temperature of 10$^{4}$ K \citep{2011piim.book.....D}. However, the less-attenuated Pa$\alpha$ flux can exceed the more-attenuated H$\alpha$ flux when $E(\bv)$ $>$ 1.12, assuming the Galactic extinction curve of \citet{1989ApJ...345..245C} with $R_V$ = 3.1. This makes the Pa$\alpha$ line well-suited for deep observations of regions obscured by interstellar dust, such as the Galactic center and star-forming regions within dense molecular clouds. However, because the Pa$\alpha$ line falls within a wavelength range heavily absorbed by water molecules in Earth's atmosphere, efficient observation of the Pa$\alpha$ line requires a space-based telescope. The {\it Hubble Space Telescope} ({\it HST}) NICMOS camera was used to observe the Pa$\alpha$ line; however, due to its limited field of view, it primarily focused on small-sized objects, such as external galaxies. A Pa$\alpha$ mosaic image covering a sky area of only $\sim$$39\times15$ arcmin$^{2}$ around the Galactic center was constructed by combining {\it HST} NICMOS data from thousands of exposure fields \citep{2010MNRAS.402..895W}.

This paper is the second in a series presenting the results of the MIPAPS Pa$\alpha$ survey. We attempt to extend the analysis conducted in Paper I for $\ell = 96\arcdeg.5$--$116\arcdeg.3$ to the entire Galactic plane by utilizing the complete MIPAPS dataset. However, the data for $-30\arcdeg \leq \ell \leq 90\arcdeg$ still contain artifacts caused by a malfunction in the filter-wheel operation, which will be described in Section \ref{subsec:pa_data}. First, we construct continuum-subtracted Pa$\alpha$ line images for the entire Galactic plane. Then, based on the Pa$\alpha$ line images, we compile a catalog of Pa$\alpha$ emission-line sources for $\ell = 90\arcdeg$--$330\arcdeg$, where the MIPAPS data are unaffected by the filter-wheel malfunction. Using the {\it WISE} \ion{H}{2} region catalog, we primarily identify Pa$\alpha$ sources originating from \ion{H}{2} regions. Subsequently, we search the Pa$\alpha$ line images for additional Pa$\alpha$ sources not included in the {\it WISE} \ion{H}{2} region catalog. To highlight the advantages of Pa$\alpha$ line observations, we also create continuum-subtracted H$\alpha$ images using the IPHAS and VPHAS+ survey data and compare them with our Pa$\alpha$ line images. We perform aperture photometry of the Pa$\alpha$ fluxes for some Pa$\alpha$ sources, including updates to the photometric results in Paper I. By combining the photometries of the MIPAPS Pa$\alpha$ and IPHAS H$\alpha$ fluxes, we also estimate the $E(\bv)$ color excesses for certain Pa$\alpha$ sources. These $E(\bv)$ color excesses, derived from extended Pa$\alpha$ and H$\alpha$ emissions, are compared with those obtained from spectrophotometry of ionizing point stars in previous studies. Utilizing our photometric results, we also provide constraints on the distances to some \ion{H}{2} regions and the spectral types of their ionizing stars.

\section{Data Reduction} \label{sec:data}

\subsection{MIPAPS Pa$\alpha$ Survey Data} \label{subsec:pa_data}

The MIRIS observational data are available on the official MIRIS website\footnote{\url{http://miris.kasi.re.kr/miris/}}. Currently, users can download data processed by the MIRIS data reduction pipeline, and flux-calibrated MIPAPS data are also available. The MIPAPS Pa$\alpha$ survey data include a total of 241 observational fields, each with a field of view of $3\arcdeg.67\times3\arcdeg.67$ and a pixel size of $51\arcsec.6$. The entire MIPAPS data cover the whole Galactic plane within the latitude range of $-3\arcdeg \la b \la +3\arcdeg$, with some fields extending to $b \approx -8.6\arcdeg$ or $+8.6\arcdeg$. Each field data comprise several FITS files, taken during individual observations using one of two narrow-band filters: the Pa$\alpha$ line (PAAL; centered at $\sim$1.875 $\mu$m) and the Pa$\alpha$ dual continuum (PAAC; centered at $\sim$1.84 $\mu$m and $\sim$1.91 $\mu$m). Each observational field was typically observed three times with the PAAL filter and three times with the PAAC filter. The main parameters of the MIPAPS Pa$\alpha$ survey are summarized in Table \ref{table:mipaps_survey}.

\begin{deluxetable}{cc}
\tablewidth{0pt}
\tablecaption{Main Parameters of the MIPAPS Pa$\alpha$ Survey\label{table:mipaps_survey}}
\tablehead{
\colhead{Parameter} & \colhead{Value}
}
\startdata
Telescope aperture & 8 cm \\
Detector array     & HgCdTe 256 $\times$ 256 \\
Field of view      & $3\arcdeg.67 \times 3\arcdeg.67$ \\
Pixel scale        & $51\arcsec.6$ \\
Filter             & PAAL (centered at $\sim$1.875 $\mu$m), \\
                   & PAAC (centered at $\sim$1.84 $\mu$m, $\sim$1.91 $\mu$m) \\
Coverage           & $0\arcdeg \leq \ell \leq 360\arcdeg$, $-3\arcdeg \la b \la +3\arcdeg$ \\
Observing period   & 2014 April--2015 May \\
Official website   & \url{http://miris.kasi.re.kr/miris/} \\
\enddata
\end{deluxetable}

First, we created one PAAL image and one PAAC image for each of the 241 MIPAPS fields by combining the same filter data using the Montage software\footnote{\url{http://montage.ipac.caltech.edu/}}. We then flux-calibrated the PAAL and PAAC images using calibration factors of 27.8 $\pm$ 2.0 mJy (ADU s$^{-1}$)$^{-1}$ for PAAL and 17.2 $\pm$ 1.2 mJy (ADU s$^{-1}$)$^{-1}$ for PAAC. In Paper I, we used calibration factors of 23.7 $\pm$ 0.1 and 14.6 $\pm$ 0.1 mJy (ADU s$^{-1}$)$^{-1}$, which were derived by comparing the fluxes of point sources identified in the PAAL and PAAC images of the Cepheus region ($96\arcdeg.5 \leq \ell \leq 116\arcdeg.3$) with those of their counterparts in the 2MASS point-source catalog \citep{2006AJ....131.1163S}. To obtain calibration factors applicable to the entire Galactic plane, we recalculated them using 2MASS point sources distributed across all 241 MIPAPS fields, following the same method as in Paper I. Point sources detected in the PAAL and PAAC images were matched with their 2MASS counterparts. The expected fluxes in the PAAL and PAAC filters were computed by applying the respective filter transmission curves to fluxes interpolated from the 2MASS H and K$_{s}$ magnitudes. We also considered the significant difference in pixel sizes between the MIPAPS ($\sim$52$\arcsec$) and 2MASS ($\sim$1$\arcsec$) images, which had been overlooked in Paper I. Since an area containing a point source in the MIPAPS images typically includes several adjacent point sources that are clearly separated in the 2MASS images, MIPAPS fluxes tend to be overestimated compared to the 2MASS fluxes. To correct for this, we revised each 2MASS flux by summing the fluxes of all 2MASS sources within the MIPAPS source area defined in the PAAL and PAAC images. By performing least-squares linear fitting between the observed fluxes (in ADU s$^{-1}$) and the expected fluxes (in mJy), we derived the updated calibration factors and their 1$\sigma$ uncertainties: 27.8 $\pm$ 2.0 for PAAL and 17.2 $\pm$ 1.2 for PAAC. These values are approximately 17\% higher than those in Paper I, as noted in the Erratum of Paper I \citep{2020ApJS..251...16K}.

\begin{figure*}[ht]
\centering
\includegraphics[scale=0.36]{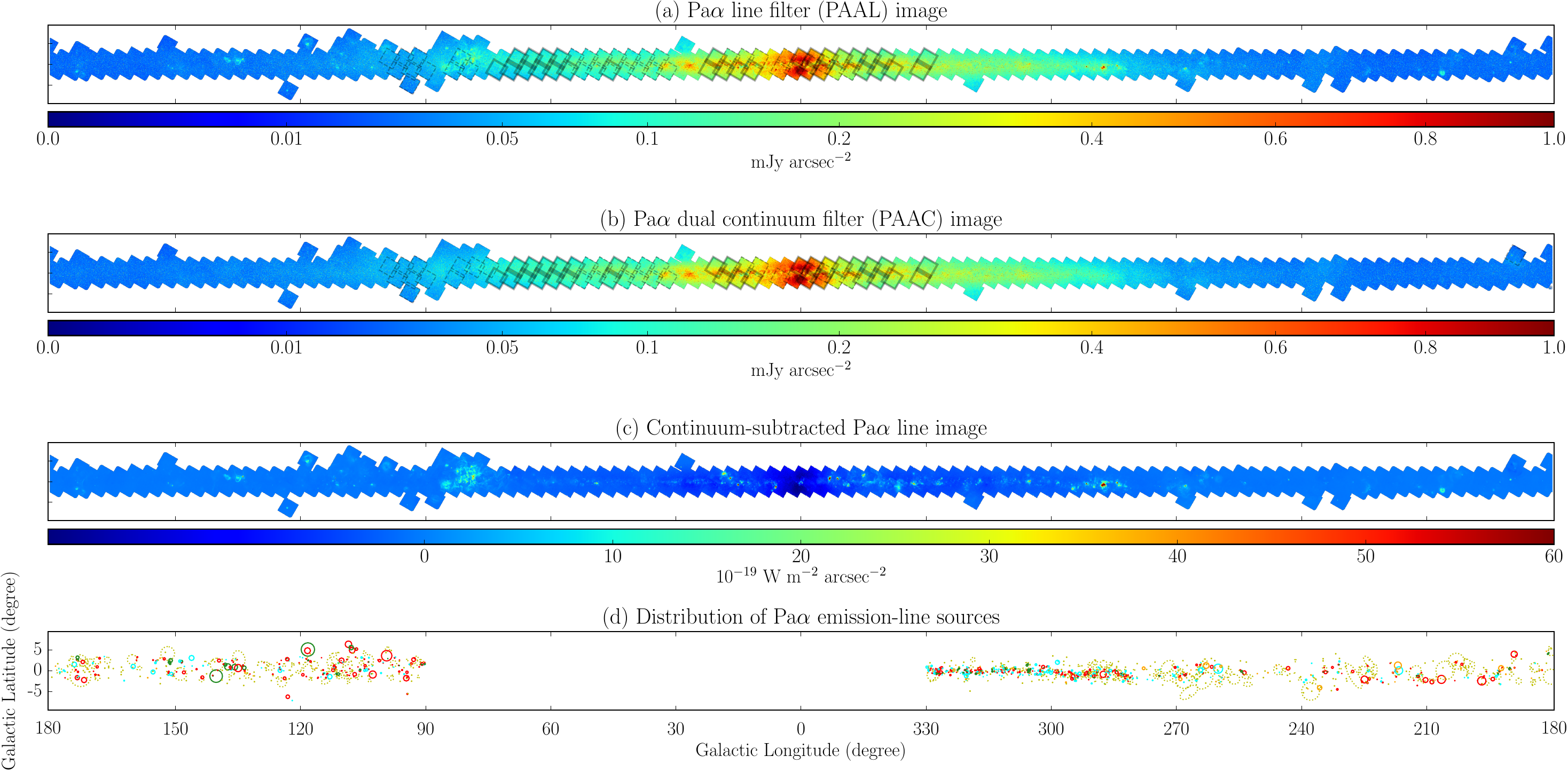}
\caption{MIPAPS mosaic images of the entire Galactic plane. Panels (a) and (b) display the calibrated MIPAPS Pa$\alpha$ line (PAAL) and dual continuum (PAAC) filter images, respectively, while panel (c) shows the continuum-subtracted Pa$\alpha$ line image. In Panels (a) and (b), the thick solid and thin dashed rectangles indicate the 66 MIPAPS fields with data obtained from highly- and slightly-shadowed observations, respectively. Panel (d) presents the Galactic distribution of 1489 Pa$\alpha$ emission-line sources cataloged in this study. The solid open circles represent 902 Pa$\alpha$ sources corresponding to {\it WISE} \ion{H}{2} region sources listed in Table \ref{table:mipaps}, with colors indicating source types: red for ``Known'', cyan for ``Candidate'', green for ``Group'', and orange for ``Radio Quiet'' {\it WISE} sources. The yellow dotted circles and ellipses mark 261 extended Pa$\alpha$ sources with no counterparts in the {\it WISE} \ion{H}{2} region catalog, also listed in Table \ref{table:mipaps}. The sizes of the circles and ellipses match the angular sizes of the individual Pa$\alpha$ sources. The yellow points represent 326 point-like Pa$\alpha$ sources, which are likewise included in Table \ref{table:mipaps}.\label{fig:whole}}
\end{figure*}

\begin{figure*}[ht]
\centering
\includegraphics[scale=0.33]{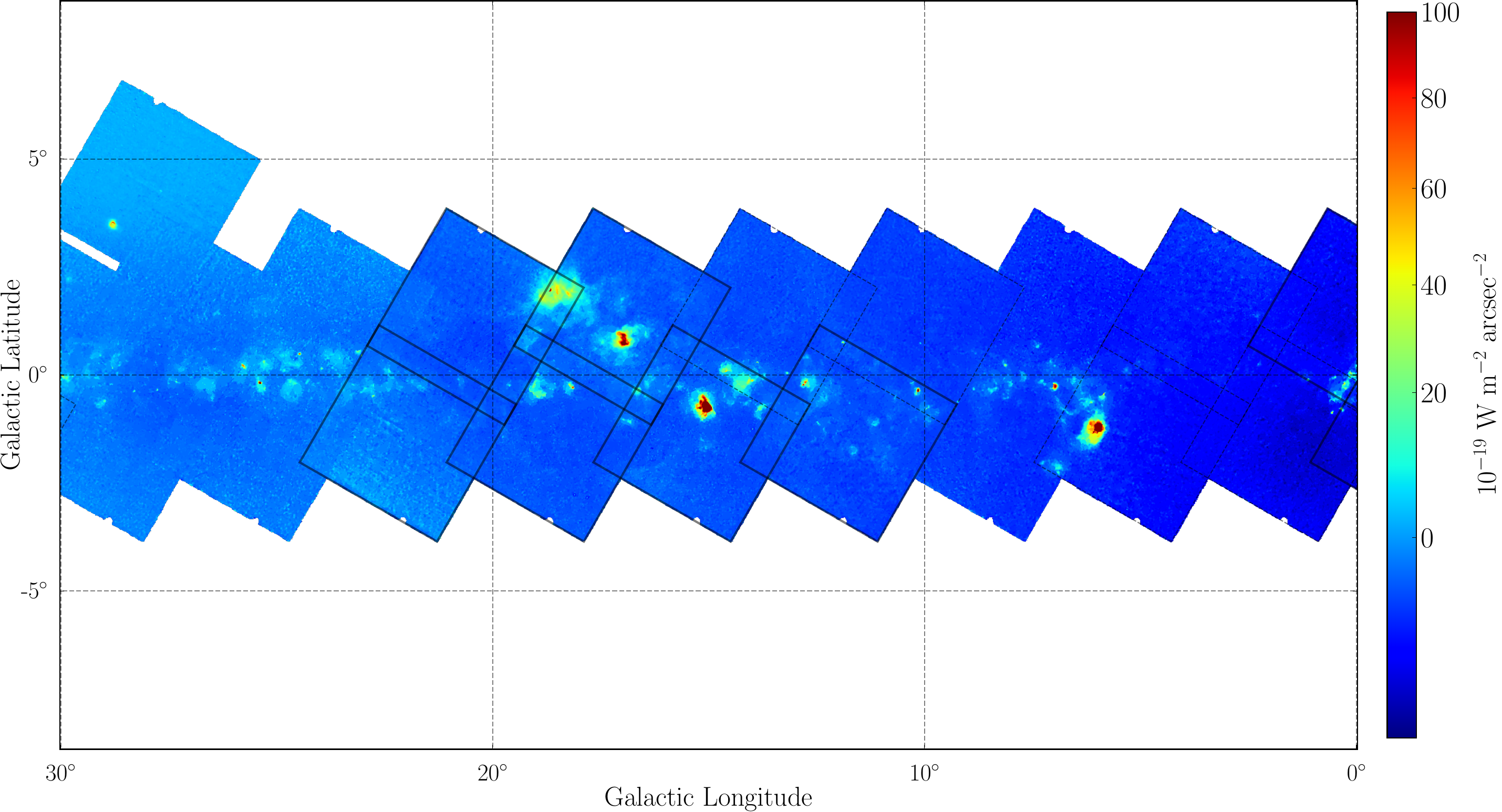}
\caption{Continuum-subtracted MIPAPS Pa$\alpha$ line image of the Galactic plane for $\ell = 0\arcdeg$--$30\arcdeg$. The thick solid and thin dashed rectangles represent MIPAPS fields with data obtained from highly- and slightly-shadowed observations, respectively.\label{fig:paa11}}
\end{figure*}

\begin{figure*}[ht]
\centering
\includegraphics[scale=0.33]{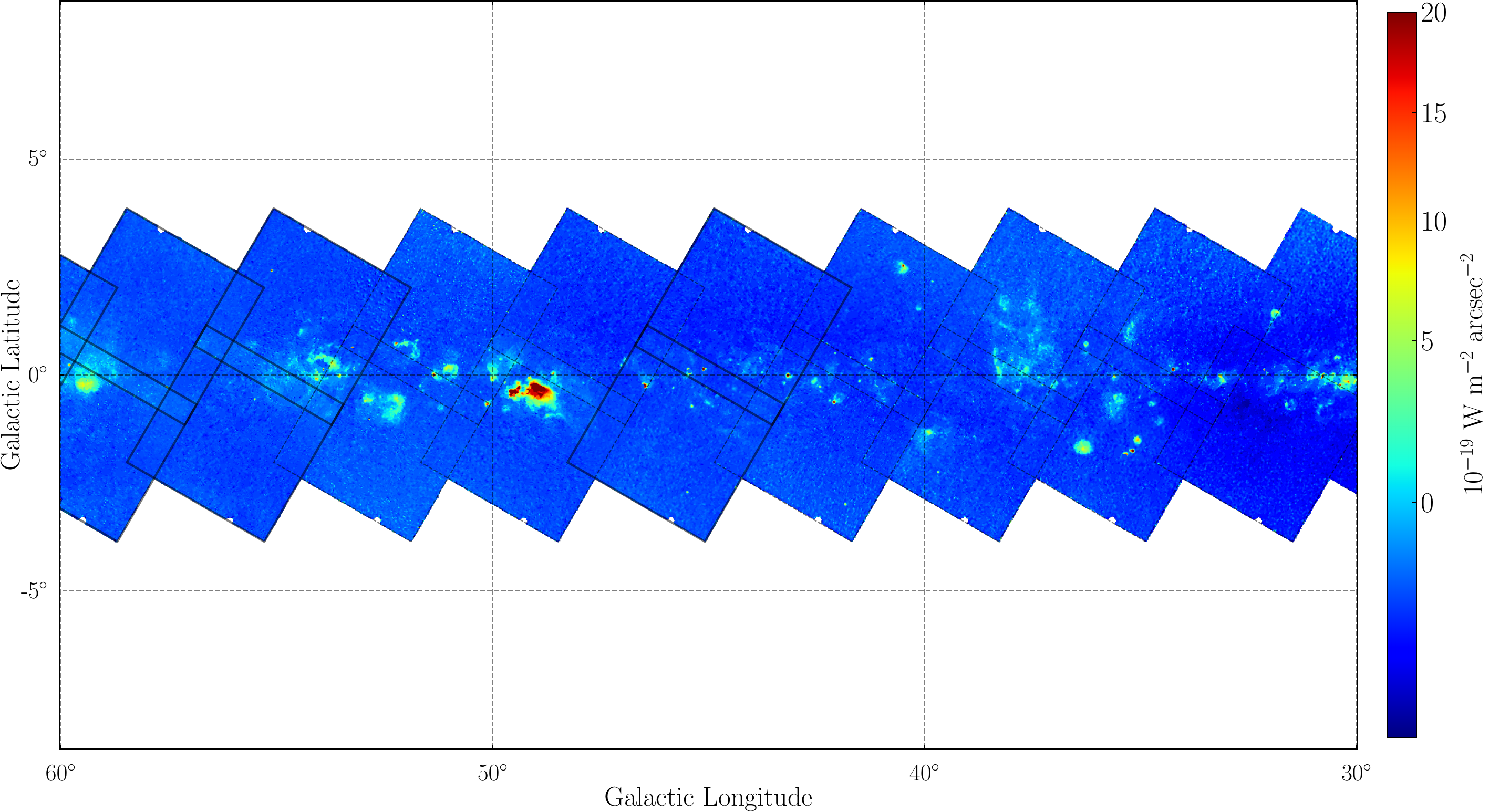}
\caption{Continuum-subtracted MIPAPS Pa$\alpha$ line image of the Galactic plane for $\ell = 30\arcdeg$--$60\arcdeg$. The thick solid and thin dashed rectangles represent MIPAPS fields with data obtained from highly- and slightly-shadowed observations, respectively.\label{fig:paa12}}
\end{figure*}

\begin{figure*}[ht]
\centering
\includegraphics[scale=0.33]{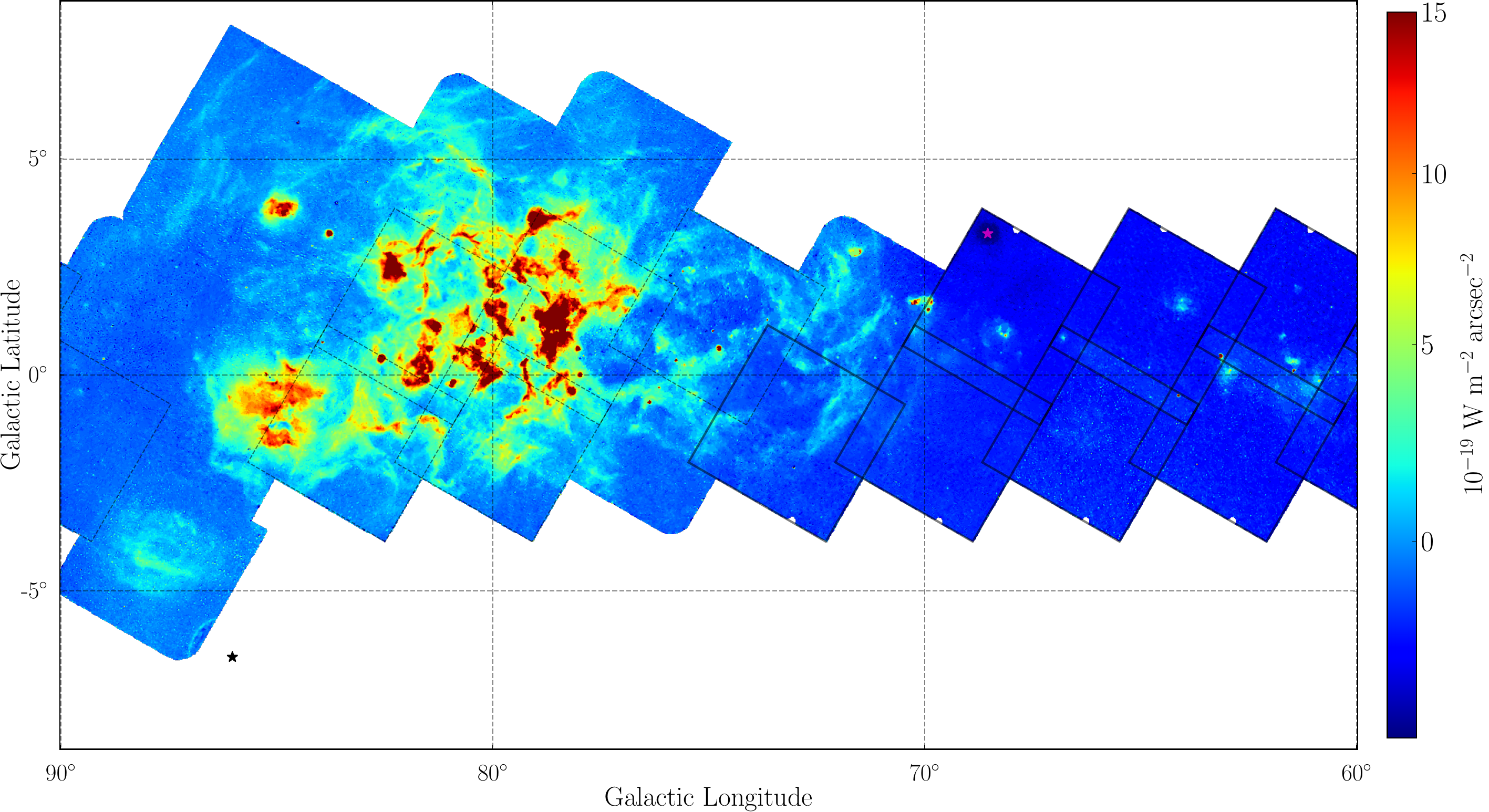}
\caption{Continuum-subtracted MIPAPS Pa$\alpha$ line image of the Galactic plane for $\ell = 60\arcdeg$--$90\arcdeg$. The thick solid and thin dashed rectangles represent MIPAPS fields with data obtained from highly- and slightly-shadowed observations, respectively. The magenta and black star symbols mark HD 187796 and HD 203712, which cause an extended dark area and an arc-like artifact centered on them, respectively.\label{fig:paa13}}
\end{figure*}

\begin{figure*}[ht]
\centering
\includegraphics[scale=0.33]{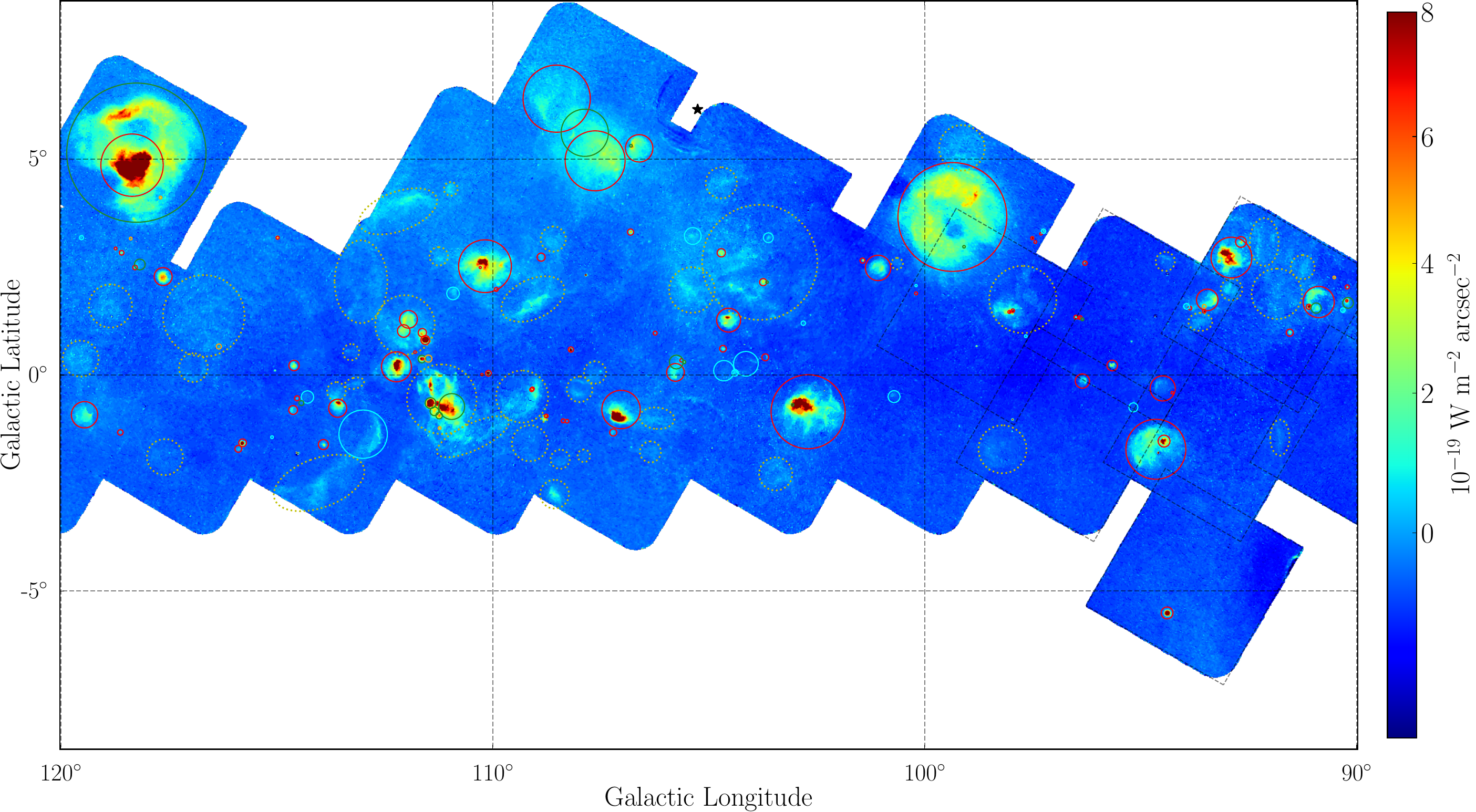}
\caption{Continuum-subtracted MIPAPS Pa$\alpha$ line image of the Galactic plane for $\ell = 90\arcdeg$--$120\arcdeg$. The thin dashed rectangles represent MIPAPS fields with data obtained from slightly-shadowed observations. The solid circles indicate 126 Pa$\alpha$ sources corresponding to {\it WISE} \ion{H}{2} region sources listed in Table \ref{table:mipaps}, with colors representing source types: red for ``Known'', cyan for ``Candidate'', green for ``Group'', and orange for ``Radio Quiet'' {\it WISE} sources. The yellow dotted circles and ellipses mark 43 extended Pa$\alpha$ sources with no counterparts in the {\it WISE} \ion{H}{2} region catalog. The black star symbol marks HD 209772, which causes arc-like artifacts centered on it.\label{fig:paa21}}
\end{figure*}

\begin{figure*}[ht]
\centering
\includegraphics[scale=0.33]{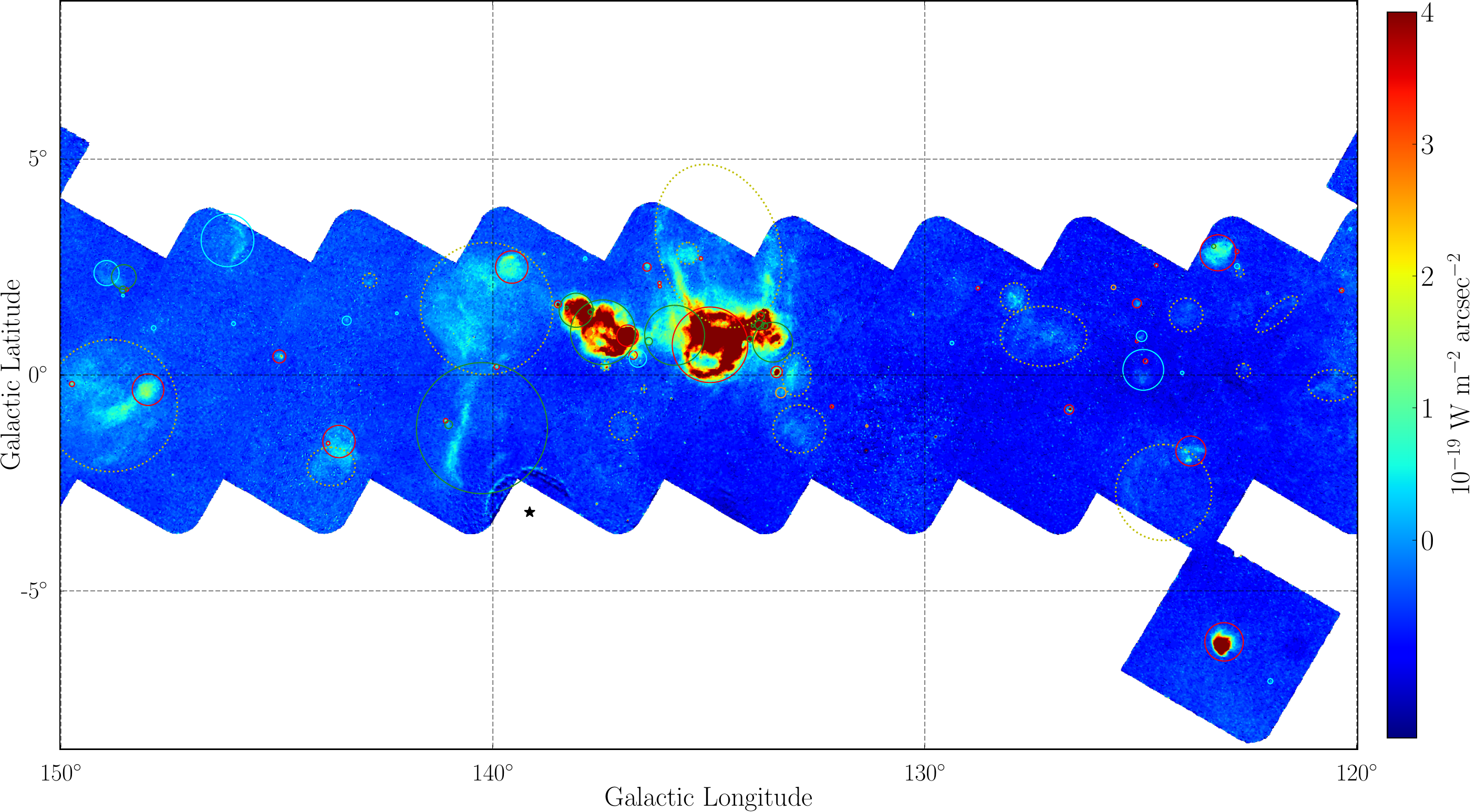}
\caption{Continuum-subtracted MIPAPS Pa$\alpha$ line image of the Galactic plane for $\ell = 120\arcdeg$--$150\arcdeg$. The solid circles indicate 70 Pa$\alpha$ sources corresponding to {\it WISE} \ion{H}{2} region sources listed in Table \ref{table:mipaps}, with colors representing source types: red for ``Known'', cyan for ``Candidate'', green for ``Group'', and orange for ``Radio Quiet'' {\it WISE} sources. The yellow dotted circles and ellipses mark 19 extended Pa$\alpha$ sources with no counterparts in the {\it WISE} \ion{H}{2} region catalog. The black star symbol marks HD 17506, which causes arc-like artifacts centered on it.\label{fig:paa22}}
\end{figure*}

\begin{figure*}[ht]
\centering
\includegraphics[scale=0.33]{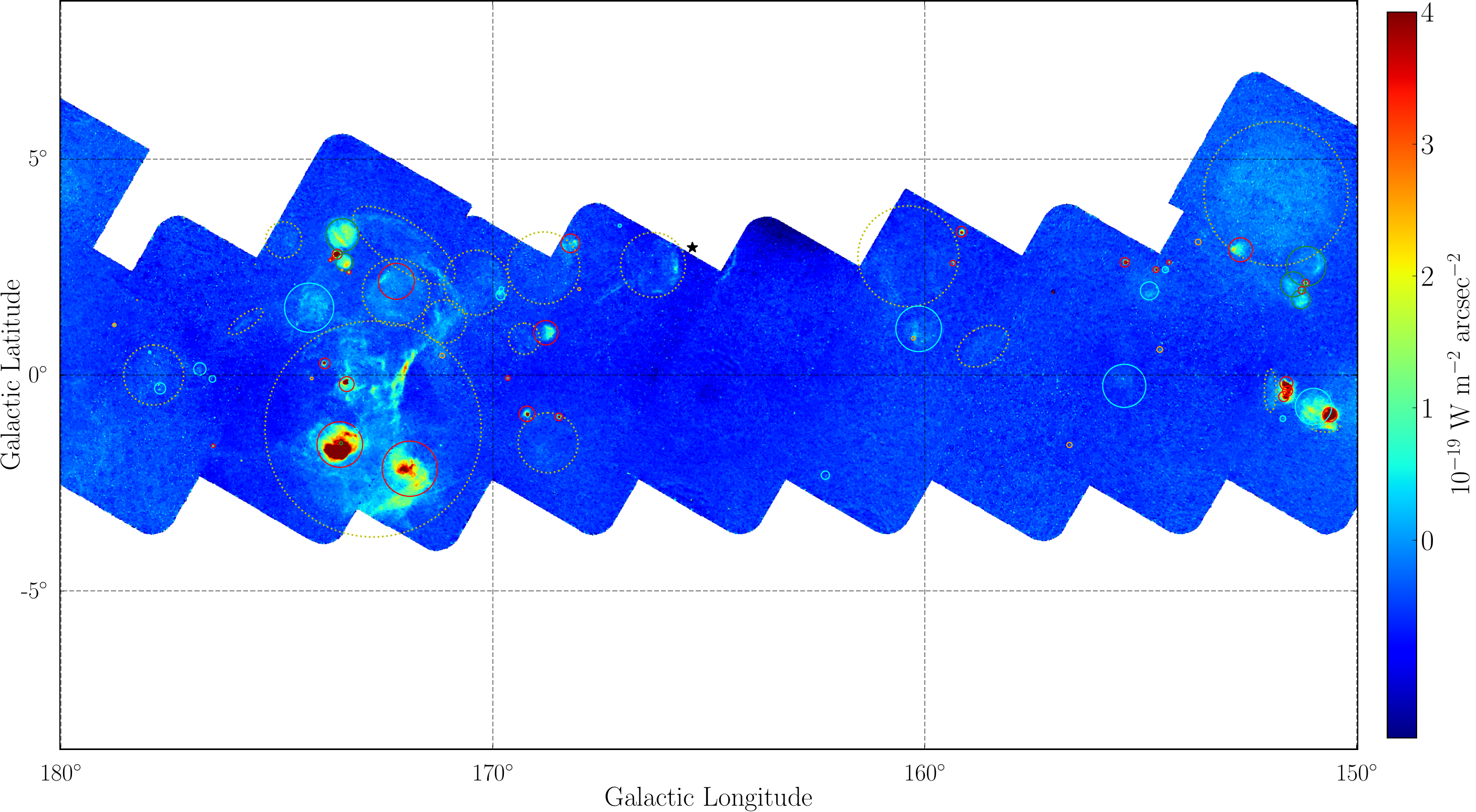}
\caption{Continuum-subtracted MIPAPS Pa$\alpha$ line image of the Galactic plane for $\ell = 150\arcdeg$--$180\arcdeg$. The solid circles indicate 60 Pa$\alpha$ sources corresponding to {\it WISE} \ion{H}{2} region sources listed in Table \ref{table:mipaps}, with colors representing source types: red for ``Known'', cyan for ``Candidate'', green for ``Group'', and orange for ``Radio Quiet'' {\it WISE} sources. The yellow dotted circles and ellipses mark 17 extended Pa$\alpha$ sources with no counterparts in the {\it WISE} \ion{H}{2} region catalog. The black star symbol marks HD 34269, which causes arc-like artifacts centered on it.\label{fig:paa23}}
\end{figure*}

\begin{figure*}[ht]
\centering
\includegraphics[scale=0.33]{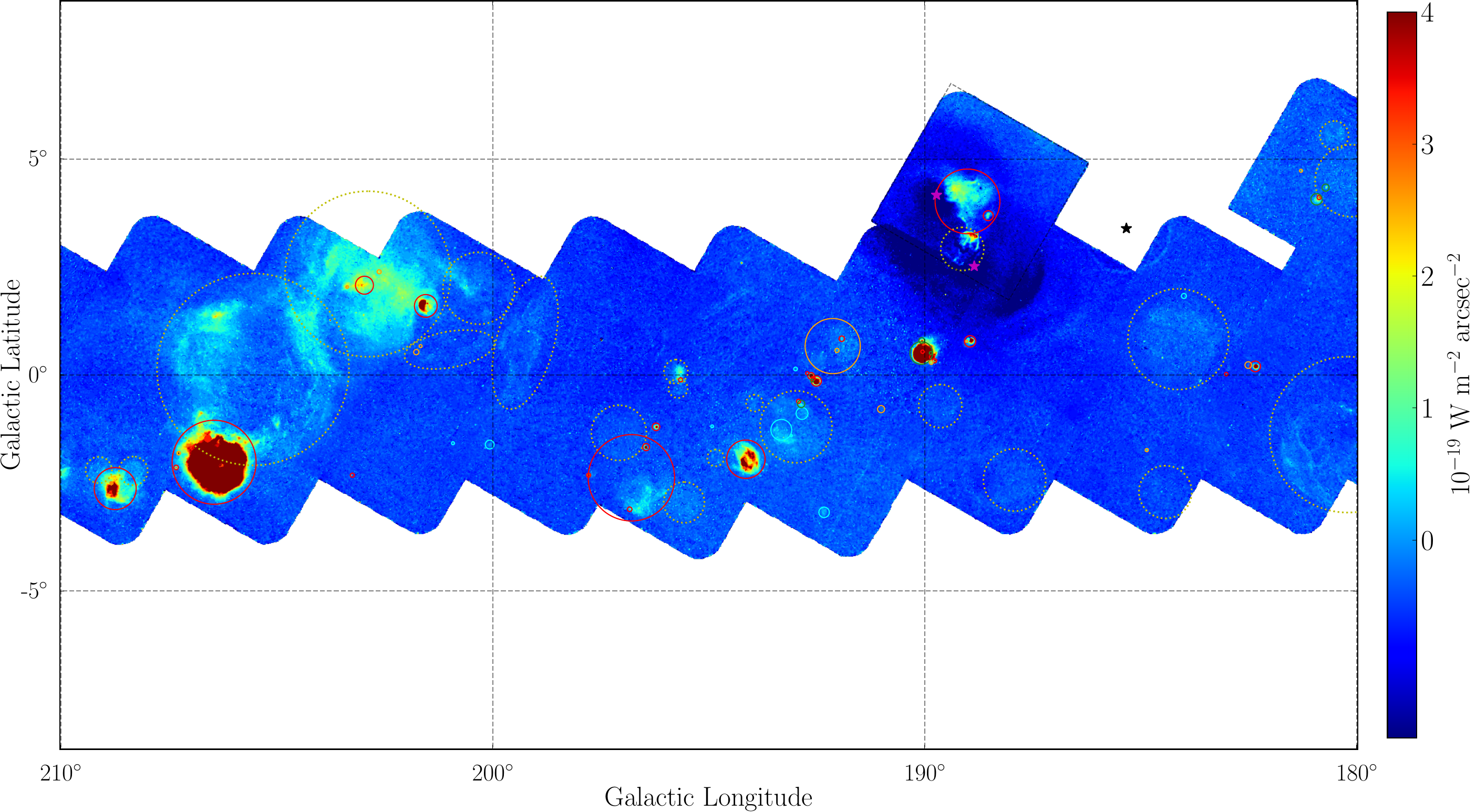}
\caption{Continuum-subtracted MIPAPS Pa$\alpha$ line image of the Galactic plane for $\ell = 180\arcdeg$--$210\arcdeg$. The solid circles indicate 57 Pa$\alpha$ sources corresponding to {\it WISE} \ion{H}{2} region sources listed in Table \ref{table:mipaps}, with colors representing source types: red for ``Known'', cyan for ``Candidate'', green for ``Group'', and orange for ``Radio Quiet'' {\it WISE} sources. The yellow dotted circles and ellipses mark 22 extended Pa$\alpha$ sources with no counterparts in the {\it WISE} \ion{H}{2} region catalog. The magenta star symbols denote HD 44478 and HD 42995, which cause extended dark areas around them, while the black star symbol marks HD 42272, which causes arc-like artifacts centered on it.\label{fig:paa31}}
\end{figure*}

\begin{figure*}[ht]
\centering
\includegraphics[scale=0.33]{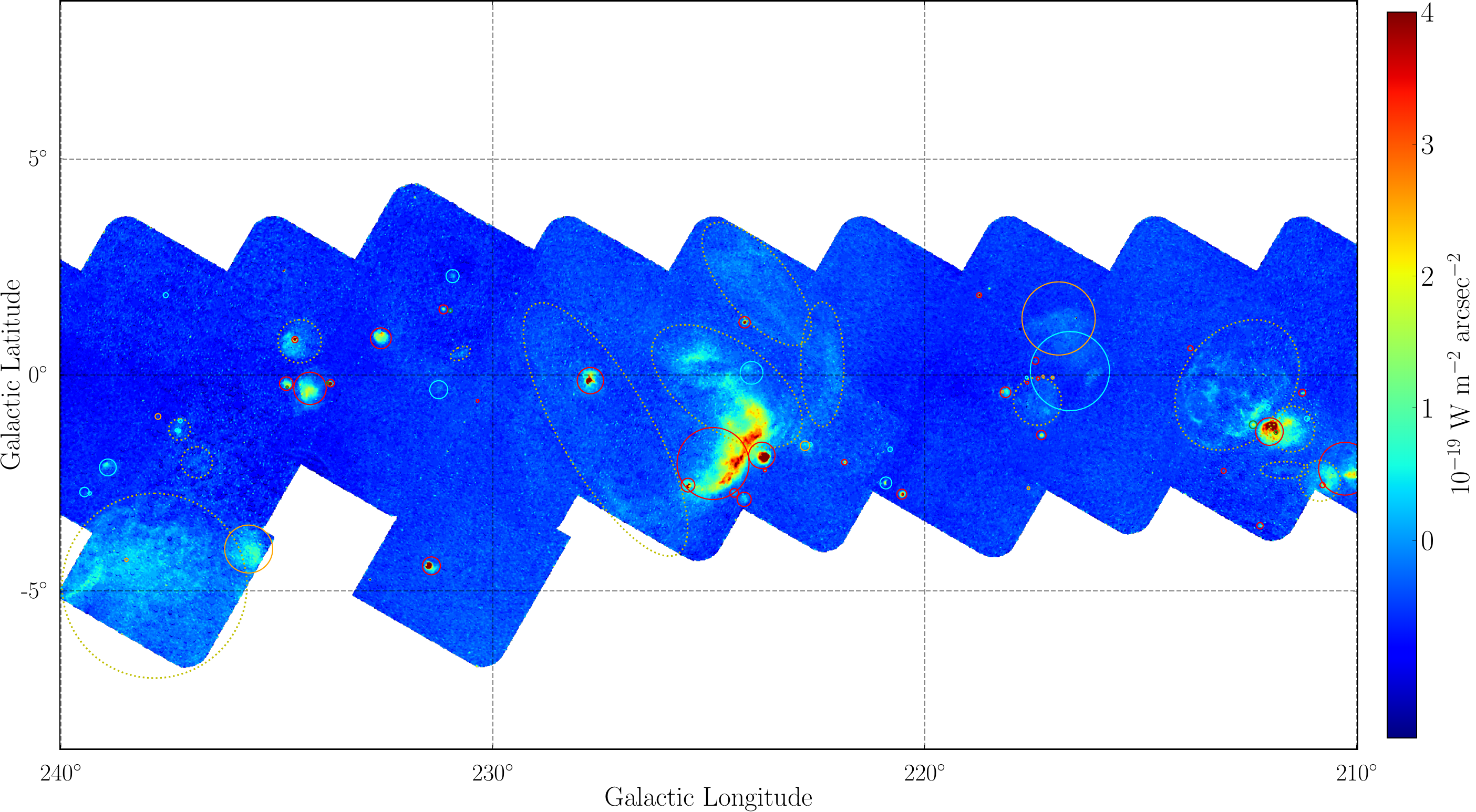}
\caption{Continuum-subtracted MIPAPS Pa$\alpha$ line image of the Galactic plane for $\ell = 210\arcdeg$--$240\arcdeg$. The solid circles indicate 62 Pa$\alpha$ sources corresponding to {\it WISE} \ion{H}{2} region sources listed in Table \ref{table:mipaps}, with colors representing source types: red for ``Known'', cyan for ``Candidate'', green for ``Group'', and orange for ``Radio Quiet'' {\it WISE} sources. The yellow dotted circles and ellipses mark 14 extended Pa$\alpha$ sources with no counterparts in the {\it WISE} \ion{H}{2} region catalog.\label{fig:paa32}}
\end{figure*}

\begin{figure*}[ht]
\centering
\includegraphics[scale=0.33]{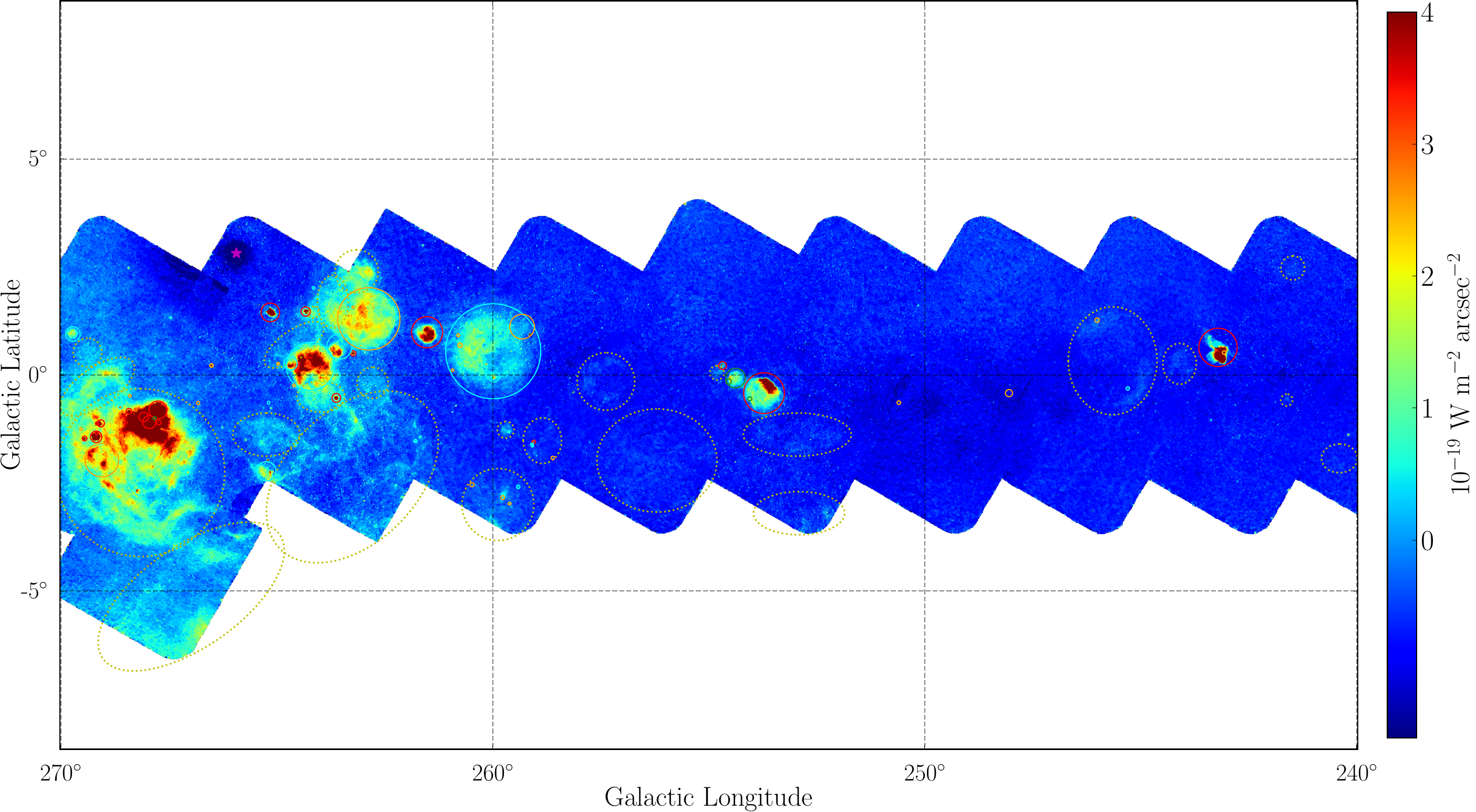}
\caption{Continuum-subtracted MIPAPS Pa$\alpha$ line image of the Galactic plane for $\ell = 240\arcdeg$--$270\arcdeg$. The solid circles indicate 65 Pa$\alpha$ sources corresponding to {\it WISE} \ion{H}{2} region sources listed in Table \ref{table:mipaps}, with colors representing source types: red for ``Known'', cyan for ``Candidate'', green for ``Group'', and orange for ``Radio Quiet'' {\it WISE} sources. The yellow dotted circles and ellipses mark 25 extended Pa$\alpha$ sources with no counterparts in the {\it WISE} \ion{H}{2} region catalog. The magenta star symbol denotes HD 78647, which causes extended dark areas around them.\label{fig:paa33}}
\end{figure*}

\begin{figure*}[ht]
\centering
\includegraphics[scale=0.33]{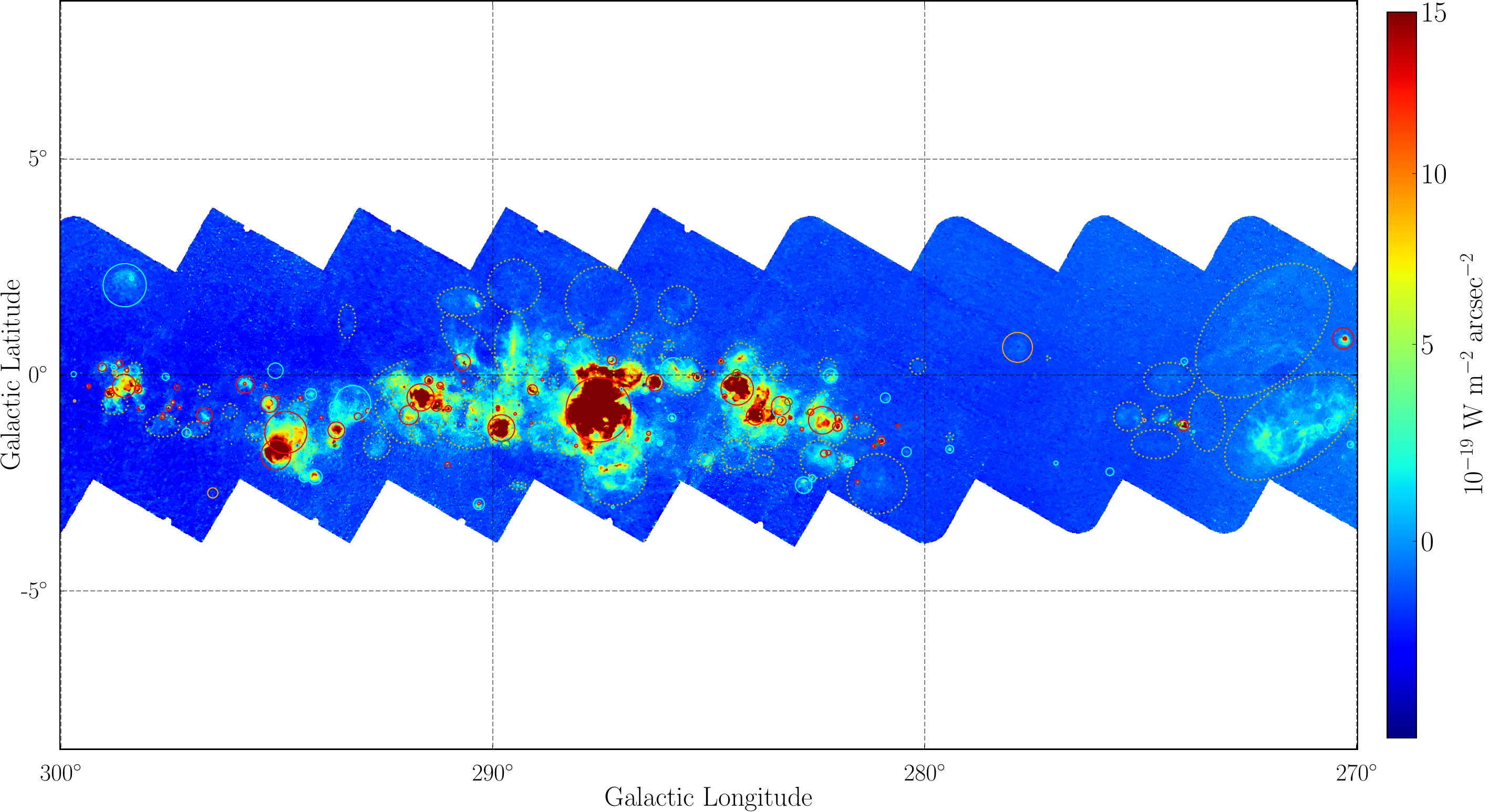}
\caption{Continuum-subtracted MIPAPS Pa$\alpha$ line image of the Galactic plane for $\ell = 270\arcdeg$--$300\arcdeg$. The solid circles indicate 166 Pa$\alpha$ sources corresponding to {\it WISE} \ion{H}{2} region sources listed in Table \ref{table:mipaps}, with colors representing source types: red for ``Known'', cyan for ``Candidate'', green for ``Group'', and orange for ``Radio Quiet'' {\it WISE} sources. The yellow dotted circles and ellipses mark 83 extended Pa$\alpha$ sources with no counterparts in the {\it WISE} \ion{H}{2} region catalog.\label{fig:paa41}}
\end{figure*}

\begin{figure*}[ht]
\centering
\includegraphics[scale=0.33]{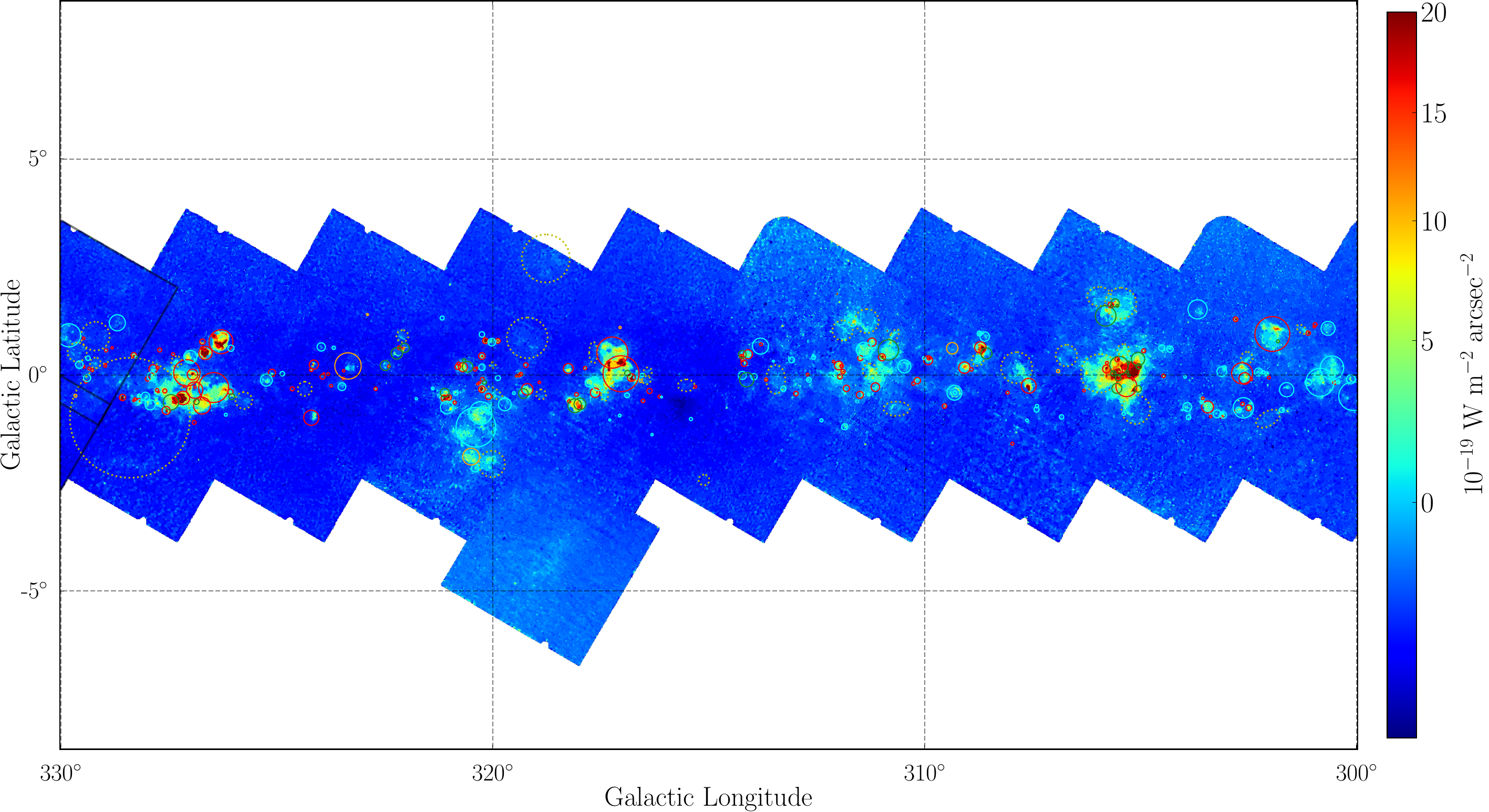}
\caption{Continuum-subtracted MIPAPS Pa$\alpha$ line image of the Galactic plane for $\ell = 300\arcdeg$--$330\arcdeg$. The thick solid rectangles represent MIPAPS fields with data obtained from highly-shadowed observations. The solid circles indicate 296 Pa$\alpha$ sources corresponding to {\it WISE} \ion{H}{2} region sources listed in Table \ref{table:mipaps}, with colors representing source types: red for ``Known'', cyan for ``Candidate'', green for ``Group'', and orange for ``Radio Quiet'' {\it WISE} sources. The yellow dotted circles and ellipses mark 38 extended Pa$\alpha$ sources with no counterparts in the {\it WISE} \ion{H}{2} region catalog.\label{fig:paa42}}
\end{figure*}

\begin{figure*}[ht]
\centering
\includegraphics[scale=0.33]{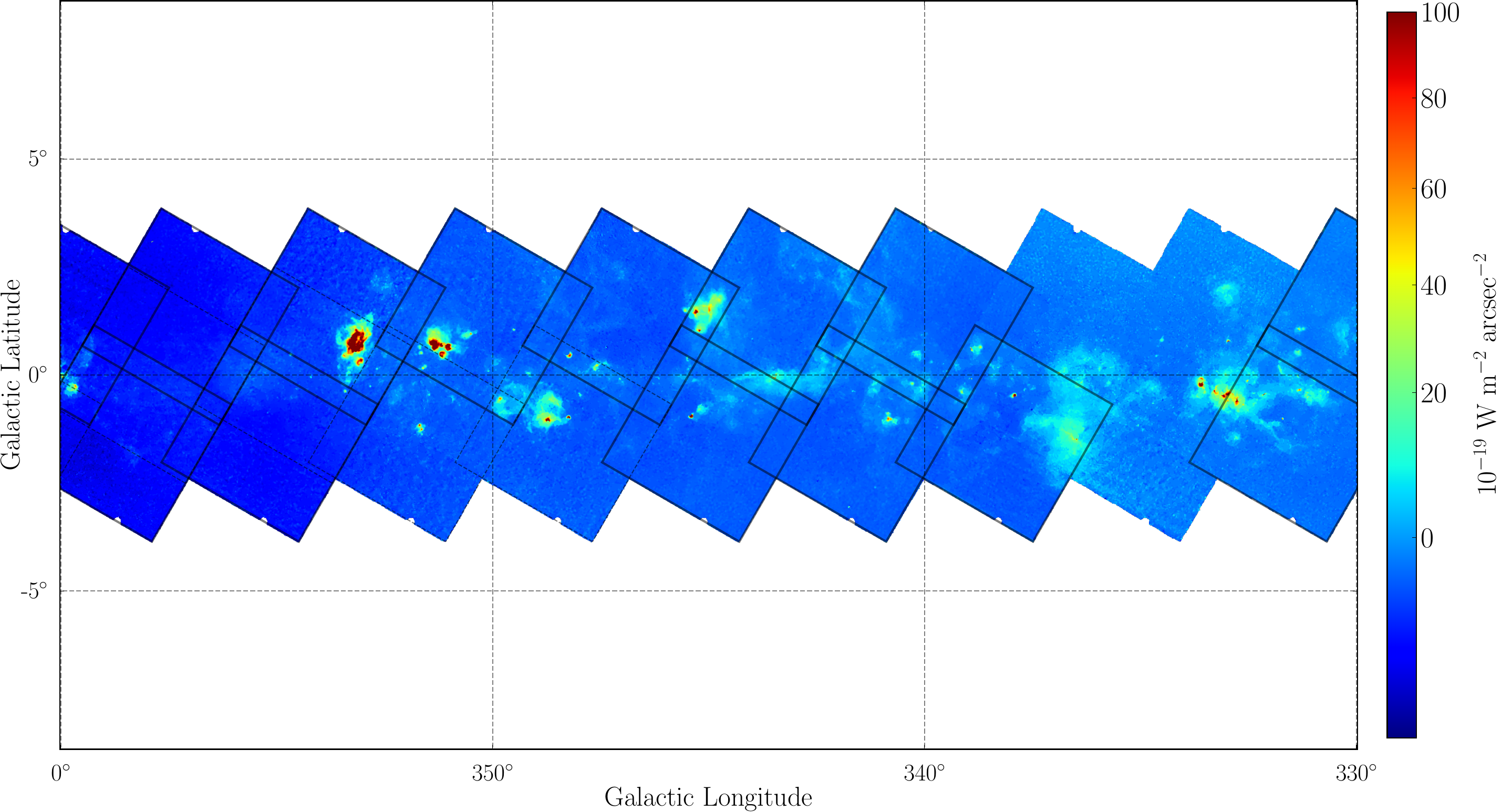}
\caption{Continuum-subtracted MIPAPS Pa$\alpha$ line image of the Galactic plane for $\ell = 330\arcdeg$--$360\arcdeg$. The thick solid and thin dashed rectangles represent MIPAPS fields with data obtained from highly- and slightly-shadowed observations, respectively.\label{fig:paa43}}
\end{figure*}

For the 241 individual MIPAPS fields, we created continuum-subtracted Pa$\alpha$ line images by subtracting the calibrated PAAC images from the calibrated PAAL images. We combined the 241 field images using the Montage software, resulting in the MIPAPS mosaic images of the entire Galactic plane. Figures \ref{fig:whole}(a)--(c) display the PAAL, PAAC, and continuum-subtracted Pa$\alpha$ line images of the entire Galactic plane. Additionally, twelve close-up Pa$\alpha$ images with $30\arcdeg$ longitude widths are shown in Figures \ref{fig:paa11}--\ref{fig:paa43}. The intensity of the Pa$\alpha$ image is expressed in in-band flux units (W m$^{-2}$) integrated over the PAAL transmission curve. As noted in Paper I, the Pa$\alpha$ dual continuum filter minimizes the difference in dust reddening effects between the PAAL and PAAC images. However, it introduces strong residual patterns around bright stars in the continuum-subtracted Pa$\alpha$ images due to significant differences in the point spread functions (PSFs) at PAAL and PAAC. To mitigate these residuals, we applied the same masking technique used in Paper I, masking out the stellar residuals and filling them with the median value from a neighboring annulus with a width equal to the masking radius. We selected 2MASS point sources with H or K$_{s}$ magnitudes of $\leq$7 as masking positions and determined the masking radii based on the individual 2MASS magnitudes. This method was applied to the Pa$\alpha$ image in Figure \ref{fig:whole}(c) and its close-up images. In the close-up images for the 2nd and 3rd Galactic quadrants ($90\arcdeg \leq \ell \leq 270\arcdeg$; Figures \ref{fig:paa21}--\ref{fig:paa33}), we extended the masking to include fainter 2MASS point sources with H or K$_{s}$ magnitudes of $\leq$8, as these regions contain relatively fewer point sources. For reference, we also provide twelve close-up Pa$\alpha$ images without masking (Figure \ref{fig:paa_origin}) and their corresponding images showing masked positions (Figure \ref{fig:paa_masking}) in the Appendix. The MIPAPS mosaic images created in this study, along with the calibrated images of the 241 individual fields, are available at \url{http://miris.kasi.re.kr/miris/pages/mipaps/}.

In Paper I, we subtracted a constant median background from each single-observation image to minimize differences in the lunar background between the PAAL and PAAC images. However, in this study, we did not account for lunar background variations, as other factors affecting the background levels of the Pa$\alpha$ images across the entire Galactic plane are more significant. These will be mentioned in Section \ref{subsec:image}.

We note that the MIPAPS Pa$\alpha$ survey data exhibit an observational artifact caused by a malfunction in the filter-wheel operation. In some observations, the MIRIS filter wheel was misaligned, producing an arc-shaped shadow along one edge of the field and resulting in localized sensitivity degradation. The degree of misalignment was estimated using MIRIS housekeeping data, which recorded the filter-wheel position during each observation. A total of 300 PAAL and 303 PAAC observations were identified as affected by this issue. The sensitivity degradation pattern within a field could be assessed if unaffected data taken with the same filter were also available for that field. For instance, in observations showing a 20\% reduction at the most shadowed edge, approximately 20\% of the field area exhibited a reduction greater than 5\%. In cases where the reduction reached 50\%, roughly 20\% and 50\% of the field area showed reductions exceeding 10\% and 5\%, respectively. However, the limited availability of unaffected counterparts made it difficult to establish consistent degradation trends as a function of the degree of misalignment. Consequently, using the relationship between brightness reduction at the most shadowed edge and the associated degree of misalignment --- derived from fields with corresponding unaffected data --- we estimated the expected sensitivity reduction for affected observations lacking such counterparts and classified them into two groups: highly-shadowed (with brightness reduction $\geq$20\%) and slightly-shadowed ($<$20\%) at the most shadowed edge, relative to unaffected observations.

For 175 MIPAPS fields with both unaffected PAAL and PAAC observations available, the affected data sets were excluded to obtain clean Pa$\alpha$ images. However, the remaining 66 MIPAPS field images were constructed using data from one or more highly- or slightly-shadowed observations. These are indicated by thick solid or thin dashed rectangles, respectively, in Figures \ref{fig:whole}(a) and (b). The continuum-subtracted Pa$\alpha$ line images for the 66 shadowed MIPAPS fields are also denoted by thick solid or thin dashed rectangles in Figures \ref{fig:paa11}--\ref{fig:paa21} and \ref{fig:paa42}--\ref{fig:paa43}.

\subsection{IPHAS and VPHAS+ H$\alpha$ Survey Data} \label{subsec:ha_data}

We used the IPHAS and VPHAS+ H$\alpha$ survey data to compare the MIPAPS Pa$\alpha$ images with the corresponding H$\alpha$ images. Since well-calibrated IPHAS data were presented by \citet{2014MNRAS.444.3230B}, we also utilized them for photometric analysis of the Pa$\alpha$ emission-line sources. The IPHAS H$\alpha$ survey covers the entire northern Galactic plane within $29\arcdeg \la \ell \la 215\arcdeg$ and $-5\arcdeg \la b \la +5\arcdeg$. Using the narrow-band H$\alpha$ and broad-band $r$ filter data, and following the same method described in Paper I, we created a continuum-subtracted IPHAS H$\alpha$ image with a pixel size of $\sim$5$\arcsec$ for the Galactic plane spanning $90\arcdeg \leq \ell \leq 215\arcdeg$. Stellar residuals in the continuum-subtracted H$\alpha$ image were initially masked out based on the IPHAS source catalog \citep{2014MNRAS.444.3230B}. For the Pa$\alpha$ sources selected for H$\alpha$ flux measurements, any remaining stellar residuals around individual sources were further masked out until the H$\alpha$ photometric results were no longer significantly affected.

The VPHAS+ H$\alpha$ survey covers the entire southern Galactic plane within $-150\arcdeg \la \ell \la 40\arcdeg$ and $-5\arcdeg \la b \la +5\arcdeg$ \citep{2014MNRAS.440.2036D}. Using the narrow-band H$\alpha$ and Sloan $r$ broad-band filter data, we created a continuum-subtracted VPHAS+ H$\alpha$ image for the Galactic plane spanning $210\arcdeg \leq \ell \leq 330\arcdeg$, following a method similar to that used for the continuum-subtracted IPHAS H$\alpha$ image. Low-confidence pixels with values below 70 in the confidence maps \citep{2014MNRAS.440.2036D} were masked out. Before merging the images using the Montage software, we subtracted the median background value from each observational image to reduce variations in the sky background. Since the VPHAS+ pixel size of $0\arcsec.21$ is too small to construct a large mosaic image efficiently, we created the VPHAS+ H$\alpha$ image with a pixel size of $\sim$5$\arcsec$ by binning 25 $\times$ 25 pixels. Due to a photometric issue related to the radial velocity dependence of the VPHAS+ H$\alpha$ flux \citep{2014MNRAS.440.2036D}, we used the VPHAS+ H$\alpha$ image only for visual inspections. Nevertheless, the pixel values were expressed in calibrated flux units of W m$^{-2}$ \AA{$^{-1}$}. We applied the photometric zero-point values provided in the image headers and adopted the mean profile of the VPHAS+ H$\alpha$ filter along with the Vega H$\alpha$ magnitude of 0.03 mag \citep{2014MNRAS.440.2036D}. For the Sloan $r$ broad-band filter profile, we assumed it to be the same as that of the IPHAS broad-band $r$ filter. The H$\alpha$ survey data used in this study are summarized in Table \ref{table:halpha_survey}, and the continuum-subtracted mosaic images created from these data are also available at \url{http://miris.kasi.re.kr/miris/pages/mipaps/}.

\begin{deluxetable*}{ccccc}
\tablewidth{0pt}
\tablecaption{H$\alpha$ Survey Data\label{table:halpha_survey}}
\tablehead{
\colhead{Survey} & \colhead{Filter used} & \colhead{Longitude used} & \colhead{Binning size used} & \colhead{Data access}
}
\startdata
IPHAS  & H$\alpha$, $r$ & $90\arcdeg < \ell < 215\arcdeg$ & 4\arcsec.95 (15 $\times$ 15 pixels) & \url{https://www.iphas.org/} \\
VPHAS+ & H$\alpha$, $r$ & $210\arcdeg < \ell < 330\arcdeg$ & 5\arcsec.25 (25 $\times$ 25 pixels) & \url{https://www.vphasplus.org/} \\
\enddata
\end{deluxetable*}

\section{Results} \label{sec:result}

\subsection{Pa$\alpha$ Line Images of the Galactic Plane} \label{subsec:image}

Using the Montage mosaic tool, we combined the images of the 241 MIPAPS fields to create the PAAL, PAAC, and continuum-subtracted Pa$\alpha$ line images for the entire Galactic plane, as shown in Figures \ref{fig:whole}(a)--(c). We did not apply the ``background match'' option in the Montage software to preserve the initial background level variations. However, the results obtained with the ``background match'' option were not found to be significantly different. Figures \ref{fig:paa11}--\ref{fig:paa43} present close-up Pa$\alpha$ images with $30\arcdeg$ longitude widths, generated using the Montage software. The projections are centered at ($\ell$, $b$) = ($15\arcdeg + n \cdot 30\arcdeg$, $0\arcdeg$), where $n = 0, 1, \cdots, 11$. Due to significant background level variations along the Galactic plane, we adjusted the color scales of the individual close-up images to enhance the visibility of Pa$\alpha$ features. Numerous Pa$\alpha$ features are clearly visible in all the close-up images, although some fields exhibit unmatched or noisy backgrounds.

\begin{figure}[ht]
\centering
\includegraphics[scale=0.28]{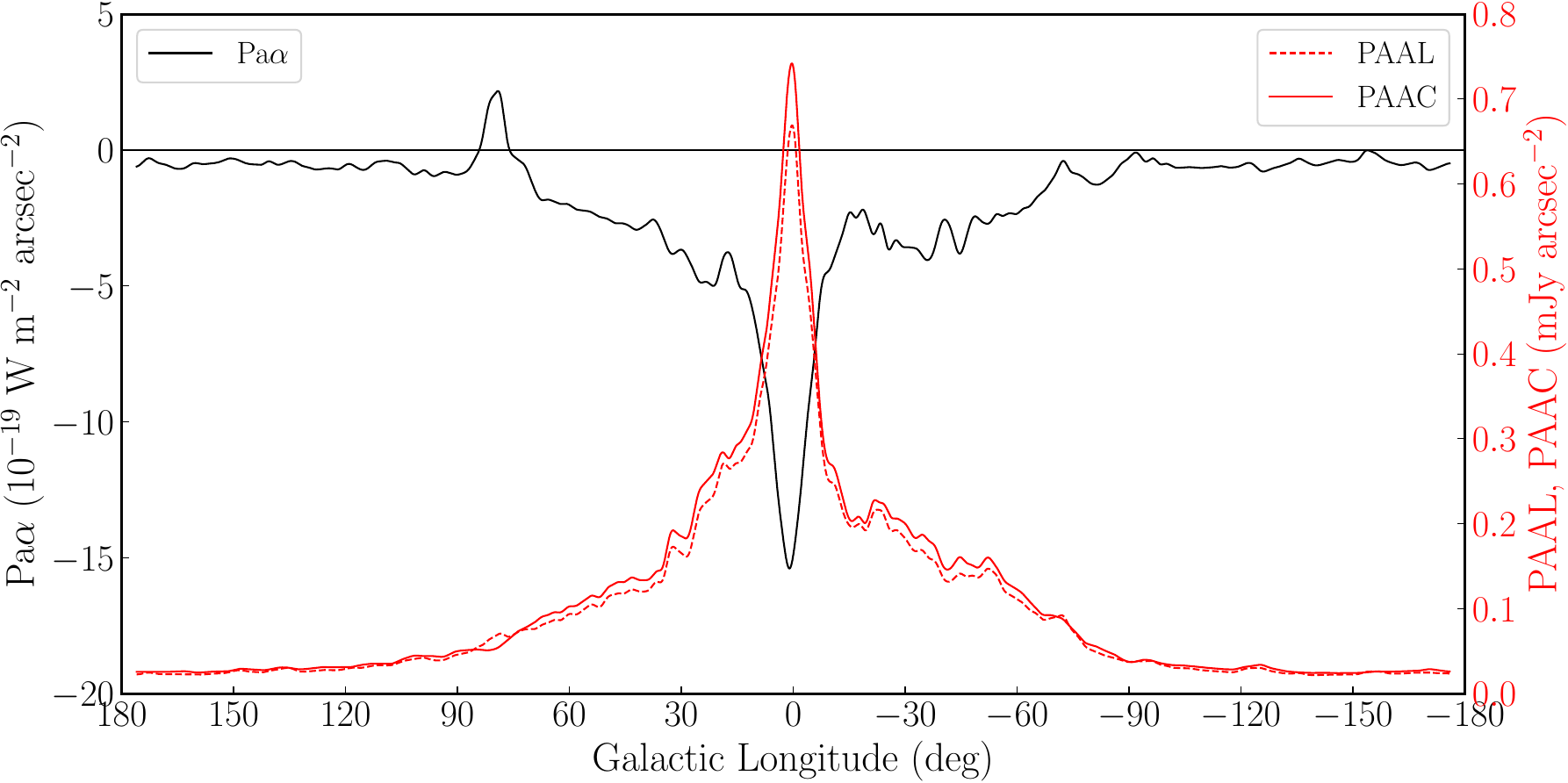}
\caption{Galactic longitude profiles of the median background levels for the PAAL, PAAC, and continuum-subtracted Pa$\alpha$ line images.\label{fig:profile}}
\end{figure}

Figure \ref{fig:whole}(c) shows that the overall level in the continuum-subtracted Pa$\alpha$ line image gradually decreases toward the Galactic center. To examine this trend in more detail, Figure \ref{fig:profile} presents the Galactic longitude profiles of the median background values in the PAAL, PAAC, and continuum-subtracted Pa$\alpha$ line images. The background level in the Pa$\alpha$ line image exhibits a strong anticorrelation with that in the PAAC image. One possible explanation is that the PSF in the PAAC image is more extended than that in the PAAL image, likely due to the double-layer structure of the MIRIS dual continuum filter. This PSF mismatch between the PAAL and PAAC images results in pronounced dark regions surrounding bright stars in the continuum-subtracted Pa$\alpha$ images. Such dark areas are observed near ($\ell$, $b$) = ($68\arcdeg.5, 3\arcdeg.2$), ($189\arcdeg.2, 2\arcdeg.8$), and ($266\arcdeg.7, 2\arcdeg.5$) in Figures \ref{fig:paa13}, \ref{fig:paa31}, and \ref{fig:paa33}, respectively. These dark areas are caused by bright stars, denoted by magenta star symbols in the figures: HD 187796, HD 44478, HD 42995, and HD 78647, with H magnitudes of $-$1.93, $-$1.68, $-$1.22, and $-$1.31, respectively. In particular, the overlapping dark regions associated with HD 44478 and HD 42995 together cover nearly the entire MIRIS field of view ($3\arcdeg.67\times3\arcdeg.67$), as shown in Figure \ref{fig:paa31}. We suggest that this effect may be related to the globally negative background levels observed in the continuum-subtracted Pa$\alpha$ images, which become increasingly prominent toward the Galactic center, where the density of bright stars is higher.

Bright stars located just outside, but near the boundaries of the observational fields, caused additional types of artifact features in the continuum-subtracted Pa$\alpha$ images. These artifacts typically appear as sharp, arc-like features centered on the stars, extending up to a few degrees away. Notable examples are found around ($\ell$, $b$) = ($86\arcdeg.8, -6\arcdeg.0$), ($105\arcdeg.5, 6\arcdeg.0$), ($139\arcdeg.3, -3\arcdeg.0$), ($164\arcdeg.5, 2\arcdeg.5$), and ($185\arcdeg.0, 3\arcdeg.0$) in Figures \ref{fig:paa13}--\ref{fig:paa31}. These artifacts are caused by bright stars, denoted by black star symbols in the figures: HD 203712, HD 209772, HD 17506, HD 34269, and HD 42272, with H magnitudes of $-$0.14, 0.37, 0.16, 0.28, and 1.24, respectively.

As indicated by rectangles in Figures \ref{fig:whole}(a) and (b), the 66 MIPAPS fields affected by filter-wheel misalignment are located within $-32\arcdeg.7 \leq \ell \leq 101\arcdeg.1$, except for one field at ($\ell$, $b$) = ($188\arcdeg, 4\arcdeg$). Notably, the 33 highly-shadowed fields (denoted by thick solid rectangles) are distributed across $-32\arcdeg.7 \leq \ell \leq 75\arcdeg.5$, including the Galactic center. As mentioned in Section \ref{subsec:pa_data}, the highly-shadowed field data resulted in sensitivity degradation exceeding 5\% over more than 20\% of the field area. For most of these fields, detailed degradation patterns could not be determined due to the absence of unaffected reference data obtained with the same filter. However, it is known that the shadowed edges in most affected fields are located along the sides nearer to the Galactic plane at $b = 0\arcdeg$. Although the shadowing effects are not visually apparent in the Pa$\alpha$ mosaic images shown in Figures \ref{fig:paa11}--\ref{fig:paa21} and \ref{fig:paa43}, we excluded the data for $-30\arcdeg \leq \ell \leq 90\arcdeg$, which include many highly-shadowed fields, when creating the MIPAPS Pa$\alpha$ catalog and performing photometry on Pa$\alpha$ sources. Instead, we briefly mention several Pa$\alpha$ features observed in this region.

Specifically, we compared the Pa$\alpha$ images within $-30\arcdeg \leq \ell \leq 90\arcdeg$ with the all-sky H$\alpha$ image from \citet{2003ApJS..146..407F}. Near the Galactic center, several Pa$\alpha$ features with angular sizes of approximately 2$\arcdeg.5$ were observed (see Figures \ref{fig:paa11} and \ref{fig:paa43}). These features exhibit a global resemblance to their counterparts in the H$\alpha$ image; however, their detailed morphologies differ significantly from the H$\alpha$ structures. The well-known large nebular regions spanning $5\arcdeg \leq \ell \leq 20\arcdeg$ display numerous bright Pa$\alpha$ features, as shown in Figure \ref{fig:paa11}. Notably, the Pa$\alpha$ features observed at ($\ell$, $b$) = ($10\arcdeg.2, -0\arcdeg.4$), ($12\arcdeg.8, -0\arcdeg.2$), ($18\arcdeg.2, -0\arcdeg.3$), and ($18\arcdeg.9, -0\arcdeg.3$) are almost invisible in the H$\alpha$ image. The region between $20\arcdeg \leq \ell \leq 60\arcdeg$ is heavily obscured by the Rift clouds, which explains the scarcity of bright H$\alpha$ features in this region. Nevertheless, numerous Pa$\alpha$ features were observed, as shown in Figures \ref{fig:paa11} and \ref{fig:paa12}. Strong Pa$\alpha$ features near ($\ell$, $b$) = ($49\arcdeg.1, -0\arcdeg.5$) are notably brighter than their counterparts in the H$\alpha$ image. In the region $60\arcdeg \leq \ell \leq 90\arcdeg$, one of the most prominent and complex regions in the Galactic plane, known as Cyg X, is located. As shown in Figure \ref{fig:paa13}, large and complex Pa$\alpha$ features were observed throughout the Cyg X region. Numerous filamentary Pa$\alpha$ features overlap near the Galactic plane and extend to Galactic latitudes greater than $5\arcdeg$. Their global morphologies are similar to their corresponding structures in the H$\alpha$ image; however, several Pa$\alpha$ local peaks near the Galactic plane within $75\arcdeg \leq \ell \leq 83\arcdeg$ have no counterparts in the H$\alpha$ image. In Figure \ref{fig:paa43}, covering $330\arcdeg \leq \ell \leq 360\arcdeg$, the Pa$\alpha$ features observed near ($\ell$, $b$) = ($331\arcdeg.0, -0\arcdeg.4$) and ($333\arcdeg.1, -0\arcdeg.4$) are significantly brighter than those in the H$\alpha$ image. Additionally, many Pa$\alpha$ features observed between $335\arcdeg \leq \ell \leq 354\arcdeg$ have bright H$\alpha$ counterparts.

\subsection{MIPAPS Catalog of Pa$\alpha$ Emission-line Sources} \label{subsec:catalog}

We compiled a catalog of Pa$\alpha$ emission-line sources in the Galactic plane within $90\arcdeg \leq \ell \leq 330\arcdeg$ (covering the 2nd, 3rd, and part of the 4th Galactic quadrants) based on the continuum-subtracted Pa$\alpha$ line images. First, we utilized the {\it WISE} \ion{H}{2} region catalog, a comprehensive catalog of Galactic \ion{H}{2} regions \citep{2014ApJS..212....1A}. Second, we identified Pa$\alpha$ sources that did not correspond to any entry in the {\it WISE} \ion{H}{2} region catalog and added them to the MIPAPS Pa$\alpha$ catalog. As a result, we present the MIPAPS catalog containing 1489 Pa$\alpha$ emission-line sources in Table \ref{table:mipaps} and show their Galactic distribution in Figure \ref{fig:whole}(d).

\begin{deluxetable*}{ccccccccccc}
\digitalasset
\tabletypesize{\scriptsize}
\tablewidth{0pt}
\tablecaption{MIPAPS Catalog of Pa$\alpha$ Emission-line Sources\label{table:mipaps}}
\tablehead{
\colhead{MIPAPS Name\tablenotemark{a}} & \multicolumn{2}{c}{Radius\tablenotemark{b}} & \colhead{Position Angle\tablenotemark{b}} & \colhead{ID\tablenotemark{c}} &
\colhead{{\it WISE} Name} & \colhead{H$\alpha$ Detection\tablenotemark{d}} & \colhead{SIMBAD Object} &
\colhead{Pa$\alpha$ Flux} & \colhead{H$\alpha$ Flux\tablenotemark{e}} & \colhead{$E(\bv)$\tablenotemark{e}} \\
\colhead{} & \multicolumn{2}{c}{(arcsec)} & \colhead{(deg)} & \colhead{} & \colhead{} &
\colhead{} & \colhead{} & \colhead{(10$^{-14}$ W m$^{-2}$)} & \colhead{(10$^{-14}$ W m$^{-2}$)} & \colhead{(mag)}
}
\startdata
G117.63+02.28   &  750    & \nodata & \nodata & WK074  & G117.639+02.275   & Y       & Sh2-170        & 30.28 $\pm$ 0.18 &  94.31 $\pm$ 0.11 & 0.64 $\pm$ 0.00 \\
G117.69+04.11   &  120    & \nodata & \nodata & WQ006  & G117.684+04.106   & Y       & \nodata        &  0.55 $\pm$ 0.03 &   0.33 $\pm$ 0.00 & 1.51 $\pm$ 0.03 \\
G118.17+02.56   &  450    & \nodata & \nodata & WG017  & G118.167+02.564   & Y       & \nodata        & \nodata          &  \nodata          & \nodata         \\
G118.25+05.15   & 5800    & \nodata & \nodata & WG018  & G118.250+05.397   & Yp      & Sh2-171        & \nodata          &  \nodata          & \nodata         \\
G118.28+02.49   &  200    & \nodata & \nodata & WK075  & G118.276+02.490   & Y       & \nodata        &  0.78 $\pm$ 0.03 &   0.85 $\pm$ 0.01 & 1.19 $\pm$ 0.02 \\
G118.35+04.86   & 2600    & \nodata & \nodata & WK076  & G118.345+04.856   & Yp      & Sh2-171        & \nodata          &  \nodata          & \nodata         \\
G118.38+03.16   &  120    & \nodata & \nodata & WQ007  & G118.376+03.165   & Y       & \nodata        &  0.07 $\pm$ 0.02 &   0.03 $\pm$ 0.01 & 1.68 $\pm$ 0.27 \\
G118.59+02.83   &  200    & \nodata & \nodata & WK077  & G118.592+02.828   & Y       & \nodata        &  1.28 $\pm$ 0.04 &   0.80 $\pm$ 0.01 & 1.48 $\pm$ 0.02 \\
G118.62$-$01.33 &  240    & \nodata & \nodata & WK078  & G118.621$-$01.329 & Y       & Sh2-172        &  0.39 $\pm$ 0.03 &   0.34 $\pm$ 0.01 & 1.32 $\pm$ 0.04 \\
G118.73+02.93   &  120    & \nodata & \nodata & WK079  & G118.735+02.919   & Y       & \nodata        &  0.25 $\pm$ 0.02 &   0.10 $\pm$ 0.00 & 1.70 $\pm$ 0.05 \\
G118.85+01.60   & 1800    & \nodata & \nodata & MPE042 & \nodata           & Y       & Du 62          & \nodata          &  \nodata          & \nodata         \\
G119.45$-$00.92 & 1100    & \nodata & \nodata & WK080  & G119.437$-$00.916 & Y       & Sh2-173        & \nodata          &  \nodata          & \nodata         \\
G119.52+03.18   &  180    & \nodata & \nodata & WC021  & G119.521+03.182   & Y       & \nodata        &  0.23 $\pm$ 0.03 &   0.06 $\pm$ 0.01 & 1.97 $\pm$ 0.08 \\
G119.54+00.38   & 1500    & \nodata & \nodata & MPE043 & \nodata           & Y       & GAL 119.6+00.4 & 34.56 $\pm$ 0.46 & 103.50 $\pm$ 0.18 & 0.66 $\pm$ 0.01 \\
G120.15+03.38   &  120    & \nodata & \nodata & WC022  & G120.152+03.375   & N       & \nodata        &  0.18 $\pm$ 0.01 &  \nodata          & \nodata         \\
G120.26+01.69   & \nodata & \nodata & \nodata & MPP033 & \nodata           & Y       & BD+63 48       &  0.51 $\pm$ 0.02 &  \nodata          & \nodata         \\
G120.35+01.96   &  200    & \nodata & \nodata & WK081  & G120.348+01.968   & Y       & Sh2-175        &  1.49 $\pm$ 0.04 &   1.99 $\pm$ 0.01 & 1.08 $\pm$ 0.01 \\
G120.57$-$00.24 & 2000    & 1300    &  90     & MPE044 & \nodata           & Y       & Sh2-177        & \nodata          &  \nodata          & \nodata         \\
G121.85+01.41   & 2200    &  700    & 130     & MPE045 & \nodata           & Y       & \nodata        & \nodata          &  \nodata          & \nodata         \\
G121.98$-$00.94 & \nodata & \nodata & \nodata & MPP034 & \nodata           & Y       & BD+61 154      &  0.59 $\pm$ 0.01 &  \nodata          & \nodata         \\
G122.00$-$07.09 &  220    & \nodata & \nodata & WC023  & G122.002$-$07.084 & No data & \nodata        & \nodata          &  \nodata          & \nodata         \\
G122.08+01.90   & \nodata & \nodata & \nodata & MPP035 & \nodata           & Y       & HD 4004        &  2.76 $\pm$ 0.03 &  \nodata          & \nodata         \\
G122.33$-$01.21 & \nodata & \nodata & \nodata & MPP036 & \nodata           & Y       & EM* AS 6       &  0.17 $\pm$ 0.02 &  \nodata          & \nodata         \\
G122.62+00.10   &  600    & \nodata & \nodata & MPE046 & \nodata           & Y       & Sh2-180        &  2.79 $\pm$ 0.10 &   3.57 $\pm$ 0.05 & 1.11 $\pm$ 0.02 \\
G122.70+02.37   &  350    & \nodata & \nodata & MPE047 & \nodata           & Y       & Sh2-181        &  0.83 $\pm$ 0.05 &   3.71 $\pm$ 0.02 & 0.45 $\pm$ 0.03 \\
\enddata
\tablenotetext{a}{MIPAPS names are assigned based on the central positions of individual Pa$\alpha$ sources.}
\tablenotetext{b}{Radius values represent the approximate angular extents of the Pa$\alpha$ sources. For source with elliptical extents, two radius values are provided, corresponding to the semi-major and semi-minor axes of the ellipse, along with the specified position angle.}
\tablenotetext{c}{The labels ``WK'', ``WC'', ``WG'', and ``WQ'' denote Pa$\alpha$ sources corresponding to ``Known'', ``Candidate'', ``Group'', and ``Radio Quiet'' {\it WISE}
\ion{H}{2} region sources, respectively. On the other hand, ``MPE'' and ``MPP'' represent extended and point-like Pa$\alpha$ sources, respectively, with no counterparts in the {\it WISE} \ion{H}{2} region catalog. Sequential numbers are assigned based on Galactic longitude.}
\tablenotetext{d}{Visually inspected in continuum-subtracted IPHAS or VPHAS+ H$\alpha$ line images. ``Y'': visually detected; ``Yp'': visually detected but partially observed; ``Y(P+E)'': point source and extended feature are detected simultaneously; ``N'': not visually detected; ``No data'': not observed.}
\tablenotetext{e}{The H$\alpha$ fluxes may include contributions from the {[}\ion{N}{2}{]} $\lambda$$\lambda$6548, 6584 \AA{} lines; however, these were corrected during the $E(\bv)$ estimations.\\}
(The complete version of this table is available in a machine-readable format in the online journal.)
\end{deluxetable*}

\subsubsection{WISE \ion{H}{2} Region Sources} \label{subsec:wise_sources}

\citet{2014ApJS..212....1A} presented the {\it WISE} \ion{H}{2} region catalog containing a total of 8399 entries. Based on the characteristic mid-infrared morphology of \ion{H}{2} regions ({\it WISE} 12 $\mu$m emission surrounding {\it WISE} 22 $\mu$m emission), they identified \ion{H}{2} region candidates through both visual and automated searches. The \ion{H}{2} region candidates were classified into four categories. Sources with counterparts at radio recombination lines (RRLs) or the H$\alpha$ line were classified as ``Known,'' indicating confirmed \ion{H}{2} regions. Sources with only radio continuum counterparts were classified as ``Candidate,'' while sources spatially associated with known \ion{H}{2} region complexes were classified as ``Group.'' Sources classified as ``Radio Quiet'' had no counterparts at any of the radio continuum, RRLs, or H$\alpha$ lines. For this study, we used version 2.2 of the {\it WISE} \ion{H}{2} region catalog, available at \url{http://astro.phys.wvu.edu/wise/}. Within the MIPAPS fields in the Galactic plane spanning $90\arcdeg \leq \ell \leq 330\arcdeg$, the catalog contains a total of 2666 entries: 578 ``Known,'' 687 ``Candidate,'' 197 ``Group,'' and 1204 ``Radio Quiet'' sources.

For the 2666 individual {\it WISE} \ion{H}{2} region sources, we visually inspected the MIPAPS Pa$\alpha$ line images. The Pa$\alpha$ detections were determined by comparing the morphology of the Pa$\alpha$ features with the corresponding structures in the {\it WISE} 12 $\mu$m and 22 $\mu$m images. The {\it WISE} 12 $\mu$m and 22 $\mu$m images were constructed using data from the NASA/IPAC Infrared Science Archive\footnote{\url{http://irsa.ipac.caltech.edu/applications/wise/}}. When a corresponding Pa$\alpha$ feature was identified for a {\it WISE} \ion{H}{2} region source, we assigned a MIPAPS name based on the central position of the Pa$\alpha$ feature and determined its radius to approximate its angular extent. For simplicity, we also assigned ID names (WK, WC, WG, and WQ), corresponding to the {\it WISE} categories of ``Known,'' ``Candidate,'' ``Group,'' and ``Radio Quiet,'' respectively, as shown in Table \ref{table:mipaps}. Additionally, we inspected corresponding H$\alpha$ detections using IPHAS or VPHAS+ H$\alpha$ images, and the results are listed in Table \ref{table:mipaps}. We also searched for matching objects in the SIMBAD database\footnote{\url{https://simbad.cfa.harvard.edu/simbad/}}, although many of them were adapted from the {\it WISE} \ion{H}{2} region catalog. Table \ref{table:mipaps} contains a total of 902 Pa$\alpha$ sources corresponding to the {\it WISE} \ion{H}{2} region sources, and their central positions and angular sizes are indicated by solid circles in Figures \ref{fig:paa21}--\ref{fig:paa42}.

\begin{figure*}[ht]
\begin{centering}
\includegraphics[scale=0.43]{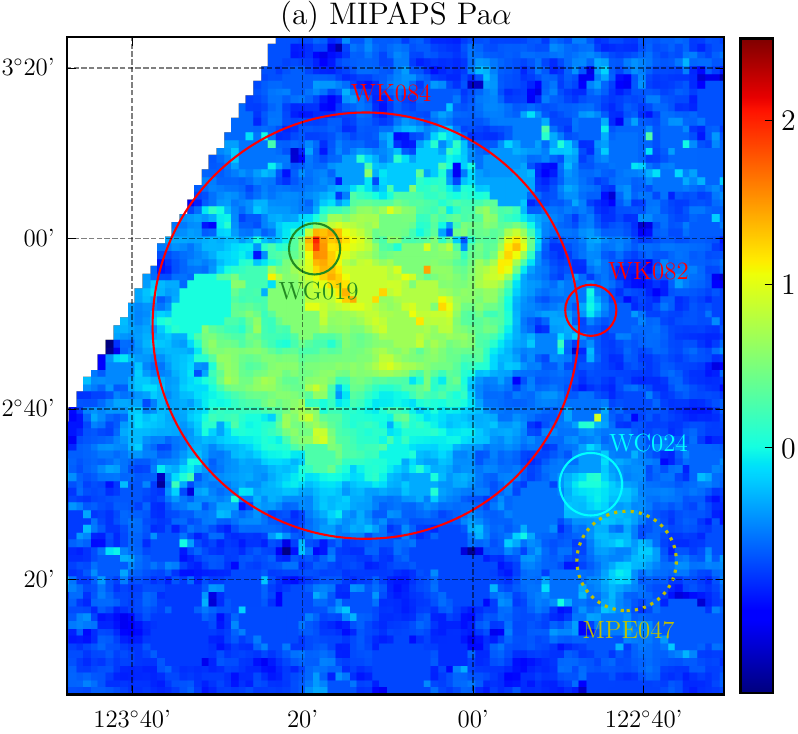}\includegraphics[scale=0.43]{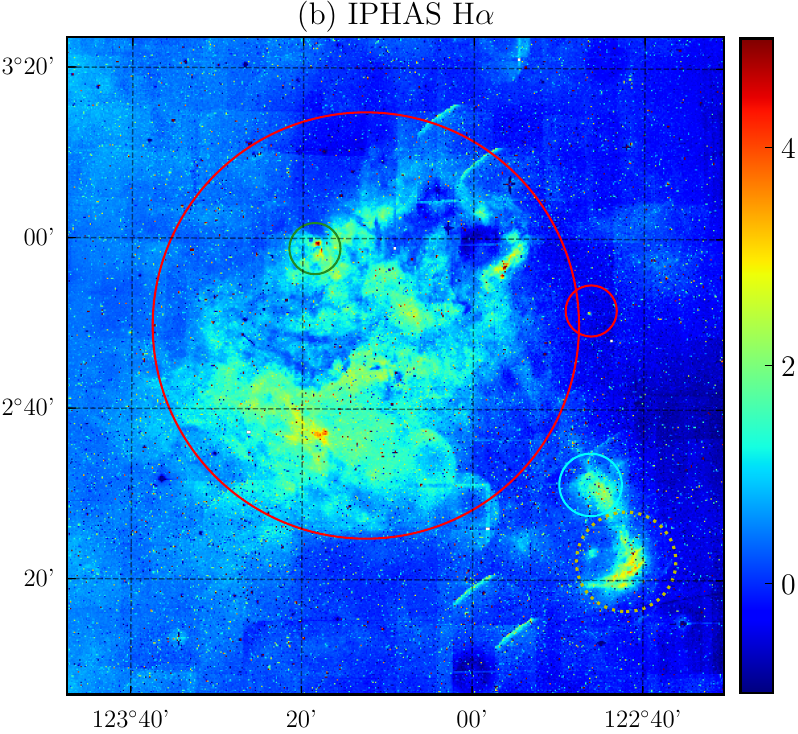}\includegraphics[scale=0.43]{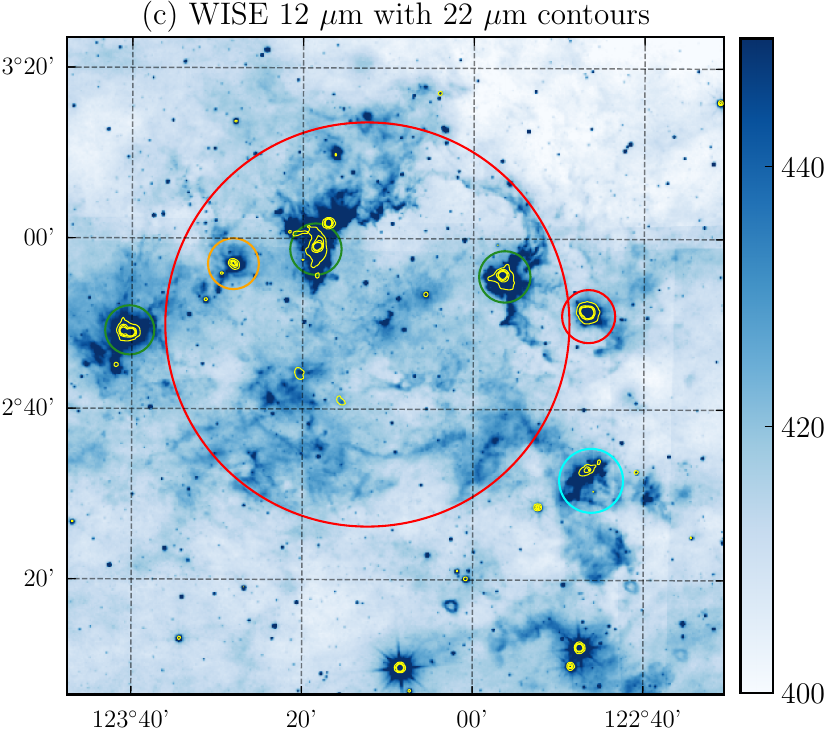}
\par\end{centering}
\begin{centering}
\includegraphics[scale=0.43]{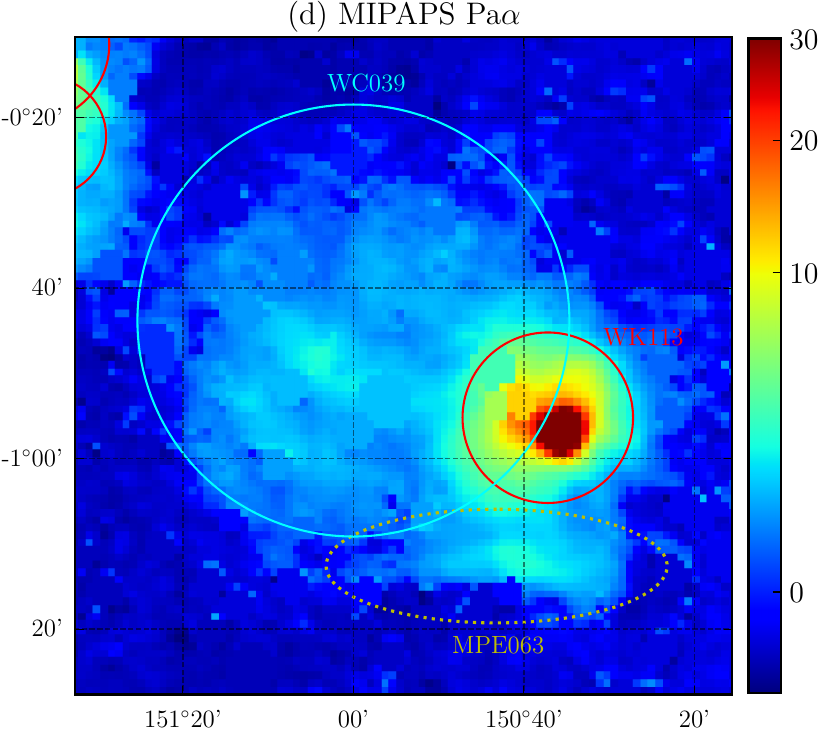}\includegraphics[scale=0.43]{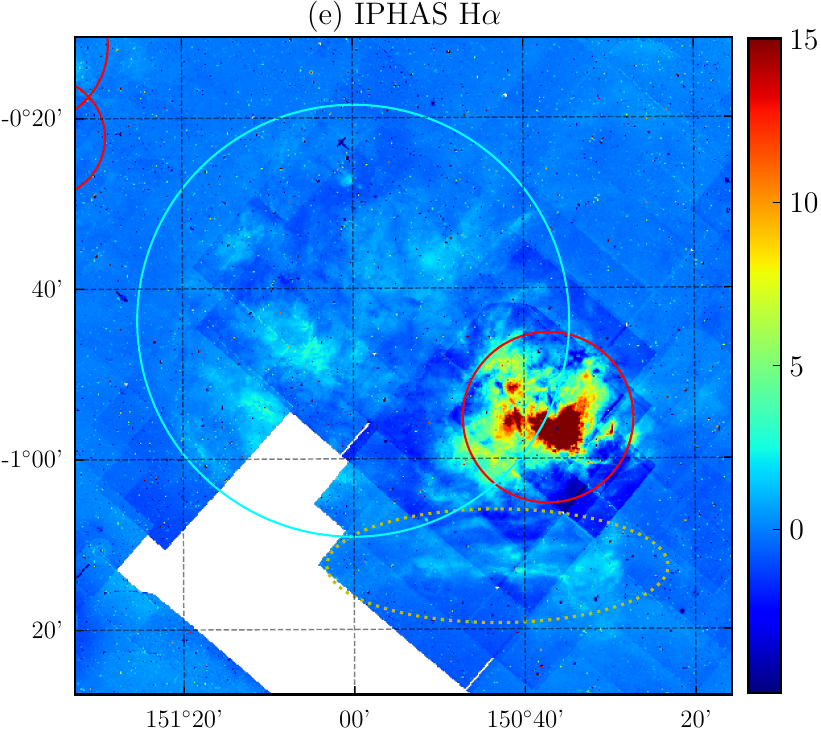}\includegraphics[scale=0.43]{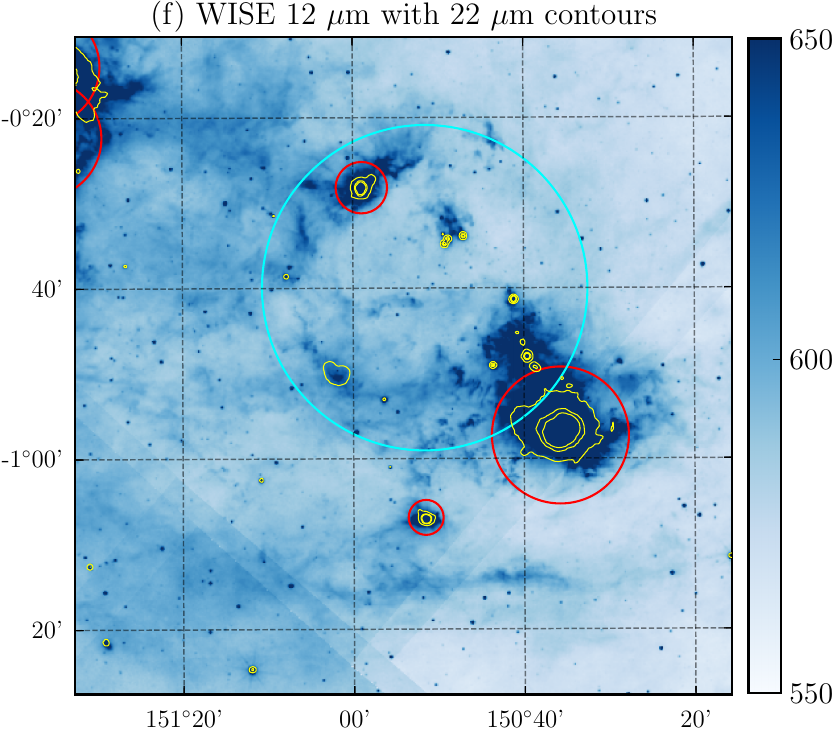}
\par\end{centering}
\begin{centering}
\includegraphics[scale=0.43]{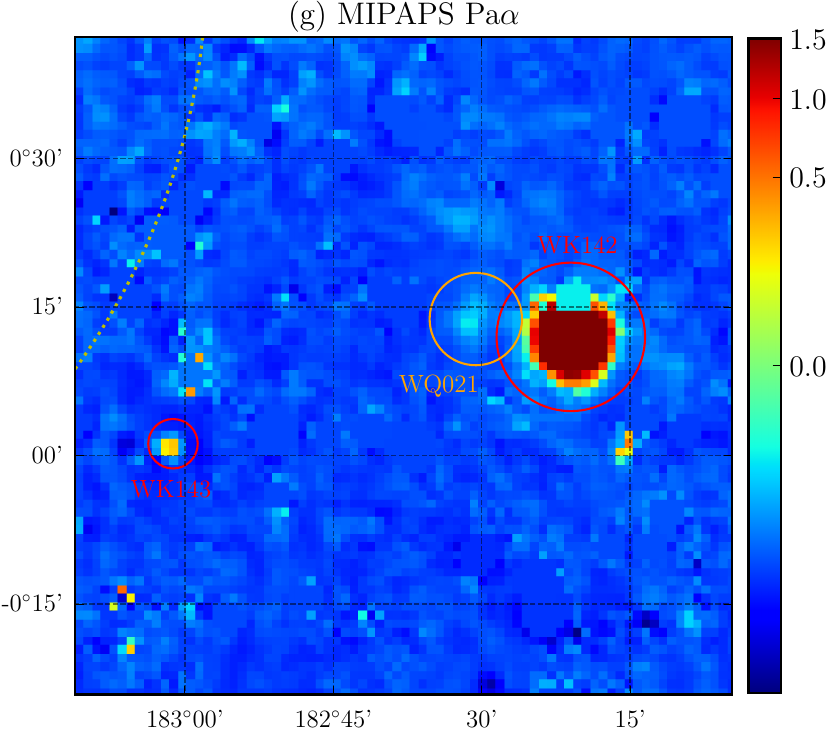}\includegraphics[scale=0.43]{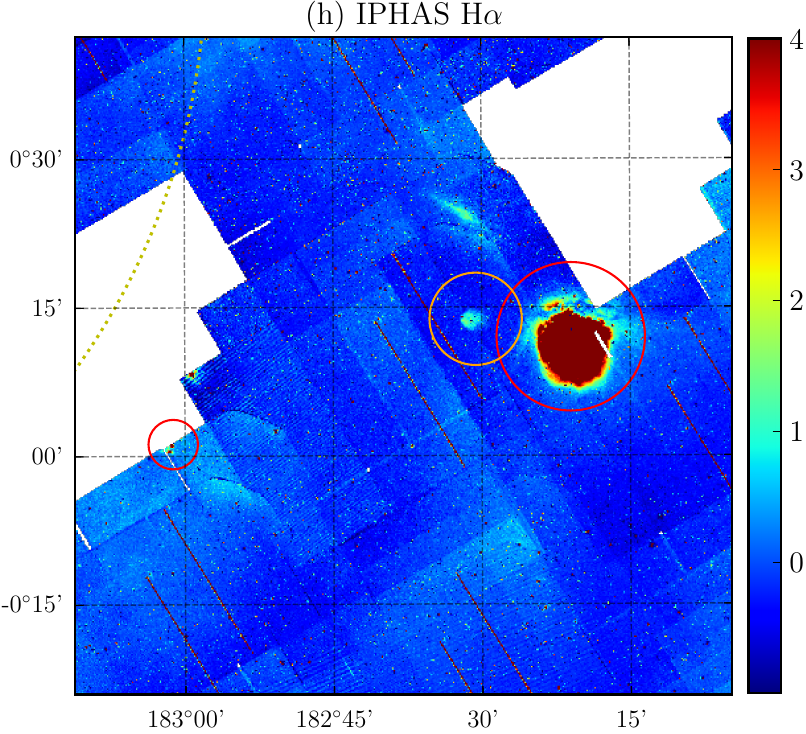}\includegraphics[scale=0.43]{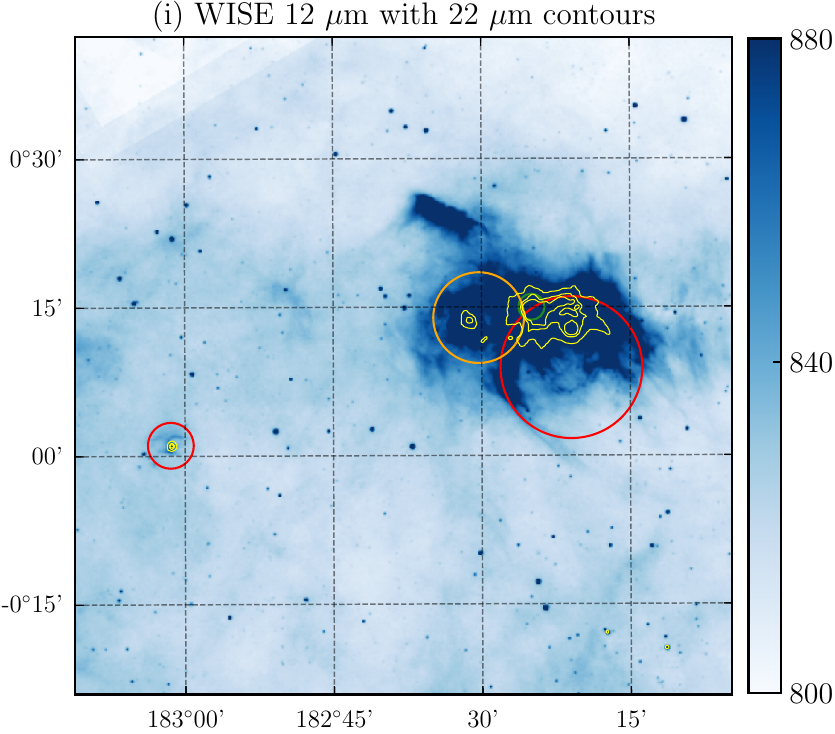}
\par\end{centering}
\caption{MIPAPS Pa$\alpha$ (left column), IPHAS H$\alpha$ (middle column), and {\it WISE} 12 $\mu$m (with 22 $\mu$m yellow contours, right column) images of three representative regions for visual inspection. The color bar units for the Pa$\alpha$ and H$\alpha$ images are 10$^{-19}$ W m$^{-2}$ arcsec$^{-2}$, while the units for the {\it WISE} 12 $\mu$m images are DN (Digital Number). The levels of the {\it WISE} 22 $\mu$m contours are 105, 115, 125, 135, and 145 DN in panel (c); 170, 190, and 210 DN in panel (f); and 270, 280, and 290 DN in panel (i). All colored circles and ellipses in the Pa$\alpha$ and H$\alpha$ images are identical to those in Figures \ref{fig:paa22}--\ref{fig:paa31}, with corresponding source names from the ``ID'' columns of Table \ref{table:mipaps}. In the {\it WISE} 12 $\mu$m images, the colored circles represent {\it WISE} \ion{H}{2} region sources, with their central positions and sizes derived from the {\it WISE} \ion{H}{2} region catalog \citep{2014ApJS..212....1A}.\label{fig:exam1}}
\end{figure*}

\begin{figure*}[ht]
\begin{centering}
\includegraphics[scale=0.43]{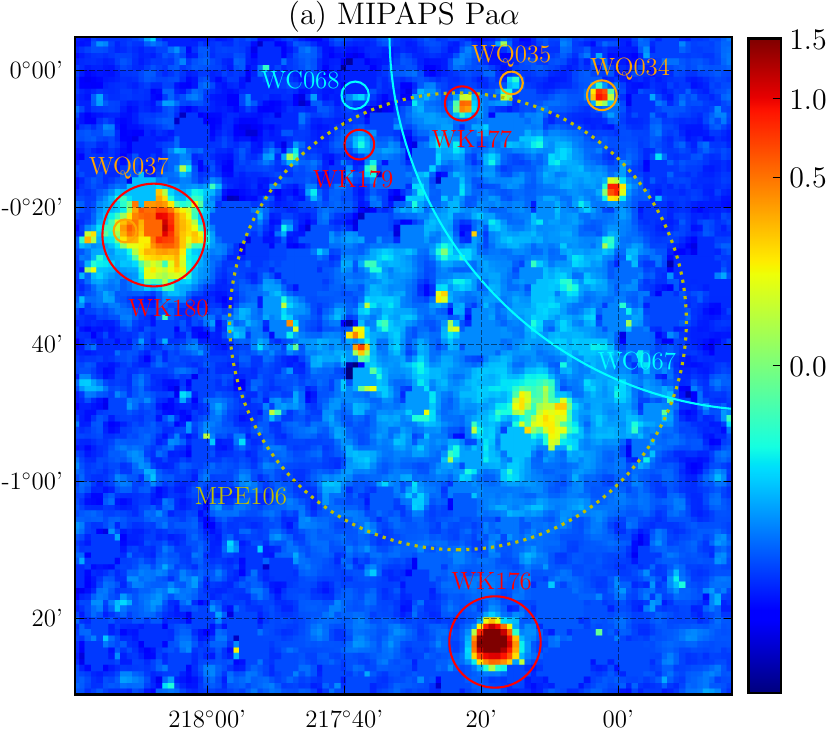}\includegraphics[scale=0.43]{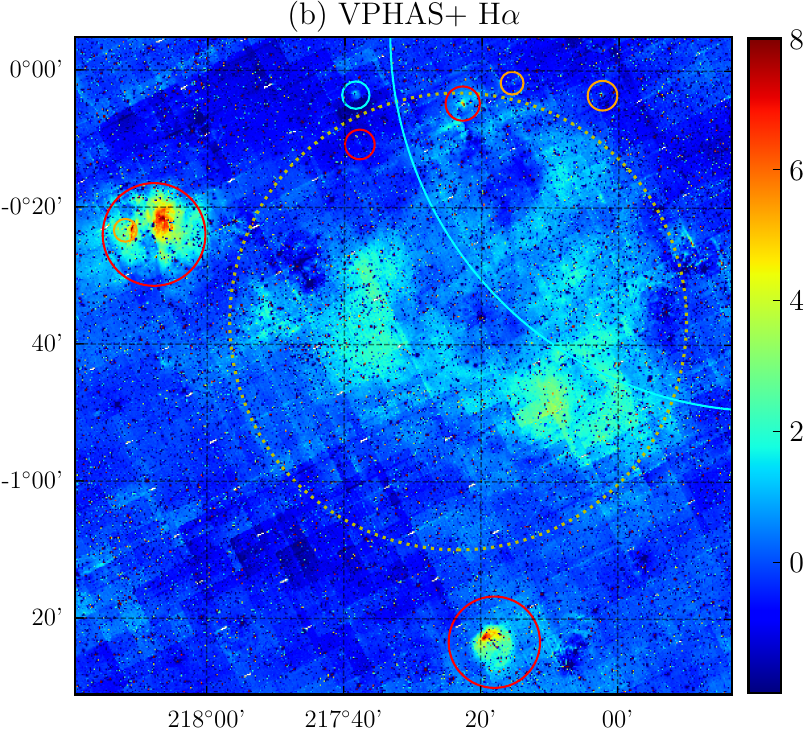}\includegraphics[scale=0.43]{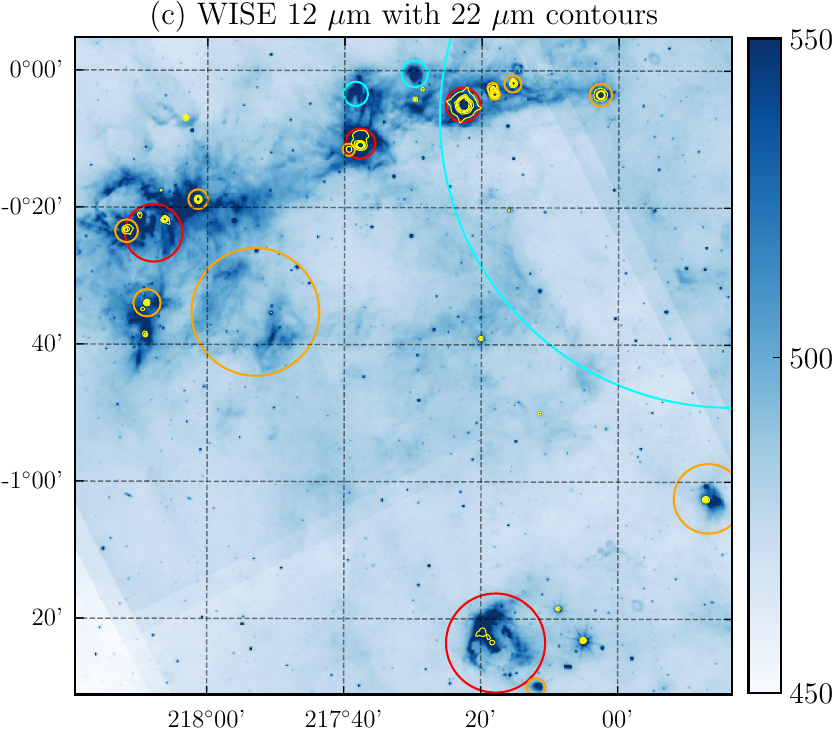}
\par\end{centering}
\begin{centering}
\includegraphics[scale=0.43]{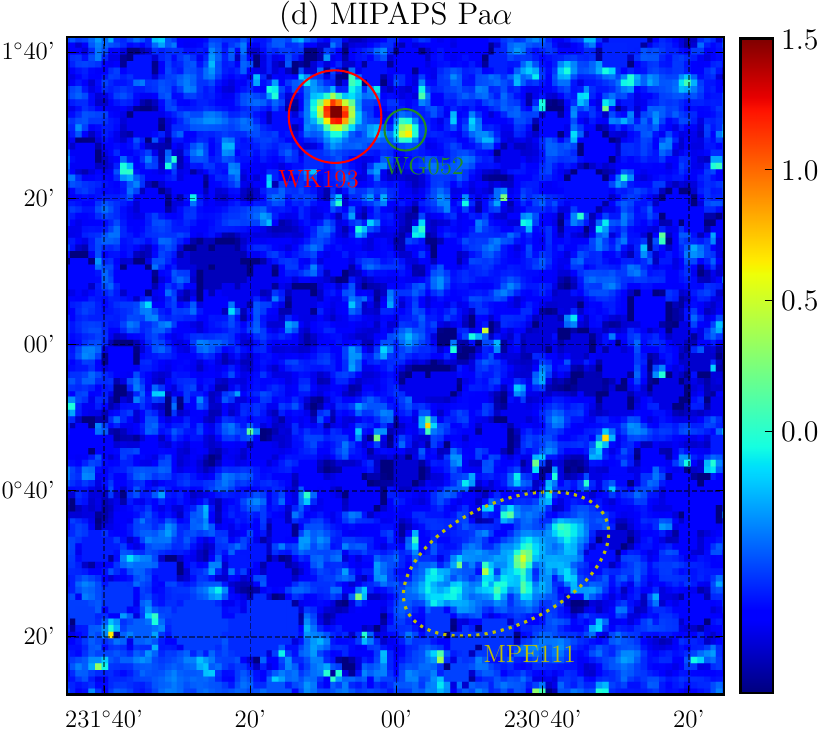}\includegraphics[scale=0.43]{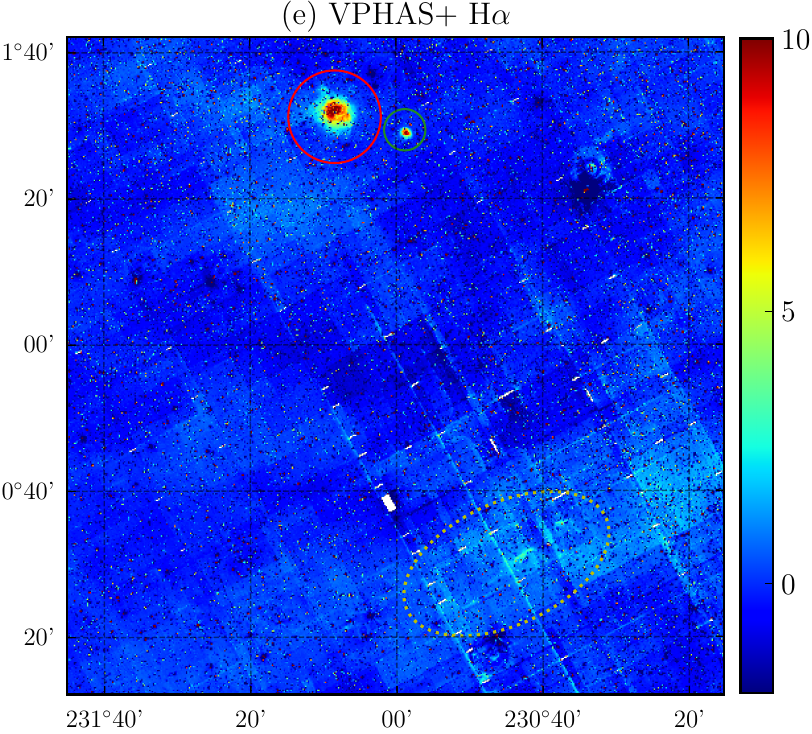}\includegraphics[scale=0.43]{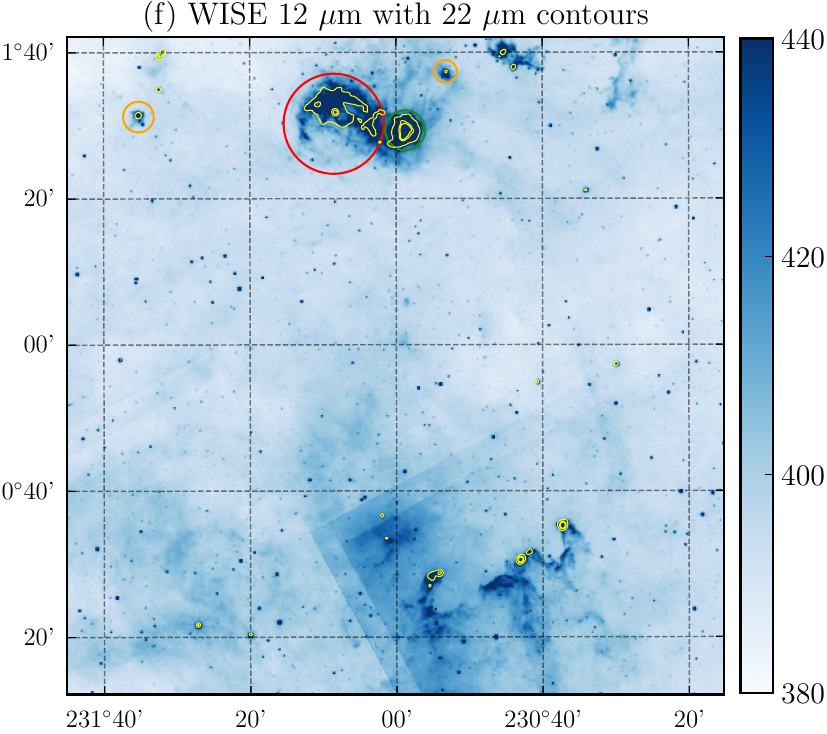}
\par\end{centering}
\begin{centering}
\includegraphics[scale=0.43]{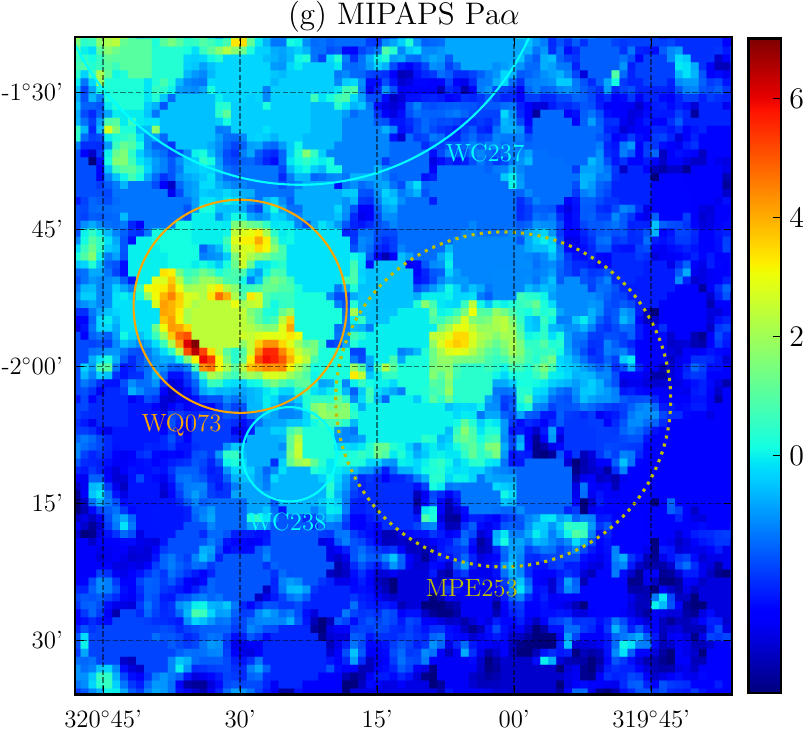}\includegraphics[scale=0.43]{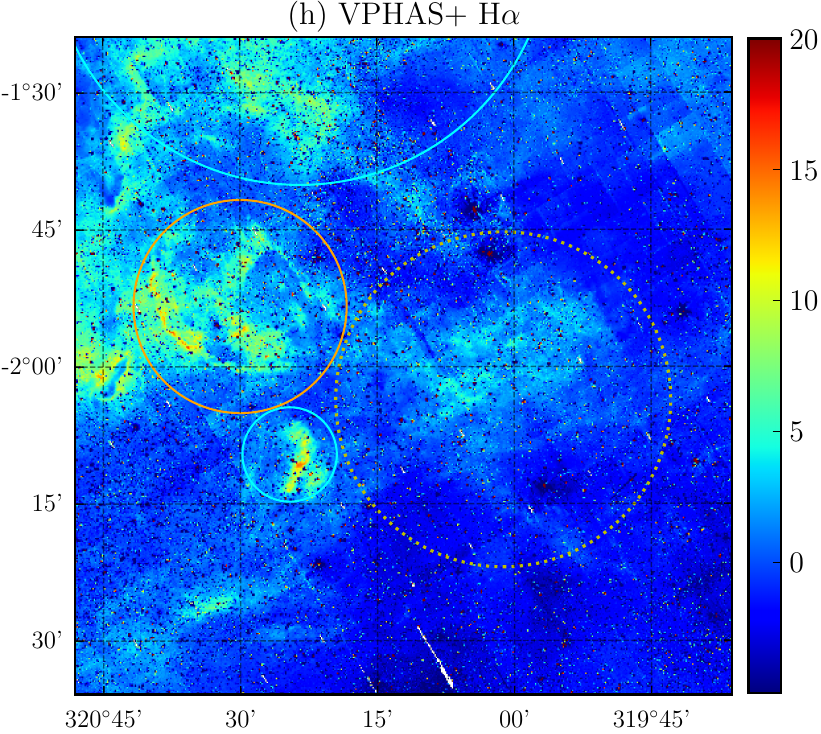}\includegraphics[scale=0.43]{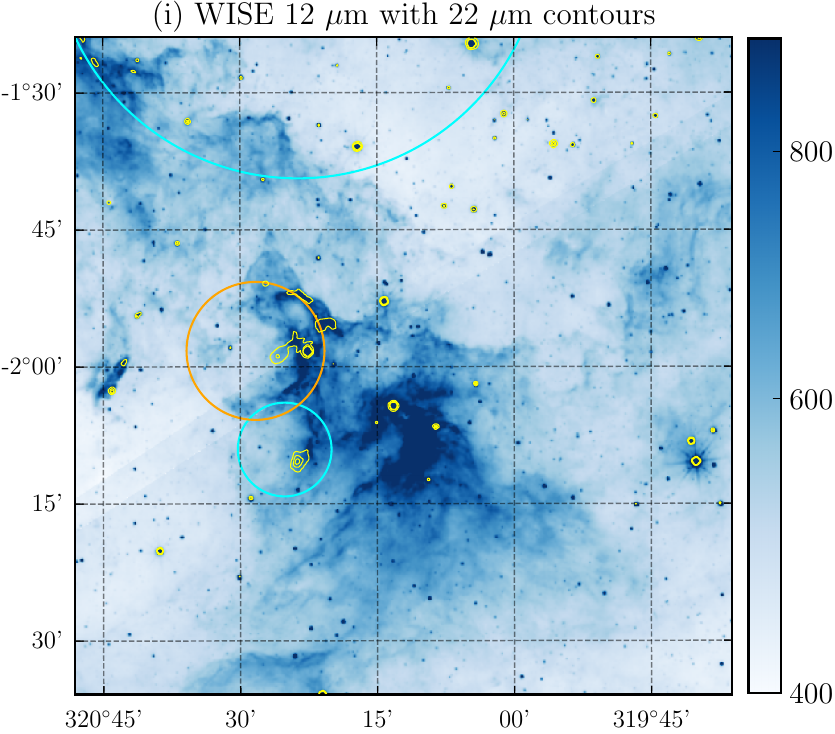}
\par\end{centering}
\caption{MIPAPS Pa$\alpha$ (left column), VPHAS+ H$\alpha$ (middle column), and {\it WISE} 12 $\mu$m (with 22 $\mu$m yellow contours, right column) images of three representative regions for visual inspection. The color bar units for the Pa$\alpha$ and H$\alpha$ images are 10$^{-19}$ W m$^{-2}$ arcsec$^{-2}$. while the units for the {\it WISE} 12 $\mu$m images are DN (Digital Number). The levels of the {\it WISE} 22 $\mu$m contours are 150, 170, 190, 210, and 230 DN in panel (c); 115, 125, and 135 DN in panel (f); and 150, 170, and 190 DN in panel (i). All colored circles and ellipses in the Pa$\alpha$ and H$\alpha$ images are identical to those in Figures \ref{fig:paa32} and \ref{fig:paa42}, with corresponding source names from the ``ID'' columns of Table \ref{table:mipaps}. In the {\it WISE} 12 $\mu$m images, the colored circles represent {\it WISE} \ion{H}{2} region sources, with their central positions and sizes derived from the {\it WISE} \ion{H}{2} region catalog \citep{2014ApJS..212....1A}.\label{fig:exam2}}
\end{figure*}

Figures \ref{fig:exam1} and \ref{fig:exam2} display six representative regions for visual inspection. In the {\it WISE} 12 $\mu$m images, the {\it WISE} \ion{H}{2} region sources are indicated by colored circles: red, cyan, green, and orange represent ``Known,'' ``Candidate,'' ``Group,'' and ``Radio Quiet'' sources, respectively. The corresponding Pa$\alpha$ sources are marked by solid circles in the MIPAPS Pa$\alpha$ and IPHAS/VPHAS+ H$\alpha$ images. The central positions and angular sizes of the Pa$\alpha$ sources can differ slightly from those of the {\it WISE} \ion{H}{2} region sources. In Figures \ref{fig:exam1} and \ref{fig:exam2}, WC039, WK113, WK180, and WQ073 exhibit larger angular sizes in the Pa$\alpha$ images compared to the {\it WISE} images. In Figures \ref{fig:exam1}(g)--(i), the central positions of the Pa$\alpha$ and H$\alpha$ features for WK142 align with a peak in the {\it WISE} 22 $\mu$m emission (indicated by yellow contours), whereas the central position of the {\it WISE} source appears to be influenced by the {\it WISE} 12 $\mu$m morphology, which extends toward the southern direction.

Table \ref{table:wise} lists 1764 {\it WISE} \ion{H}{2} region sources with no MIPAPS Pa$\alpha$ detection. In the continuum-subtracted Pa$\alpha$ line images, we examined whether individual source areas overlap with other bright Pa$\alpha$ sources or stellar residuals, which could explain the lack of Pa$\alpha$ detection. A total of 1107 sources were found to have some overlaps with either bright Pa$\alpha$ sources or stellar residuals. Additionally, in the IPHAS or VPHAS+ H$\alpha$ images, 195 {\it WISE} \ion{H}{2} region sources with no Pa$\alpha$ detection were found to have corresponding H$\alpha$ detections.

\begin{deluxetable*}{cccc}
\digitalasset
\tablewidth{0pt}
\tablecaption{{\it WISE} \ion{H}{2} Region Sources with No MIPAPS Pa$\alpha$ Detection\label{table:wise}}
\tablehead{
\colhead{{\it WISE} Name} & \colhead{Bright Source Overlap\tablenotemark{a}} &
\colhead{Stellar Residual Overlap\tablenotemark{b}} & \colhead{H$\alpha$ Detection\tablenotemark{c}}
}
\startdata
G298.924+00.473   & N & largely   & Y \\
G298.928$-$00.526 & N & N         & N \\
G298.991$-$00.443 & Y & N         & N \\
G299.013+00.129   & Y & N         & N \\
G299.153+00.008   & N & entirely  & N \\
G299.290$-$00.289 & Y & entirely  & Y \\
G299.326$-$00.311 & Y & largely   & Y \\
G299.336$-$00.329 & Y & N         & N \\
G299.384+01.064   & N & largely   & N \\
G299.399$-$00.235 & Y & largely   & Y \\
G299.402+01.111   & N & entirely  & N \\
G299.429+01.081   & N & largely   & N \\
G299.439$-$00.039 & N & N         & N \\
G299.451$-$01.096 & N & entirely  & N \\
G299.474$-$01.101 & N & entirely  & N \\
G299.505+00.025   & N & partially & N \\
G299.606+00.177   & N & N         & N \\
G299.610+00.169   & N & N         & N \\
G299.688$-$00.593 & Y & largely   & N \\
G299.770$-$00.008 & N & entirely  & N \\
G299.773$-$00.286 & N & largely   & N \\
G299.922+00.006   & N & N         & N \\
G300.051$-$00.399 & N & largely   & N \\
G300.238$-$00.758 & N & N         & N \\
G300.322+01.644   & N & entirely  & N \\
\enddata
\tablenotetext{a}{Visually inspected whether the individual source areas overlap with other bright Pa$\alpha$ sources in continuum-subtracted MIPAPS Pa$\alpha$ line images.}
\tablenotetext{b}{Visually inspected whether, and to what extent, the individual source areas overlap with stellar residuals in continuum-subtracted MIPAPS Pa$\alpha$ line images. ``N'': the entire source area overlaps with no stellar residuals; ``partially'': source area partially overlaps; ``largely'': source area largely overlaps; ``entirely'': source area entirely overlaps with stellar residuals.}
\tablenotetext{c}{Visually inspected in continuum-subtracted IPHAS or VPHAS+ H$\alpha$ line images. ``Y'': visually detected; ``N'': not visually detected; ``No data'': not observed.\\}
(The complete version of this table is available in a machine-readable format in the online journal.)
\end{deluxetable*}

The results of the visual inspection for the 2666 {\it WISE} \ion{H}{2} region sources are summarized in Table \ref{table:vis}. Pa$\alpha$ and H$\alpha$ detections were identified in 902 and 823 {\it WISE} sources, respectively, and hydrogen recombination lines from either Pa$\alpha$ or H$\alpha$ were observed in a total of 1097 {\it WISE} \ion{H}{2} region sources. Excluding 478 ``Known'' sources with previously confirmed hydrogen recombination line detections, we newly confirmed 619 \ion{H}{2} region candidates (comprising 322 ``Candidate,'' 129 ``Group,'' and 168 ``Radio Quiet'' sources) through Pa$\alpha$ or H$\alpha$ detections.

\begin{deluxetable*}{cccccccccc}
\tabletypesize{\scriptsize}
\tablewidth{0pt}
\tablecaption{Summary of Pa$\alpha$ and H$\alpha$ Detections for {\it WISE} \ion{H}{2} Region Sources\label{table:vis}}
\tablehead{
\colhead{Category} & \colhead{{\it WISE} Source} & \multicolumn{3}{c}{MIPAPS Pa$\alpha$ Detection} &&
\multicolumn{3}{c}{IPHAS or VPHAS+ H$\alpha$ Detection} & \colhead{Pa$\alpha$ and/or H$\alpha$ Detection} \\
\cline{3-5} \cline{7-9}
\colhead{} & \colhead{} & \multicolumn{2}{c}{Yes} & \colhead{No} &&
\multicolumn{2}{c}{Yes} & \colhead{No} & \colhead{}
}
\startdata
``Known''       &  578 & 459 & (79.4\%) &  119 && 343 & (59.3\%) &  235 &  478 \\
``Candidate''   &  687 & 265 & (38.6\%) &  422 && 224 & (32.5\%) &  463 &  322 \\
``Group''       &  197 & 102 & (51.8\%) &   95 && 108 & (54.8\%) &   89 &  129 \\
``Radio Quiet'' & 1204 &  76 &  (6.3\%) & 1128 && 148 & (12.3\%) & 1056 &  168 \\
\hline
Total           & 2666 & 902 & (33.8\%) & 1764 && 823 & (30.9\%) & 1843 & 1097 \\
\enddata
\tablecomments{All numbers represent the counts of sources except for those in parentheses, which indicate detection percentages.}
\end{deluxetable*}

\subsubsection{Pa$\alpha$ Sources without WISE Counterparts} \label{subsec:mipaps_sources}

In the MIPAPS Pa$\alpha$ line images, we searched for Pa$\alpha$ emission-line sources that are not associated with any entry in the {\it WISE} \ion{H}{2} region catalog. Within the Galactic plane spanning $90\arcdeg \leq \ell \leq 330\arcdeg$, we identified a total of 587 Pa$\alpha$ sources with no corresponding {\it WISE} sources, which were added to the MIPAPS Pa$\alpha$ catalog in Table \ref{table:mipaps}. These sources are divided into two categories: MIPAPS Pa$\alpha$ extended sources and MIPAPS Pa$\alpha$ point-like sources, denoted by the ID prefixes MPE and MPP, respectively.

We identified a total of 261 MIPAPS Pa$\alpha$ extended (MPE) sources. The MIPAPS names were assigned based on the central positions of the individual Pa$\alpha$ features, and their radii and position angles were determined according to their spatial extents, as listed in Table \ref{table:mipaps}. The MPE sources are indicated by circles and ellipses with yellow dotted lines in Figures \ref{fig:paa21}--\ref{fig:paa42}. In the IPHAS or VPHAS+ H$\alpha$ images, 231 MPE sources were found to have corresponding H$\alpha$ features, while 29 MPE sources had no H$\alpha$ counterparts. One remaining MPE source lacks H$\alpha$ data from both the IPHAS and VPHAS+ surveys. We searched the SIMBAD database for previously known extended objects within the radii of individual MPE sources and identified matching objects based on their central positions and angular sizes. As a result, matching objects were found for 168 MPE sources, including 30 with multiple matches, and their object types are summarized in Table \ref{table:mpe}. Among the 138 MPE sources with a single matching object, 67 are associated with \ion{H}{2} regions, 38 with molecular or dark clouds, 12 with supernova remnants, five with interstellar medium objects, two with open clusters, and one with an external galaxy. The remaining 13 MPE sources were matched with only infrared or radio sources that lack a clearly identified object type.

\begin{deluxetable}{ccc}
\tablewidth{0pt}
\tablecaption{Matches with SIMBAD for MPE Sources\label{table:mpe}}
\tablehead{
\colhead{Matching Object Type} & \colhead{MPE Source} & \colhead{H$\alpha$ Detection}
}
\startdata
\ion{H}{2} region          &  67 &  60 \\
Molecular cloud            &  33 &  28 \\
Supernova remnant          &  12 &  10 \\
Dark cloud                 &   5 &   3 \\
Interstellar medium object &   5 &   5 \\
Open cluster               &   2 &   2 \\
External galaxy            &   1 &   1 \\
Infrared/Mid-IR source     &  12 &  11 \\
Radio source               &   1 &   0 \\
Multiple matching          &  30 &  29 \\
No matching                &  93 &  82 \\
\hline
Total                      & 261 & 231 \\
\enddata
\tablecomments{All numbers represent the counts of sources.}
\end{deluxetable}

Figures \ref{fig:exam1} and \ref{fig:exam2} present five examples of MPE sources: MPE047, MPE063, MPE106, MPE111, and MPE253. Sources MPE047 and MPE106 were identified as \ion{H}{2} regions, Sh2-181 and Siv 6, respectively, in SIMBAD, and their bright H$\alpha$ features are also visible in Figures \ref{fig:exam1}(b) and \ref{fig:exam2}(b). Source MPE063 has no matching object in SIMBAD, and its Pa$\alpha$ feature appears to overlap somewhat with WK113 and WC039, suggesting that MPE063 may be a part of either WK113 or WC039. However, MPE063 shows distinct H$\alpha$ and {\it WISE} 12 $\mu$m features, as seen in Figures \ref{fig:exam1}(e) and (f). In Figure \ref{fig:exam2}, MPE111 and MPE253 have fainter VPHAS+ H$\alpha$ counterparts compared to their appearances in the Pa$\alpha$ images. Notably, MPE111 corresponds to prominent {\it WISE} features, characterized by three 22 $\mu$m peaks accompanied by extended 12 $\mu$m emission, as shown in Figure \ref{fig:exam2}(f). In SIMBAD, MPE111 is associated only with {\it IRAS} infrared sources, while MPE253 is matched with two molecular clouds.

We identified a total of 326 MIPAPS Pa$\alpha$ point-like (MPP) sources. In the IPHAS or VPHAS+ H$\alpha$ images, 301 MPP sources were found to have corresponding H$\alpha$ counterparts, while 15 MPP sources had no H$\alpha$ counterparts. The remaining MPP sources lack available IPHAS and VPHAS+ data. Notably, the H$\alpha$ counterparts of 16 MPP sources exhibit both central H$\alpha$ point sources and extended H$\alpha$ features surrounding them. These are marked as ``Y(P+E)'' in the ``H$\alpha$ Detection'' column of Table \ref{table:mipaps}. We searched the SIMBAD database for previously known objects within a radius of 180$\arcsec$. When multiple objects were identified, matching objects were selected based on their central positions and whether their object types are known to exhibit emission lines. As a result, we found matching objects for 306 MPP sources, including seven with multiple matches, and their object types are summarized in Table \ref{table:mpp}. Most MPP sources are matched with stellar objects, while 49 MPP sources are associated with extended objects such as planetary nebulae and an external galaxy.

\begin{deluxetable}{ccc}
\tablewidth{0pt}
\tablecaption{Matches with SIMBAD for MPP Sources\label{table:mpp}}
\tablehead{
\colhead{Matching Object Type} & \colhead{MPP Source} & \colhead{H$\alpha$ Detection}
}
\startdata
Be/Emission-line star  & 118 & 116 \\
Planetary nebula       &  48 &  47 \\
Wolf-Rayet star        &  38 &  36 \\
Carbon star            &  23 &  18 \\
Long-period variable   &  18 &  13 \\
Star                   &  15 &  15 \\
Herbig Ae/Be star      &  14 &  13 \\
Mira variable          &   7 &   5 \\
Blue supergiant        &   4 &   4 \\
Eclipsing binary       &   3 &   3 \\
Symbiotic star         &   3 &   2 \\
High-mass X-ray binary &   2 &   2 \\
Variable star          &   2 &   2 \\
Classical nova         &   1 &   0 \\
External galaxy        &   1 &   1 \\
Red giant branch star  &   1 &   1 \\
Young stellar object   &   1 &   1 \\
Multiple matching      &   7 &   7 \\
No matching            &  20 &  15 \\
\hline
Total                  & 326 & 301 \\
\enddata
\tablecomments{All numbers represent the counts of sources.}
\end{deluxetable}

\subsubsection{Photometry of Pa$\alpha$ Sources} \label{subsec:photometric}

For the Pa$\alpha$ sources in the MIPAPS catalog, we performed aperture photometry to measure their Pa$\alpha$ fluxes. The Pa$\alpha$ fluxes of extended sources (WK, WC, WG, WQ, and MPE) were calculated using their radius or semi-major axis values as the radii for circular aperture photometry in the Pa$\alpha$ line images. The aperture sizes were determined to be large enough to fully cover each source, ensuring that the calculated fluxes can be regarded as the total Pa$\alpha$ fluxes of the extended sources. For MPP sources, we conducted rough aperture photometry of their Pa$\alpha$ fluxes, although PSF-fitted photometry would be the most suitable method for point-like sources. Since many MPP sources are associated with 2MASS point sources, they were collectively masked during the removal of stellar residuals, as described in Section \ref{subsec:pa_data}. Therefore, we performed photometry of MPP sources using the continuum-subtracted Pa$\alpha$ line images without applying the masking method. To estimate the total Pa$\alpha$ fluxes, we applied a sufficiently large aperture size of 360$\arcsec$ ($\sim$7 MIPAPS pixels) to the MPP sources.

The Pa$\alpha$ sources that largely overlap with other bright sources or stellar residuals were excluded from the photometry. We also excluded Pa$\alpha$ sources whose brightness was affected by artifact effects mentioned in Section \ref{subsec:image}. To subtract the background value from each photometric result, we first defined the background region around each source as an annulus with a thickness of twice the aperture radius. Then, other bright sources and stellar residuals shown in the background region were masked out, and a median in the remaining background region was taken as the background value. The 1$\sigma$ uncertainty of flux photometry was defined as a standard deviation in the background region. Table \ref{table:mipaps} gives the Pa$\alpha$ photometric results of 631 Pa$\alpha$ sources, together with the 1$\sigma$ uncertainties.

Among the extended sources (WK, WC, WG, WQ, and MPE) with measured Pa$\alpha$ fluxes, those fully covered by the calibrated IPHAS data and with H$\alpha$ detections were selected for aperture photometry of the H$\alpha$ fluxes. The total H$\alpha$ fluxes were calculated by applying the same aperture and background region to the continuum-subtracted IPHAS H$\alpha$ image as those used for the Pa$\alpha$ photometry of each source. Some IPHAS mosaic images of Pa$\alpha$ sources exhibited significantly mismatched background levels between observational fields, resulting in their exclusion from the H$\alpha$ photometry. IPHAS images with strong artifact features were also excluded. As noted in the Erratum of Paper I \citep{2020ApJS..251...16K}, the contribution of the {[}\ion{N}{2}{]} $\lambda$$\lambda$6548, 6584 \AA{} lines to the IPHAS H$\alpha$ fluxes should be considered. Assuming a line ratio of {[}\ion{N}{2}{]}/H$\alpha$ $\approx$ 0.25 for classical \ion{H}{2} regions \citep{2009RvMP...81..969H}, we corrected the H$\alpha$ fluxes by eliminating the contribution of the {[}\ion{N}{2}{]} lines.

By comparing the observed Pa$\alpha$-to-H$\alpha$ flux ratio with that predicted by the theory of radiative recombination of hydrogen, we derived the $E(\bv)$ color excess averaged over each Pa$\alpha$ source. We assumed case B conditions at a temperature of 10$^{4}$ K, as listed in Table 14.2 of \citet{2011piim.book.....D}, although some MPE sources may not correspond to true \ion{H}{2} regions. The differential extinction, $A_{Pa\alpha} - A_{H\alpha}$, was converted into $E(\bv)$ using the extinction curve of \citet{1989ApJ...345..245C} with $R_V$ = 3.1. Table \ref{table:mipaps} provides the resulting H$\alpha$ fluxes and $E(\bv)$ color excesses for 138 Pa$\alpha$ sources, along with their 1$\sigma$ uncertainties. Hereafter, the $E(\bv)$ color excess calculated in this manner will be referred to as the Pa$\alpha$-H$\alpha$ $E(\bv)$ when differentiation from values derived through other methods is necessary.

As shown in Figures \ref{fig:paa21} and \ref{fig:paa42}, some Pa$\alpha$ sources are located within MIPAPS fields affected by shadowing due to filter-wheel misalignment. However, none of the Pa$\alpha$ sources included in this photometric study were significantly impacted by this artifact, as they are either positioned far from the shadowed edges of the fields or have angular sizes small enough relative to the MIPAPS field size.

\section{Discussion} \label{sec:discussion}

\subsection{Pa$\alpha$ and H$\alpha$ Detections for WISE \ion{H}{2} Regions} \label{subsec:disc_wise}

As mentioned in Paper I, the Pa$\alpha$ flux from \ion{H}{2} regions, which experiences less dust attenuation than H$\alpha$, exceeds the H$\alpha$ flux when $E(\bv)$ $>$ 1.12. As shown in Table \ref{table:vis}, the overall detection rates of Pa$\alpha$ and H$\alpha$ for {\it WISE} \ion{H}{2} region sources are comparable: 33.8\% (902 of 2666) and 30.9\% (823 of 2666), respectively. However, among the 902 Pa$\alpha$-detected and 823 H$\alpha$-detected sources, 462 (51.2\% of the 902) and 282 (34.3\% of the 823) are located in the 4th Galactic quadrant ($270\arcdeg \leq \ell \leq 330\arcdeg$), where dust extinction is relatively high. In contrast, 117 (13.0\% of the 902) and 155 (18.8\% of the 823) lie in the anticenter direction ($150\arcdeg \leq \ell \leq 210\arcdeg$), where dust extinction is relatively low. These results are consistent with the expectation that Pa$\alpha$ observations are more effective than H$\alpha$ for detecting \ion{H}{2} regions in areas with high dust extinction. On the other hand, of the 195 {\it WISE} sources with H$\alpha$ detections but no Pa$\alpha$ detections, 152 (77.9\%) have small angular sizes of $\leq$360$\arcsec$ ($\sim$7 MIPAPS pixels). Even within the two largest {\it WISE} sources, the areas with H$\alpha$ detections were entirely overlapped by the masking positions of stellar residuals in the continuum-subtracted Pa$\alpha$ line images. This indicates that IPHAS and VPHAS+, with their higher spatial resolutions, provided clear advantages over MIPAPS in detecting small-sized {\it WISE} sources. By combining the strengths of the MIPAPS Pa$\alpha$ data and the IPHAS/VPHAS+ H$\alpha$ data, we detected hydrogen recombination lines from a total of 1097 {\it WISE} \ion{H}{2} region sources and newly confirmed 619 \ion{H}{2} region candidates (322 ``Candidate,'' 129 ``Group,'' and 168 ``Radio Quiet'' sources) as definitive \ion{H}{2} regions through Pa$\alpha$ or H$\alpha$ detections.

Figures \ref{fig:exam1} and \ref{fig:exam2} show several cases where there are differences in angular sizes and central positions between the {\it WISE} \ion{H}{2} region sources and the corresponding Pa$\alpha$ sources. Among the 902 {\it WISE} sources with Pa$\alpha$ detections, 17.0\% and 5.4\% exhibit angular sizes in the Pa$\alpha$ images that are more than 20\% larger and smaller, respectively, than those in the {\it WISE} images. Additionally, 12.0\% of the 902 {\it WISE} sources have Pa$\alpha$ central positions located more than 0$\arcdeg$.02 away from their corresponding {\it WISE} central positions. In many cases, these discrepancies are due to the ambiguous boundaries of overlapping \ion{H}{2} regions, as seen in WK113 and WK180 in Figures \ref{fig:exam1} and \ref{fig:exam2}. The significant difference in spatial resolution between MIPAPS and {\it WISE} also contributes to these discrepancies. However, certain \ion{H}{2} regions, such as WK142 and WQ073 in Figures \ref{fig:exam1} and \ref{fig:exam2}, exhibit clear differences in central positions and angular sizes between MIPAPS and {\it WISE}. The Pa$\alpha$ emission line directly traces ionized hydrogen gas in \ion{H}{2} regions, whereas the 12 $\mu$m and 22 $\mu$m emissions originate from polycyclic aromatic hydrocarbon molecules and small dust grains heated within the \ion{H}{2} regions \citep{2014ApJS..212....1A}. Therefore, such spatial discrepancies are expected, particularly for \ion{H}{2} regions embedded in a non-uniform medium.

\subsection{Identities of MPE and MPP Sources} \label{subsec:disc_mipaps}

In the MIPAPS Pa$\alpha$ line images, we identified 261 MPE sources with extended sizes. Although these sources do not exhibit the typical {\it WISE} morphological characteristics of \ion{H}{2} regions required in \citet{2014ApJS..212....1A}, many of them display extended features in the {\it WISE} 12 $\mu$m and/or 22 $\mu$m images that are likely associated with those in the Pa$\alpha$ images, as seen in the MPE sources in Figures \ref{fig:exam1} and \ref{fig:exam2}. As listed in Table \ref{table:mpe}, \ion{H}{2} regions are the most commonly associated objects with the MPE sources, with a total of 85 MPE sources, including 18 with multiple matching objects. Thirty-three MPE sources matched with molecular clouds may be linked to star-forming regions. Hydrogen recombination lines can also originate from supernova remnants. Indeed, 14 MPE sources, including two with multiple matching objects, appear to be associated with supernova remnants. Notably, ten MPE sources identified within $250\arcdeg \leq \ell \leq 275\arcdeg$ are likely part of two well-known, large-scale H$\alpha$-emitting structures: the Gum Nebula and the Vela supernova remnant.

We could not find matching objects in the SIMBAD database for 93 MPE sources. However, most of them (82 sources) show morphological correspondence between the MIPAPS Pa$\alpha$ and the IPHAS or VPHAS+ H$\alpha$ images, supporting their origins from ionized hydrogen gas. Sources located near other bright objects, such as MPE063 in Figure \ref{fig:exam1}, and large sources encompassing several {\it WISE} \ion{H}{2} region sources, such as MPE030, MPE032, and MPE258, might not represent independent objects. Sources with elongated shapes may be part of nearby diffuse H$\alpha$-emitting structures, such as the Gum Nebula. Notably, ten MPE sources (MPE019, MPE021, MPE029, MPE031, MPE087, MPE125, MPE149, MPE161, MPE172, and MPE254) showing separate circular Pa$\alpha$ and H$\alpha$ features could represent previously unidentified \ion{H}{2} regions or supernova remnants. Among them, MPE029 and MPE172 have been found to host central early B-type stars as ionizing source candidates: BD+63 1907 and HD 92383, respectively. There are 11 MPE sources lacking both SIMBAD matching objects and H$\alpha$ counterparts. Most of these are located near other bright objects, suggesting possible associations.

We identified 326 MPP sources with point-like sizes in the MIPAPS Pa$\alpha$ line images. As listed in Table \ref{table:mpp}, the most commonly matching objects in SIMBAD are Be or emission-line stars, comprising a total of 118 MPP sources. Other matching objects, such as Wolf-Rayet stars (38 sources) and Herbig Ae/Be stars (14 sources), are also classified as types of emission-line stars. An intriguing result is that 23 MPP sources were matched with carbon stars. Among these, 18 MPP sources were found to have H$\alpha$ counterparts in the IPHAS or VPHAS+ H$\alpha$ images. \citet{2008MNRAS.384.1277W} reported that only a few sources in the IPHAS catalog of H$\alpha$ emission-line objects were matched with carbon stars in SIMBAD. The small angular sizes of 49 MPP sources, associated with extended objects like planetary nebulae and an external galaxy, caused them to appear as point-like sources under the spatial resolution of MIPAPS.

Among the 20 MPP sources without SIMBAD matching objects, 15 have point-like H$\alpha$ counterparts in the IPHAS or VPHAS+ H$\alpha$ images. Since all of these also show point-like counterparts in the {\it WISE} 12 $\mu$m and 22 $\mu$m images, they are likely previously unidentified emission-line stars. In contrast, MPP251, which has an extended H$\alpha$ counterpart, may correspond to a small \ion{H}{2} region or part of a larger structure. The remaining four sources, which lack H$\alpha$ counterparts, are also absent in the {\it WISE} 12 $\mu$m and 22 $\mu$m images. Therefore, we note that these sources (MPP111, MPP116, MPP125, and MPP207) may result from residual stellar contamination or artifact features in the continuum-subtracted Pa$\alpha$ line images.

In the IPHAS or VPHAS+ H$\alpha$ images, 16 MPP sources exhibit surrounding extended H$\alpha$ features along with central H$\alpha$ point sources. In SIMBAD, the central H$\alpha$ point sources are matched with six Be or emission-line stars, four Herbig Ae/Be stars, two Wolf-Rayet stars, one blue supergiant, one long-period variable, one young stellar object, and one \ion{H}{2} region. Five surrounding extended H$\alpha$ features also have matching objects in SIMBAD: two interstellar medium objects, one reflection nebula, one planetary nebula, and one external galaxy. Source MPP320, matched with a long-period variable and a planetary nebula, and source MPP325, matched with an \ion{H}{2} region and an external galaxy, are likely cases of multiple objects overlapping along the line of sight. However, for the remaining 14 MPP sources, the surrounding extended H$\alpha$ features are likely associated with the central H$\alpha$ point sources. Following the reporting of three sources (MPP020, MPP023, and MPP029) in Paper I, we performed high-resolution near-infrared spectroscopic follow-up observations to investigate the origins of the surrounding extended H$\alpha$ features. \citet{2021AJ....162...24K} presented the results for MPP029, revealing that the H$\alpha$ nebulosity around the central Herbig star, MWC 1080, primarily originates from the stellar emission of MWC 1080A scattered by dust. Additionally, H$\alpha$ components arising from shock-excited ionized hydrogen gas driven by stellar outflows were identified. This follow-up study supports the associations between the central H$\alpha$ point sources and the surrounding extended H$\alpha$ features in the other MPP sources.

Similar to MPP027, which was reported in Paper I, we found that several MPP sources appear to be associated with MIPAPS extended sources that lack matching objects in SIMBAD. Source MPP157, matched with HD 55439, a B2Ve-type star in SIMBAD, is located near the brightest region of WC072 in the Pa$\alpha$, H$\alpha$, and {\it WISE} 12 $\mu$m images. Additionally, sources MPP192, MPP236, and MPP253, all matched with Wolf-Rayet stars, are situated near the centers of MPE120, MPE151, and MPE175, respectively. Notably, the VPHAS+ H$\alpha$ images of MPE120, MPE151, and MPE175 reveal several prominent filamentary features.

\subsection{$E(\bv)$ derived from Extended Emissions} \label{subsec:photo}

In this study, the $E(\bv)$ color excesses for 138 Pa$\alpha$ sources were derived from the photometry of the MIPAPS Pa$\alpha$ and IPHAS H$\alpha$ fluxes, as presented in Table \ref{table:mipaps}. These Pa$\alpha$-H$\alpha$ $E(\bv)$ values represent the color excesses directly obtained from the extended emission of ionized hydrogen gas, assuming the Pa$\alpha$ sources are \ion{H}{2} regions with a characteristic temperature of 10$^{4}$ K. Among the ten MPE sources with derived $E(\bv)$ values, seven are matched with \ion{H}{2} regions in SIMBAD. Sources MPE019 and MPE037 have no matching objects in SIMBAD but show clear IPHAS H$\alpha$ counterparts. Notably, as discussed in Section \ref{subsec:disc_mipaps}, MPE019, with its isolated circular morphology, could be a previously unidentified \ion{H}{2} region. On the other hand, MPE056 is matched with an external galaxy (Sh2-197) in SIMBAD but was not excluded from our photometric study.

\begin{figure*}[ht]
\centering
\includegraphics[scale=0.8]{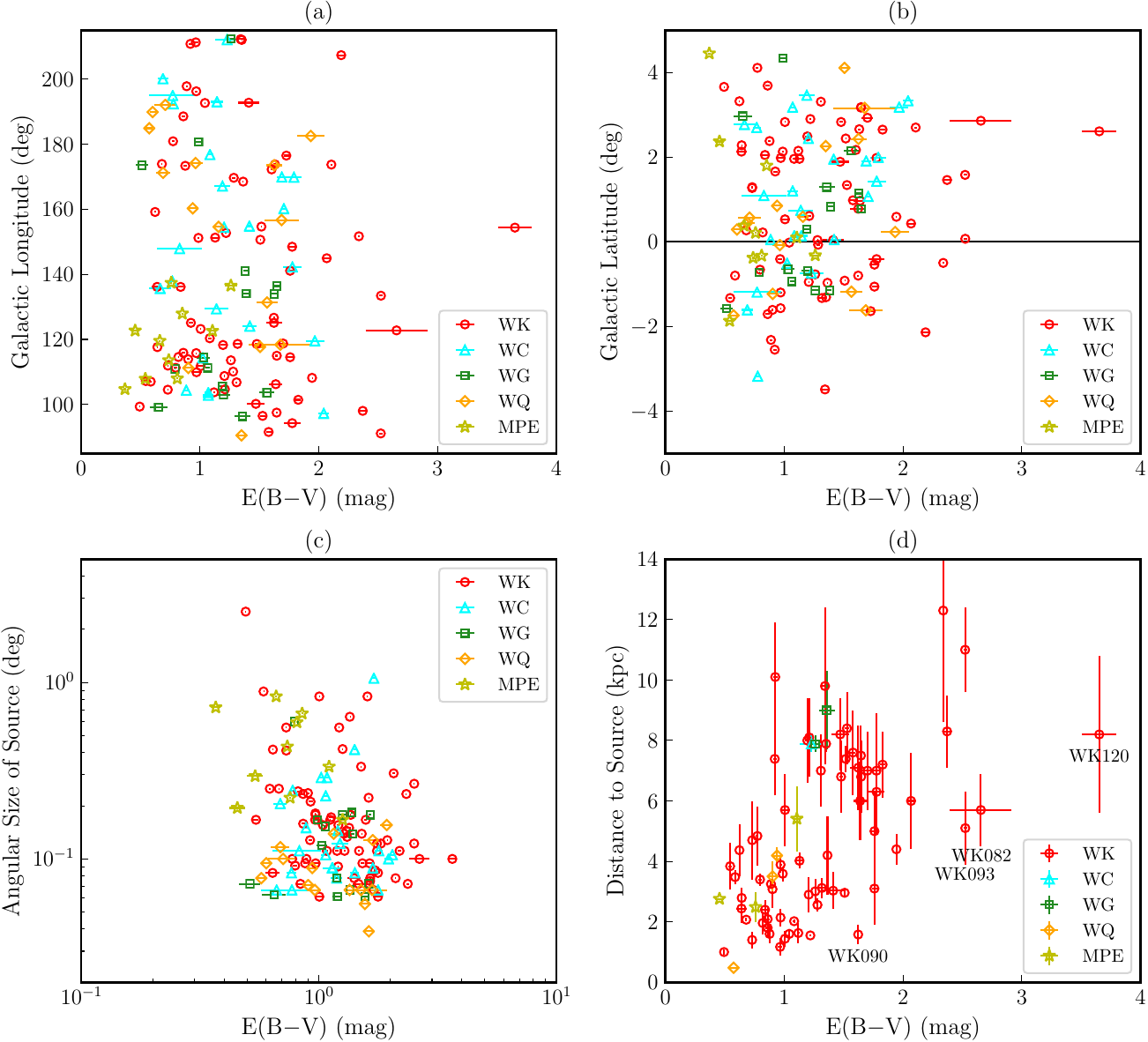}
\caption{$E(\bv)$ vs. (a) Galactic longitude, (b) Galactic latitude, (c) angular size of source, and (d) distance to source for 138 Pa$\alpha$ sources, with $E(\bv)$ values available in Table \ref{table:mipaps}. In panel (c), both $E(\bv)$ and the angular size of source are displayed on a logarithmic scale. In panel (d), only 76 sources with known distances in Table \ref{table:dis} are shown, along with vertical error bars for the distances. Symbols without visible error bars indicate sources with uncertainties smaller than the symbol sizes.\label{fig:cecorr}}
\end{figure*}

In Figures \ref{fig:cecorr}(a)--(c), the $E(\bv)$ values of the 138 Pa$\alpha$ sources are compared with their Galactic coordinates and angular sizes. In Figure \ref{fig:cecorr}(d), the $E(\bv)$ values of 76 Pa$\alpha$ sources with known distances are also compared with the distances to the sources. Kinematic, spectrophotometric, and parallax distances for these 76 sources were collected from several references, as listed in Table \ref{table:dis}. For sources with multiple distance values, the value with the highest distance-to-uncertainty ratio (those without parentheses in Table \ref{table:dis}) was applied. No significant correlation with Galactic coordinates is observed in Figures \ref{fig:cecorr}(a) and (b). However, weak negative and positive correlations with angular size and distance, respectively, are shown in Figures \ref{fig:cecorr}(c) and (d). In principle, these correlations are consistent with the fact that more distant \ion{H}{2} regions generally appear smaller in angular size and are more attenuated by interstellar dust. However, young \ion{H}{2} regions still embedded within dense molecular clouds can exhibit high extinction due to their local environments, even when they are relatively nearby and minimally affected by interstellar dust along the line of sight. These young \ion{H}{2} regions likely occupy the lower-right part of Figure \ref{fig:cecorr}(d) (characterized by high $E(\bv)$ values relative to distance), where several WK sources (WK082, WK090, WK093, WK120, etc.) are located. Among them, WK090, matched with Sh2-187, has been extensively studied across various wavelengths. \citet{1992ApJ...387..591J} concluded that this \ion{H}{2} region remains embedded in its parental cloud and is relatively young ($\sim$1--2 $\times 10^{5}$ yr).

\citet{2015AJ....150..147F} estimated the $E(\bv)$ values for 103 \ion{H}{2} regions based on the spectrophotometric information of ionizing stars associated with the \ion{H}{2} regions. Among the Pa$\alpha$ sources with derived Pa$\alpha$-H$\alpha$ $E(\bv)$ values, we found 35 Pa$\alpha$ sources corresponding to the \ion{H}{2} regions listed in \citet{2015AJ....150..147F}. Two MPE sources, MPE046 and MPE058, which are included among the Pa$\alpha$ sources, are also matched with \ion{H}{2} regions in SIMBAD (Sh2-180 and Sh2-198, respectively). Additionally, we applied the $E(\bv)$ information of Sh2-283 from \citet{2007AA...470..161R} to WK170. For a total of 36 Pa$\alpha$ sources, the $E(\bv)$ values obtained from previous studies are summarized in Table \ref{table:dis}. For six of these sources, we excluded ionizing stars located outside their angular boundaries and adopted the values from the remaining stars, as noted in Table \ref{table:dis}. The $E(\bv)$ values provided by \citet{2015AJ....150..147F} and \citet{2007AA...470..161R} represent color excesses estimated from the photometry of ionizing point stars.

\begin{figure}[ht]
\centering
\includegraphics[scale=0.7]{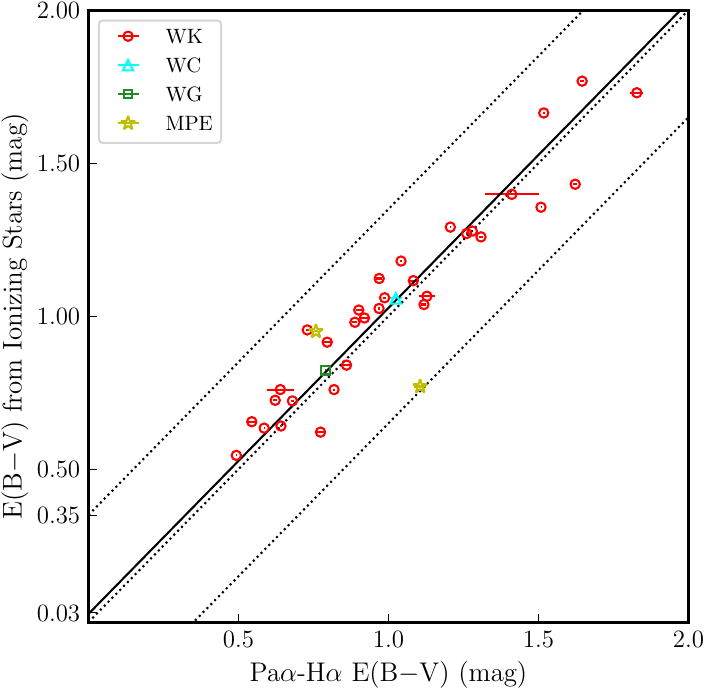}
\caption{Comparison of the Pa$\alpha$-H$\alpha$ $E(\bv)$ values from extended emissions with the $E(\bv)$ values from ionizing stars for 36 Pa$\alpha$ sources. The Pa$\alpha$-H$\alpha$ $E(\bv)$ values (with 1$\sigma$ error bars) are taken from Table \ref{table:mipaps}. Symbols without visible error bars indicate sources where the Pa$\alpha$-H$\alpha$ $E(\bv)$ uncertainties are smaller than the symbol sizes. The $E(\bv)$ values from ionizing stars (without uncertainties) come from \citet{2015AJ....150..147F} and \citet{2007AA...470..161R}, as noted in Table \ref{table:dis}. The diagonal dotted lines indicate agreement between the two $E(\bv)$ estimates within $\sim$0.35 mag, while the solid line represents the best-fit line with an offset of 0.03 $\pm$ 0.02 mag.\label{fig:cecomp}}
\end{figure}

In Figure \ref{fig:cecomp}, we compare the Pa$\alpha$-H$\alpha$ $E(\bv)$ values derived from extended emission with the $E(\bv)$ values obtained from ionizing stars for 36 Pa$\alpha$ sources. The three diagonal dotted lines indicate a global agreement between them within approximately 0.35 mag, while the diagonal solid line represents the result of a least-squares fit (0.03 $\pm$ 0.02 mag). In the Erratum of Paper I \citep{2020ApJS..251...16K}, after accounting for the {[}\ion{N}{2}{]} line contribution to the IPHAS H$\alpha$ fluxes, the Pa$\alpha$-H$\alpha$ $E(\bv)$ values for Pa$\alpha$ sources in the Cepheus region showed good agreement with those reported by \citet{2015AJ....150..147F}, without any systematic offset. In this paper, we obtained similar results when comparing the two types of $E(\bv)$ values for the Galactic plane within $90\arcdeg \leq \ell \leq 330\arcdeg$. As discussed in Paper I, the random deviations between the Pa$\alpha$-H$\alpha$ $E(\bv)$ values and those derived from ionizing stars can be attributed to differences in the geometries of the target sources. The distances to various locations within an extended \ion{H}{2} region likely differ from those to the ionizing stars used to estimate $E(\bv)$. In addition, if dense molecular clouds lie along certain lines of sight, the corresponding $R_V$ values can be larger than the standard $R_V$ = 3.1 for the diffuse interstellar medium \citep{1989ApJ...345..245C}, further contributing to the deviations described above. The uncertainties in the {[}\ion{N}{2}{]} line contribution to the H$\alpha$ fluxes would also influence the deviations.

\begin{figure*}[ht]
\centering
\includegraphics[scale=0.7]{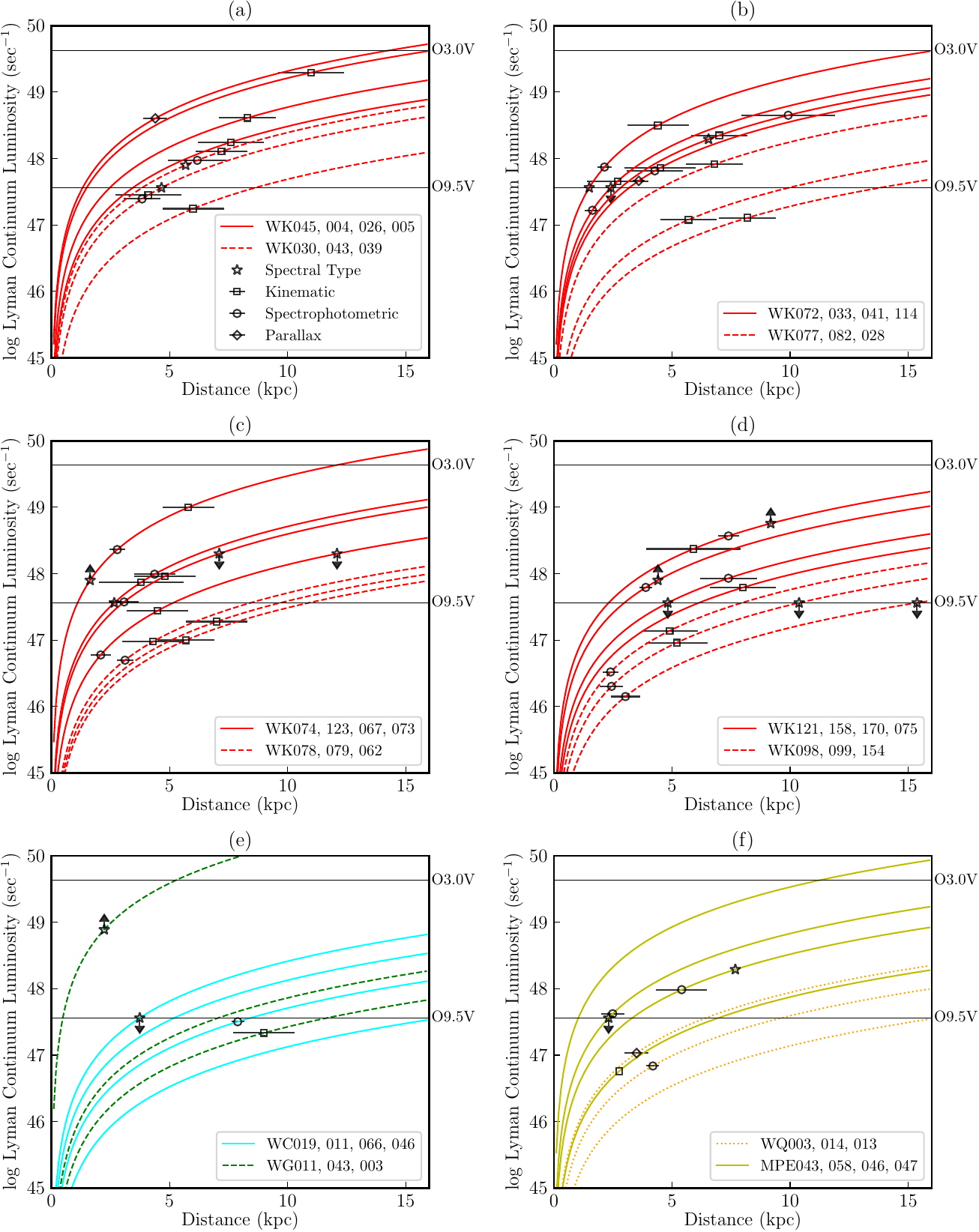}
\caption{Total Lyman continuum luminosity as a function of distance for 42 representative Pa$\alpha$ sources. In each panel, several source IDs (as listed in Table \ref{table:mipaps}) correspond to the same color and line type, with the curves for each source arranged from left to right. The curve for each source is calculated using its Pa$\alpha$ total flux and $E(\bv)$ provided in Table \ref{table:mipaps}. Based on Table 1 of \citet{2005A&A...436.1049M}, the Lyman continuum luminosities for O3.0V- and O9.5V-type stars are indicated by two horizontal lines. Star symbols on the curves represent positions corresponding to Lyman continuum luminosities calculated using Tables 1--3 of \citet{2005A&A...436.1049M} and the spectral types of the ionizing stars listed in Table \ref{table:dis}. Stars with upper- (or lower-) limit symbols denote sources with only one B-type ionizing star (or additional B-type ionizing stars alongside O-type ionizing stars). Squares, circles, and diamonds (with error bars) on the curves mark positions corresponding to the kinematic, spectrophotometric, and parallax distances provided in Table \ref{table:dis}, respectively.\label{fig:lyc}}
\end{figure*}

By applying the estimated $E(\bv)$ value to the observed Pa$\alpha$ total flux, the reddening-corrected Pa$\alpha$ total flux for each \ion{H}{2} region can be determined. Assuming the relative luminosity ratios between hydrogen recombination lines under the case B condition (as given in Table 14.2 of \citet{2011piim.book.....D}), the Lyman continuum flux emitted by its ionizing star(s) can be derived from the intrinsic Pa$\alpha$ total flux. The total luminosity of the Lyman continuum can then be estimated if the distance to the \ion{H}{2} region is known. The spectral type(s) of the star(s) can also be inferred by comparing the derived total luminosity with the theoretically predicted values \citep{2005A&A...436.1049M}. Conversely, if the spectral types of the ionizing stars are known, their distances can be estimated. In Figure \ref{fig:lyc}, each curve represents the total Lyman continuum luminosity as a function of distance for one of 42 representative Pa$\alpha$ sources, for which Pa$\alpha$-H$\alpha$ $E(\bv)$ values were derived in this study. These sources were selected based on the availability of distance or spectral type information in the literature, except for several cases shown in panels (e) and (f). Two horizontal lines in the figure indicate the Lyman continuum luminosities of O3V- and O9.5V-type stars. The positions on the curves corresponding to the kinematic, spectrophotometric, and parallax distances given in Table \ref{table:dis} are denoted by square, circle, and diamond symbols, respectively. Star symbols indicate the positions on the curves corresponding to the Lyman continuum luminosities calculated using Tables 1--3 of \citet{2005A&A...436.1049M} and the spectral types of the ionizing stars listed in Table \ref{table:dis}. For the calculation of Lyman continuum luminosities, we considered only O-type ionizing stars, as \citet{2005A&A...436.1049M} provided Lyman continuum luminosities only for spectral types between O3 and O9.5. For \ion{H}{2} regions with B-type ionizing stars, we instead present upper or lower limits for the Lyman continuum luminosities. In Figure \ref{fig:lyc}, star symbols with upper limits are used for \ion{H}{2} regions with only one B-type ionizing star, with the upper limits corresponding to the values for the latest O-type star (O9.5) provided by \citet{2005A&A...436.1049M}. On the other hand, star symbols with lower limits are used for \ion{H}{2} regions that have additional B-type ionizing stars along with O-type ionizing stars, with the lower limits derived solely from the O-type ionizing stars. Additionally, we assumed luminosity class V for stars classified as IV since \citet{2005A&A...436.1049M} provided Lyman continuum luminosities only for luminosity classes I, III, and V.

Figure \ref{fig:lyc} can be used to constrain either the distance to each \ion{H}{2} region or the spectral type(s) of its ionizing star(s), provided that the other parameter is known. Using this approach, we assess the consistency between the distances and spectral types reported in the literature Among the 16 \ion{H}{2} regions with both distances and spectral types indicated in the figure, WK030, WK033, WK041, WK043, WK123, and MPE058 show good agreement between the known distances and spectral types. For WK072 and WK170, either their distances may have been overestimated or the luminosities corresponding to their spectral types may have been underestimated. As shown in Table \ref{table:dis}, several ionizing stars previously suggested as responsible for these \ion{H}{2} regions were excluded because they are located near the angular boundaries of WK072 and WK170. However, if the true extents of these \ion{H}{2} regions are slightly larger, or if the excluded stars had past positions nearer to the centers due to proper motion, their possible associations cannot be ruled out. In the case of WK072 (identified as Sh2-168, the uppermost curve in panel (b)), four B-type stars listed by \citet{2015AJ....150..147F} were excluded. Including these stars may help reconcile the total Lyman continuum luminosity inferred solely from the O9.5V spectral type with that derived from kinematic or spectrophotometric distances, assuming these distances are accurate. For WK170 (identified as Sh2-283, the third uppermost curve in panel (d)), four stars --- including O7V- and B3V-type stars listed in \citet{2007AA...470..161R} --- were excluded, which may have led to an underestimation of the total Lyman continuum luminosity. Conversely, in the cases of WK121, WK158, and MPE046, either their distances may have been underestimated or the luminosities corresponding to their spectral types may have been overestimated. \citet{2015AJ....150..147F} reported O9V-, O9.5Ib-, and B0V-type stars as the ionizing stars of WK121 (identified as Sh2-211); O9V- and B0.5V-type stars for WK158 (Sh2-267); and an O8V-type star for MPE046 (Sh2-180). Excluding certain stars or revising the known spectral types could help reconcile the discrepancies in total Lyman continuum luminosities.

Among the 1489 Pa$\alpha$ sources listed in the MIPAPS catalog, Pa$\alpha$ fluxes were obtained for only 631 sources. In the MIPAPS Pa$\alpha$ images, which have a moderate spatial resolution of $\sim$52$\arcsec$, many Pa$\alpha$ sources could not be resolved from nearby bright sources and/or stellar residuals, making precise photometry challenging. Additionally, the photometry of large Pa$\alpha$ sources was affected by gradient variations in the surrounding background level, as shown in Figure \ref{fig:profile}. Among the 349 extended Pa$\alpha$ sources (WK, WC, WG, WQ, and MPE) with measured Pa$\alpha$ fluxes, Pa$\alpha$-H$\alpha$ $E(\bv)$ values were calculated for only 138 sources. This is primarily because we utilized only the well-calibrated IPHAS H$\alpha$ data in the range of $90\arcdeg \la \ell \la 215\arcdeg$, excluding Pa$\alpha$ sources in the range of $215\arcdeg \la \ell \la 330\arcdeg$ from the H$\alpha$ photometry. Additionally, the IPHAS H$\alpha$ mosaic images of large Pa$\alpha$ sources exhibit significant mismatches in background levels between observational fields, leading to their exclusion from the H$\alpha$ photometry. This issue arises from the use of numerous IPHAS fields, each covering a relatively small sky area of $\sim$0.29 deg$^2$ \citep{2014MNRAS.444.3230B}, for constructing the large mosaic images. Pa$\alpha$ sources whose IPHAS H$\alpha$ images contain numerous stellar residuals and artifact features that are difficult to fully eliminate were also excluded from the H$\alpha$ photometry.

In Paper I, we presented Pa$\alpha$ fluxes for 104 Pa$\alpha$ sources and Pa$\alpha$-H$\alpha$ $E(\bv)$ values for 78 Pa$\alpha$ sources within the Cepheus region ($96\arcdeg.5 \leq \ell \leq 116\arcdeg.3$). However, those photometric results were affected by imprecise background estimation. In this study, we improved the background treatment by thoroughly masking nearby sources and stellar residuals, and by using background values only from unaffected regions in both the Pa$\alpha$ and H$\alpha$ images. Sources with unreliable photometry were excluded from the analysis. As a result, we recalculated Pa$\alpha$ fluxes for 59 sources and Pa$\alpha$-H$\alpha$ $E(\bv)$ values for 41 sources in the Cepheus region, thereby updating the photometric results from Paper I.

\section{Summary and Conclusions} \label{sec:summary}

Using the whole MIPAPS data, we completed a continuum-subtracted Pa$\alpha$ line image for the entire Galactic plane and presented close-up Pa$\alpha$ line images with $30\arcdeg$ longitude widths. Although these continuum-subtracted Pa$\alpha$ images display overall negative background levels, numerous Pa$\alpha$ features are clearly visible along the Galactic plane. Based on these Pa$\alpha$ images, we compiled the MIPAPS catalog of Pa$\alpha$ emission-line sources for the Galactic plane of $90\arcdeg \leq \ell \leq 330\arcdeg$, where the MIPAPS observational data are unaffected by shadowing effects caused by filter-wheel misalignment. Using the {\it WISE} \ion{H}{2} region catalog, a comprehensive catalog of Galactic \ion{H}{2} regions, we primarily identified 902 Pa$\alpha$ sources originating from \ion{H}{2} regions. Subsequently, we searched for additional Pa$\alpha$ sources not coincident with any entry in the {\it WISE} \ion{H}{2} region catalog and added 587 Pa$\alpha$ sources to the MIPAPS catalog. As a result, the MIPAPS catalog was completed with a total of 1489 Pa$\alpha$ emission-line sources. Additionally, we created continuum-subtracted H$\alpha$ images using IPHAS and VPHAS+ survey data and compared them with our Pa$\alpha$ line images. The main results of our study can be summarized as follows.

1. For the 2666 individual {\it WISE} \ion{H}{2} region sources, we visually inspected the MIPAPS Pa$\alpha$ and IPHAS/VPHAS+ H$\alpha$ line images, identifying Pa$\alpha$ and H$\alpha$ detections in 902 and 823 {\it WISE} sources, respectively. Among these, 462 (51.2\% of 902) and 282 (34.3\% of 823) {\it WISE} sources are located within $270\arcdeg \leq \ell \leq 330\arcdeg$ (the 4th Galactic quadrant), a region with relatively high dust extinction, while 117 (13.0\% of 902) and 155 (18.8\% of 823) {\it WISE} sources are found in the anticenter direction ($150\arcdeg \leq \ell \leq 210\arcdeg$), where dust extinction is relatively low. These results are consistent with the expectation that Pa$\alpha$ observations are more effective than H$\alpha$ for detecting \ion{H}{2} regions in areas with high dust extinction. Conversely, of the 195 {\it WISE} sources with H$\alpha$ detections but no Pa$\alpha$ detections, 152 (77.9\%) sources have small angular sizes ($\leq$360$\arcsec$). This indicates that the higher spatial resolution of IPHAS and VPHAS+ offers a better advantage over MIPAPS for detecting small \ion{H}{2} regions. By combining the complementary strengths of the MIPAPS and IPHAS/VPHAS+ data, hydrogen recombination lines, either Pa$\alpha$ or H$\alpha$, were detected in a total of 1097 {\it WISE} \ion{H}{2} region sources. Excluding 478 ``Known'' sources previously identified as \ion{H}{2} regions, a total of 619 \ion{H}{2} region candidates (322 ``Candidate,'' 129 ``Group,'' and 168 ``Radio Quiet'' sources) were newly confirmed as definitive \ion{H}{2} regions through Pa$\alpha$ or H$\alpha$ detections. Additionally, we cataloged 1764 {\it WISE} \ion{H}{2} region sources with no MIPAPS Pa$\alpha$ detection.

2. In the MIPAPS Pa$\alpha$ line images, we identified 261 MPE sources with extended sizes. Although these sources were not included in the {\it WISE} \ion{H}{2} region catalog, many exhibit corresponding features in the {\it WISE} 12 $\mu$m and/or 22 $\mu$m images. Particularly, 231 MPE sources were found to have H$\alpha$ counterparts in the IPHAS or VPHAS+ images. We searched the SIMBAD database for previously known extended objects matching the MPE sources and found matches for 168 of them. Among these, \ion{H}{2} regions were the most commonly matching objects (85 sources), followed by molecular clouds (33 sources) and supernova remnants (14 sources). No SIMBAD matches were found for 93 MPE sources; however, the majority of these (82 sources) have H$\alpha$ counterparts in the IPHAS or VPHAS+ images. We proposed that ten MPE sources with distinct circular Pa$\alpha$ and H$\alpha$ features could be previously unidentified \ion{H}{2} regions or supernova remnants.

3. We also identified 326 MPP sources characterized by point-like sizes. Among these, 301 MPP sources were found to have H$\alpha$ counterparts in the IPHAS or VPHAS+ images. A search of the SIMBAD database revealed matches for 306 MPP sources. The most commonly matching objects are Be or emission-line stars (118 sources). Wolf-Rayet stars and Herbig Ae/Be stars, which are also types of emission-line stars, were matched with 38 and 14 MPP sources, respectively. Notably, carbon stars were matched with a significant number of MPP sources (23 in total). Additionally, 48 planetary nebulae were observed as MPP sources under the spatial resolution of MIPAPS. Among the 20 MPP sources without SIMBAD matches, 15 have point-like H$\alpha$ counterparts in the IPHAS or VPHAS+ images, suggesting that they are likely previously unidentified emission-line stars. Furthermore, we found that 16 MPP sources exhibit surrounding extended H$\alpha$ features along with central H$\alpha$ point sources in the IPHAS or VPHAS+ image. We proposed that the associations between the central H$\alpha$ point sources and the surrounding extended H$\alpha$ features are genuine for 14 MPP sources, including MPP029, as confirmed by a high-resolution spectroscopic follow-up study.

4. For 631 Pa$\alpha$ sources in the MIPAPS catalog, we performed aperture photometry of the Pa$\alpha$ fluxes, including updates for the sources presented in Paper I. Furthermore, we carried out photometry of the IPHAS H$\alpha$ fluxes for 138 Pa$\alpha$ sources and estimated the $E(\bv)$ color excesses by combining the measured Pa$\alpha$ and H$\alpha$ fluxes. The Pa$\alpha$-H$\alpha$ $E(\bv)$ values show no significant correlation with Galactic coordinates but exhibit negative and positive correlations with angular size and distance, respectively. These trends are consistent with the fact that more distant \ion{H}{2} regions are generally smaller in angular size and more attenuated by interstellar dust. These $E(\bv)$ color excesses, derived from extended Pa$\alpha$ and H$\alpha$ emissions of ionized hydrogen gas, were compared with those obtained from spectrophotometry of ionizing point stars. The comparisons for 36 Pa$\alpha$ sources demonstrate good agreement with no systematic offset. Additionally, we presented the total Lyman continuum luminosities as a function of distance for 42 Pa$\alpha$ sources by applying the estimated $E(\bv)$ values to the observed Pa$\alpha$ total fluxes. These results provide constraints on the distances to the \ion{H}{2} regions and the spectral types of their ionizing stars.

Our results highlight the scientific potential of Pa$\alpha$ line observations. Despite the limitations imposed by moderate spatial resolution, the MIPAPS Pa$\alpha$ survey successfully provided a complete Pa$\alpha$ line image for the entire Galactic plane, revealing numerous Pa$\alpha$ features that are not visible in the IPHAS or VPHAS+ H$\alpha$ images. Furthermore, the study demonstrated the usefulness of combining multiple recombination lines from ionized hydrogen gas, for instance, in providing quantitative information such as $E(\bv)$ color excesses. In future work, if reliable corrections for the sensitivity degradation caused by the filter-wheel misalignment in the data for $-30\arcdeg \leq \ell \leq 90\arcdeg$ can be established, completing the MIPAPS Pa$\alpha$ catalog for the entire Galactic plane will become possible.

\begin{acknowledgements}
This work was supported by the National Research Foundation of Korea (NRF) grant funded by the Korea government (MSIT) (No. 2019R1F1A1064112). We sincerely appreciate the efforts of the KASI, SaTReC, and KARI staff in the development and operation of STSAT-3 and MIRIS. We are also grateful to the ISAS/JAXA and Genesia Co. staff in Japan for their significant contributions to the development of the MIRIS instrument. MIRIS was supported by a National Research Foundation of Korea grant funded by the Korean government and by ISAS/JAXA. K. I. S. was partially supported by a National Research Foundation of Korea (NRF) grant funded by the Korea government (MSIT) (No. 2020R1A2C1005788). T. N. acknowledges the support by JSPS Kakenhi JSPS KAKENHI grant Nos. JP21H04496, JP23H05441, and 23K17695.

This paper makes use of data obtained as part of the IPHAS (http://www.iphas.org/) survey, carried out at the Isaac Newton Telescope (INT), and data products from observations made with ESO Telescopes at the La Silla Paranal Observatory under programme ID 177.D-3023, as part of the VPHAS+ survey. This research has also utilized the NASA/IPAC Infrared Science Archive, operated by the Jet Propulsion Laboratory, California Institute of Technology, under contract with the National Aeronautics and Space Administration (NASA).

Additionally, we used the online {\it WISE} catalog of Galactic \ion{H}{2} regions (http://astro.phys.wvu.edu/wise/) and data products from the Two Micron All Sky Survey (2MASS), a joint project of the University of Massachusetts and the Infrared Processing and Analysis Center/California Institute of Technology, funded by NASA and the National Science Foundation. This research has made use of the SIMBAD database, operated at CDS, Strasbourg, France.
\end{acknowledgements}

\facility{IRSA}
\software{Astropy \citep{2013A&A...558A..33A,2018AJ....156..123A}, Matplotlib \citep{2007CSE.....9...90H}, Montage \citep{2003ASPC..295..343B,2010ascl.soft10036J}, Numpy \citep{2007CSE.....9c..10O}}

\appendix
\section{P\MakeLowercase{a}$\alpha$ Images: Original and Masked Versions}

\begin{figure*}[ht]
\digitalasset
\centering
\includegraphics[scale=0.33]{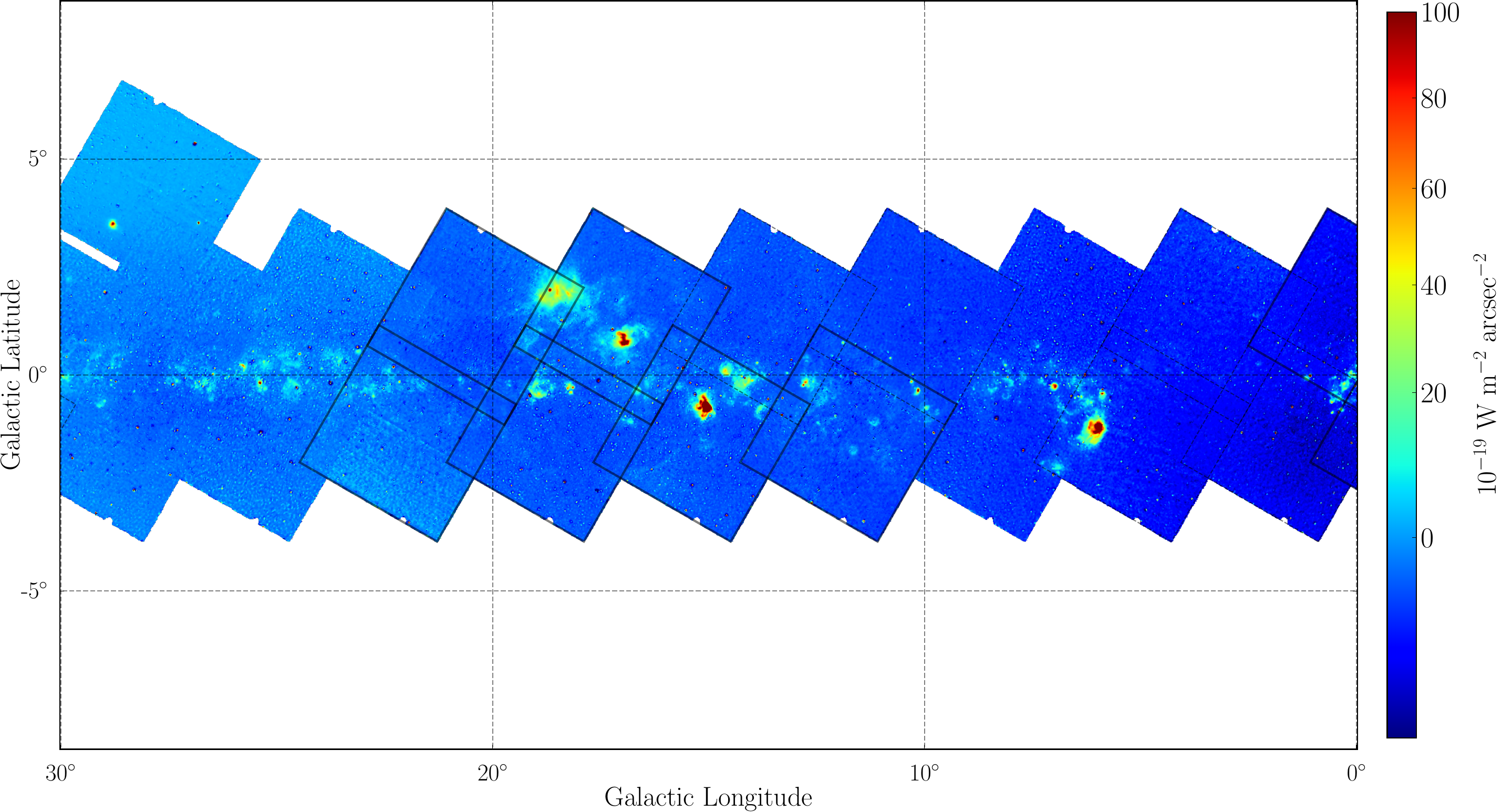}
\caption{MIPAPS Pa$\alpha$ images without masking, corresponding to Figures \ref{fig:paa11}--\ref{fig:paa43}.\\
(The complete figure set (12 images) is available in the online article.)\label{fig:paa_origin}}
\end{figure*}

\begin{figure*}[ht]
\digitalasset
\centering
\includegraphics[scale=0.33]{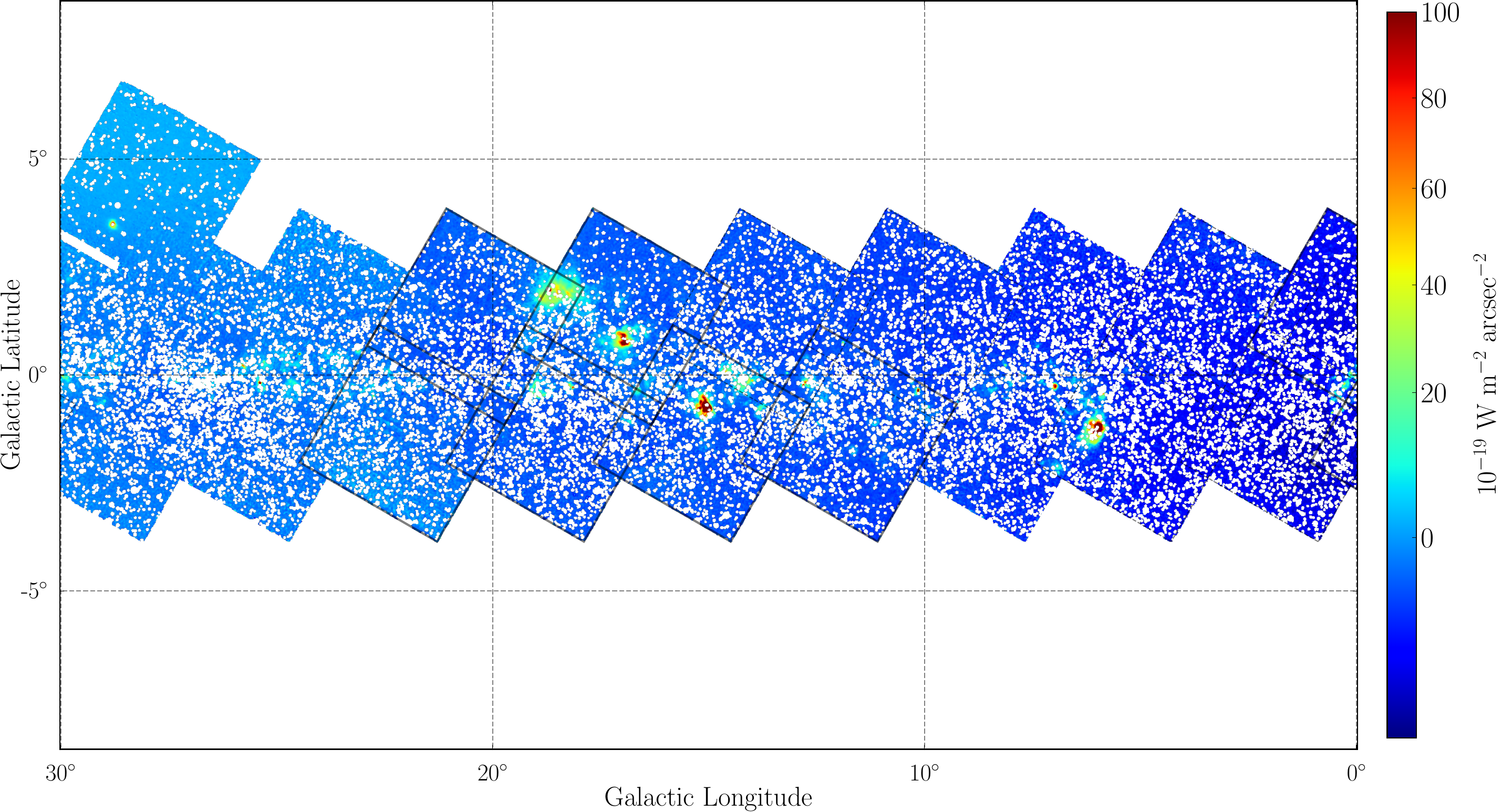}
\caption{MIPAPS Pa$\alpha$ images with masked positions, corresponding to Figures \ref{fig:paa11}--\ref{fig:paa43}.\\
(The complete figure set (12 images) is available in the online article.)\label{fig:paa_masking}}
\end{figure*}

For reference regarding the twelve close-up Pa$\alpha$ images in Figures \ref{fig:paa11}--\ref{fig:paa43}, where the masking method has been applied, we provide the original Pa$\alpha$ images without masking in Figure \ref{fig:paa_origin} and their corresponding images showing masked positions in Figure \ref{fig:paa_masking}.

\figsetstart
\figsetnum{20}
\figsettitle{MIPAPS Pa$\alpha$ images without masking}

\figsetgrpstart
\figsetgrpnum{20.1}
\figsetgrptitle{$\ell = 0\arcdeg$--$30\arcdeg$}
\figsetplot{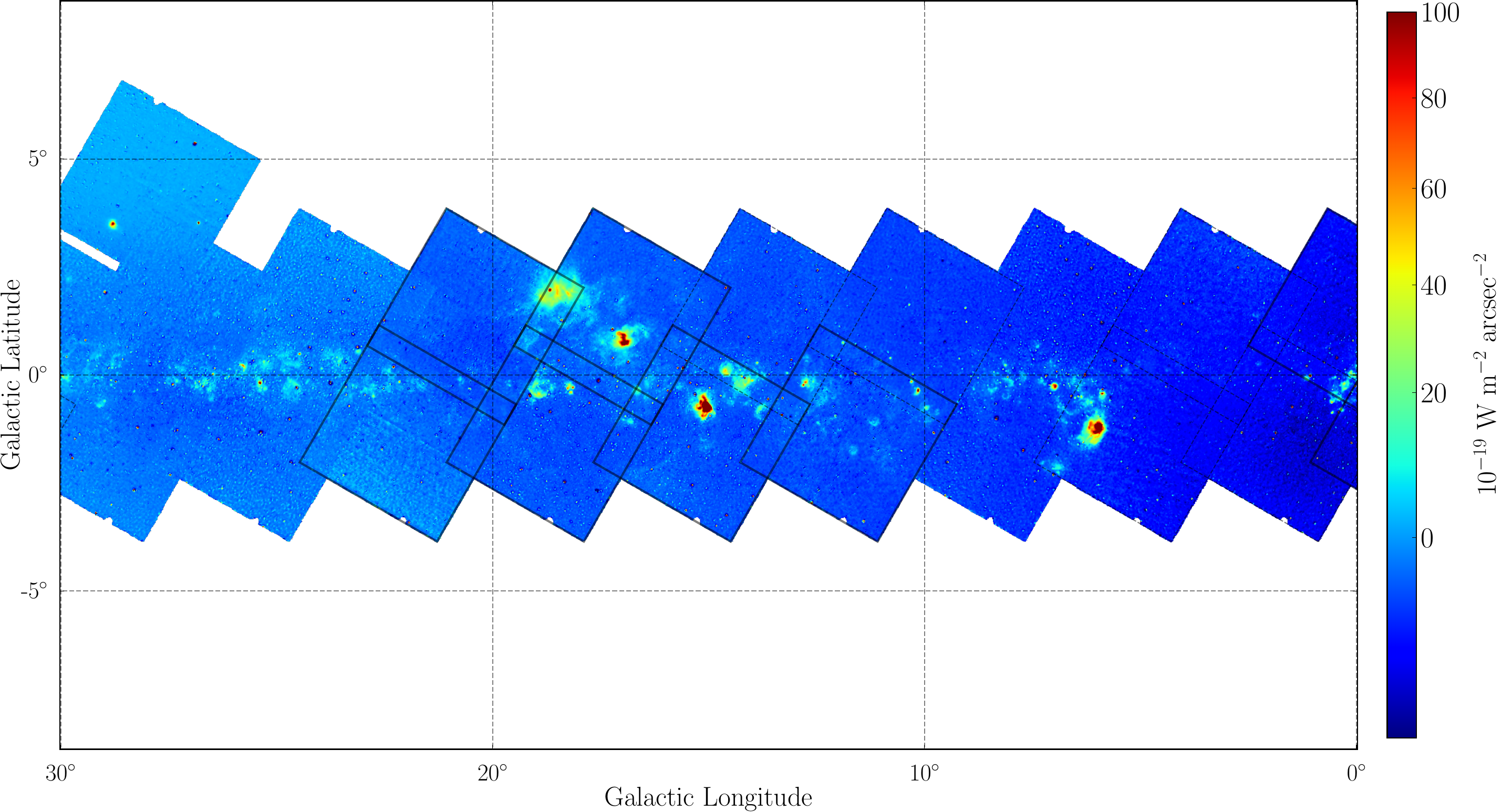}
\figsetgrpnote{Continuum-subtracted MIPAPS Pa$\alpha$ line image (without masking) of the Galactic plane for $\ell = 0\arcdeg$--$30\arcdeg$. The thick solid and thin dashed rectangles represent MIPAPS fields with data obtained from highly- and slightly-shadowed observations, respectively.}
\figsetgrpend

\figsetend

\figsetstart
\figsetnum{21}
\figsettitle{MIPAPS Pa$\alpha$ images with masked positions}

\figsetgrpstart
\figsetgrpnum{21.1}
\figsetgrptitle{$\ell = 0\arcdeg$--$30\arcdeg$}
\figsetplot{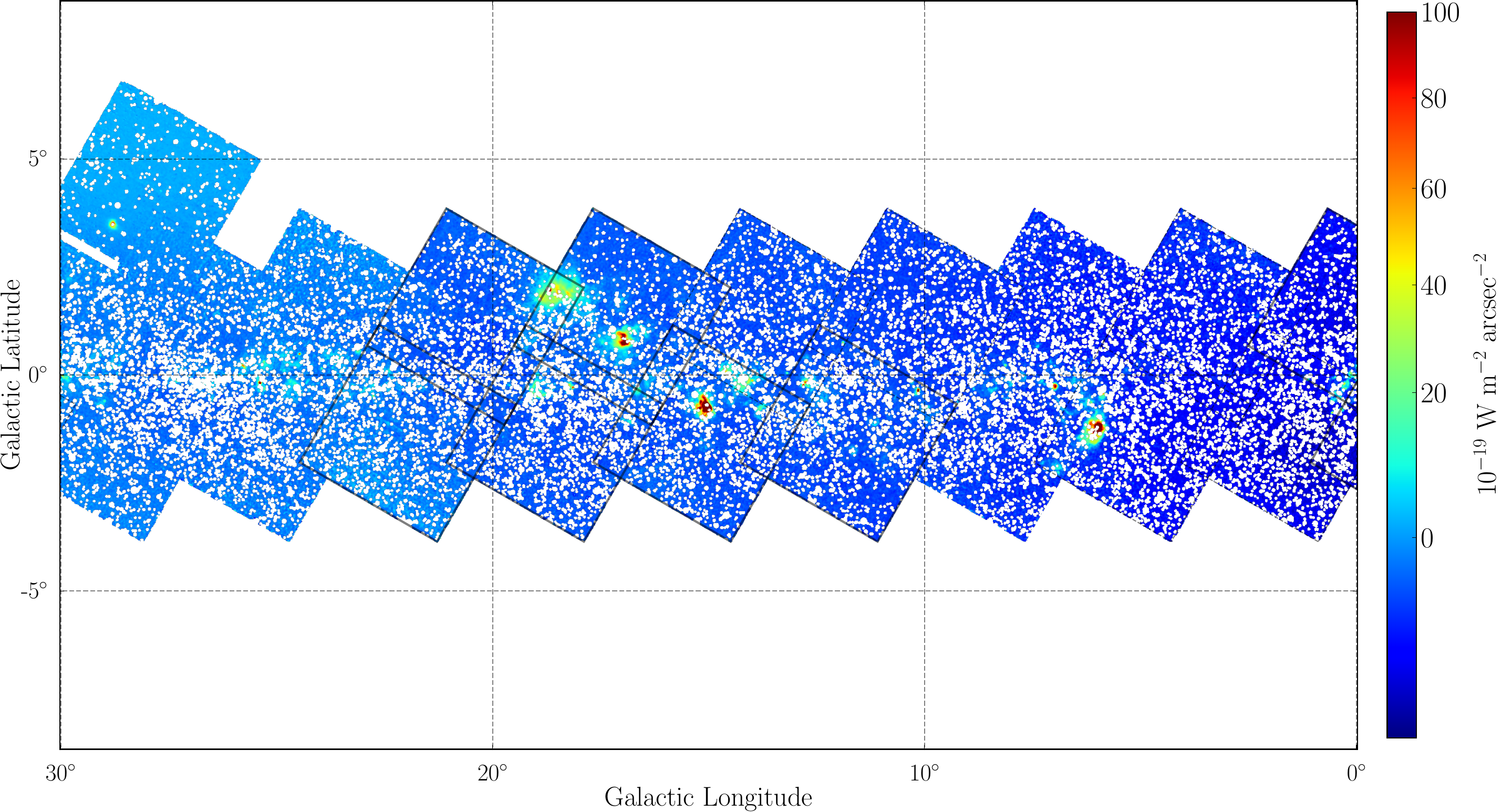}
\figsetgrpnote{Continuum-subtracted MIPAPS Pa$\alpha$ line image (with masked positions) of the Galactic plane for $\ell = 0\arcdeg$--$30\arcdeg$. The thick solid and thin dashed rectangles represent MIPAPS fields with data obtained from highly- and slightly-shadowed observations, respectively.}
\figsetgrpend

\figsetend

\bibliography{references}{}
\bibliographystyle{aasjournal}

\begin{longrotatetable}
\begin{deluxetable*}{cccccccc}
\tabletypesize{\scriptsize}
\tablewidth{0pt}
\tablecaption{Information for 78 MIPAPS Pa$\alpha$ Sources\label{table:dis}}
\tablehead{
\colhead{ID} & \multicolumn{3}{c}{Distance} && \multicolumn{3}{c}{Corresponding \ion{H}{2} Region} \\
\cline{2-4} \cline{6-8}
\colhead{} & \colhead{Kinematic} & \colhead{Spectrophotometric} & \colhead{Parallax} && \colhead{Name} & \colhead{$E(\bv)$} & \colhead{Spectral Type of Ionizing Star} \\
\colhead{} & \colhead{(kpc)} & \colhead{(kpc)} & \colhead{(kpc)} && \colhead{} & \colhead{(mag)} & \colhead{}
}
\startdata
WK004 & 11.0 $\pm$ 1.4  & \nodata           & \nodata && \nodata & \nodata                  & \nodata \\
WK005 &  7.6 $\pm$ 1.4  & \nodata           & \nodata && \nodata & \nodata                  & \nodata \\
WK011 &  6.3 $\pm$ 1.4  & \nodata           & \nodata && \nodata & \nodata                  & \nodata \\
WG003 &  9.0 $\pm$ 1.3  & \nodata           & \nodata && \nodata & \nodata                  & \nodata \\
WK021 &  8.4 $\pm$ 1.2  & \nodata           & \nodata && \nodata & \nodata                  & \nodata \\
WK024 & \nodata   & (8.06 $\pm$ 1.61) &  7.5 $\pm$ 1.0 && Sh2-128 &  1.768                  & O7V \\
WK026 &  8.3 $\pm$ 1.2  & \nodata           & \nodata && \nodata & \nodata                  & \nodata \\
WK027 & \nodata          &  1.00 $\pm$ 0.08  & \nodata && Sh2-131 &  0.545                  & O6.5V; O9.5IV; ten B-types \\
WK028 &  8.2 $\pm$ 1.2  & \nodata           & \nodata && \nodata & \nodata                  & \nodata \\
WK030 &  7.2 $\pm$ 1.1  & (6.18 $\pm$ 1.24) & \nodata && BFS 10  &   1.730                  & O9V \\
WK033 & (2.7 $\pm$ 1.3) &  1.63 $\pm$ 0.33  & \nodata && Sh2-134 &   1.038                  & B1(V) \\
WK034 & (3.1 $\pm$ 1.3) &  1.40 $\pm$ 0.28  & \nodata && Sh2-135 &   0.955                  & O9.5V \\
WK035 &  8.1 $\pm$ 1.3  & \nodata           & \nodata && \nodata & \nodata                  & \nodata \\
WK039 &  6.0 $\pm$ 1.3  & \nodata           & \nodata && \nodata & \nodata                  & \nodata \\
WK041 &  7.0 $\pm$ 1.2  & (9.92 $\pm$ 1.98) & \nodata && Sh2-141 &   1.259                  & O8V \\
WK042 & (4.3 $\pm$ 1.4) &  3.48 $\pm$ 0.24  & \nodata && Sh2-142 &   0.634                  & O5V; O8V; seven B-types \\
WK043 & (4.1 $\pm$ 1.4) &  3.84 $\pm$ 0.77  & \nodata && Sh2-143 &   0.655                  & O9.5V \\
WK045 & \nodata         & \nodata           & 4.4 $\pm$ 0.5 && Sh2-146 & \nodata            & \nodata \\
WK049 & (5.6 $\pm$ 1.3) &  2.90 $\pm$ 0.58  & (4.45 $\pm$ 1.56\tablenotemark{a}) && Sh2-152 & 1.291 & O8.5V \\
WK054 & (5.5 $\pm$ 1.3) & (2.68 $\pm$ 0.54) & 2.56 $\pm$ 0.21\tablenotemark{a} && Sh2-156 & 1.278 & O6.5V \\
WG011 & \nodata         & \nodata           & \nodata && Sh2-157 (BD+59 2677; BD+59 2673; BD+59 2681) & 0.823 & O9.5III; O8III; one B-type \\
WQ003 & \nodata         & \nodata           & 3.5 $\pm$ 0.5 && \nodata & \nodata            & \nodata \\
WK058 & \nodata         & (2.93 $\pm$ 0.28) & 3.4 $\pm$ 0.2 && Sh2-157 (ALS 19704) & 0.915  & O9.5V \\
WK062 &  5.7 $\pm$ 1.2  & \nodata           & \nodata && \nodata & \nodata                  & \nodata \\
WK063 &  4.7 $\pm$ 1.3  & \nodata           & \nodata && \nodata & \nodata                  & \nodata \\
WK066 & (4.4 $\pm$ 1.3) &  3.01 $\pm$ 0.41  & \nodata && Sh2-163 &   1.270                  & O9.5V; O9V; O8V; three B-types \\
WK067 & (4.8 $\pm$ 1.3) &  3.08 $\pm$ 0.62  & \nodata && Sh2-164 &   1.020                  & B0.2III \\
WC019 & \nodata         & \nodata      & \nodata && Sh2-166 (ALS 13072) & 1.058             & B1V \\
WK068 &  5.0 $\pm$ 1.3  & \nodata           & \nodata && \nodata & \nodata                  & \nodata \\
WK069 & (4.9 $\pm$ 1.3) &  1.96 $\pm$ 0.39  & \nodata && Sh2-165 &   0.760                  & B0V \\
WK071 &  6.8 $\pm$ 1.2  & \nodata           & \nodata && Sh2-167 & \nodata                  & \nodata \\
WK072 & (4.4 $\pm$ 1.3) &  2.14 $\pm$ 0.30  & \nodata && Sh2-168 (LS I +60 50) & 1.025      & O9.5V \\
WK073 & (4.5 $\pm$ 1.3) &  2.09 $\pm$ 0.42  & \nodata && Sh2-169 &   0.840                  & B0.5III \\
WK074 & (5.8 $\pm$ 1.1) &  2.79 $\pm$ 0.33  & \nodata && Sh2-170 &   0.641                  & O9V; five B-types \\
WK075 &  8.0 $\pm$ 1.4  & \nodata           & \nodata && \nodata & \nodata                  & \nodata \\
WK077 &  6.8 $\pm$ 1.2  & \nodata           & \nodata && \nodata & \nodata                  & \nodata \\
WK078 & (4.3 $\pm$ 1.3) &  3.12 $\pm$ 0.34\tablenotemark{b} & \nodata && Sh2-172 & \nodata  & \nodata \\
WK079 &  7.0 $\pm$ 1.3  & \nodata           & \nodata && \nodata & \nodata                  & \nodata \\
WK081 & (5.4 $\pm$ 1.1) & (2.67 $\pm$ 0.53) &  2.02 $\pm$ 0.05\tablenotemark{a} && Sh2-175 & 1.116 & B1III \\
MPE046& \nodata         &  5.41 $\pm$ 1.08  & \nodata && Sh2-180 &   0.770                  & O8V \\
MPE047&  2.76\tablenotemark{c} & \nodata    & \nodata && Sh2-181 & \nodata                  & \nodata \\
WK082 &  5.7 $\pm$ 1.2  & \nodata           & \nodata && \nodata & \nodata                  & \nodata \\
WK084 & (6.5 $\pm$ 1.3) &  1.44 $\pm$ 0.26\tablenotemark{b} & \nodata && Sh2-183 & \nodata  & \nodata \\
WK088 &  7.1 $\pm$ 1.4  & \nodata           & \nodata && \nodata & \nodata                  & \nodata \\
WK089 & 10.1 $\pm$ 1.8  & \nodata           & \nodata && \nodata & \nodata                  & \nodata \\
WK090 & \nodata         &  1.58 $\pm$ 0.32  & \nodata && Sh2-187 &   1.431                  & B2.5V \\
WK093 &  5.1 $\pm$ 1.2  & \nodata           & \nodata && \nodata & \nodata                  & \nodata \\
WK098 & (4.9 $\pm$ 1.2) &  2.40 $\pm$ 0.32\tablenotemark{b} & \nodata && Sh2-194 & \nodata  & \nodata \\
WK099 & (5.2 $\pm$ 1.3) &  2.44 $\pm$ 0.49  & \nodata && Sh2-193 &   0.760                  & B2.5V \\
MPE058& \nodata         &  2.49 $\pm$ 0.50  & \nodata && Sh2-198 &   0.950                  & B0V \\
WK106 &  3.1 $\pm$ 1.2  & \nodata           & \nodata && \nodata & \nodata                  & \nodata \\
WK109 &  6.0 $\pm$ 1.6  & \nodata           & \nodata && \nodata & \nodata                  & \nodata \\
WK111 &  7.0 $\pm$ 1.9  & \nodata           & \nodata && \nodata & \nodata                  & \nodata \\
WK113 & (3.4 $\pm$ 1.4) & (3.02 $\pm$ 0.60) & 2.96 $\pm$ 0.16\tablenotemark{a} && Sh2-206 & 1.356 & O4V \\
WK114 & (4.5 $\pm$ 1.5) & (4.27 $\pm$ 1.19) & 3.59 $\pm$ 0.25\tablenotemark{a} && Sh2-207 & 1.060 & B2.5(V); B4(V) \\
WK115 & (3.9 $\pm$ 1.4) & \nodata           & 4.02 $\pm$ 0.26\tablenotemark{a} && Sh2-208 (the central O9.5V star) & 1.065 & O9.5V \\
WK118 & 12.3 $\pm$ 3.7  & \nodata           & \nodata && \nodata & \nodata                  & \nodata \\
WK119 &  1.55\tablenotemark{c} & \nodata    & \nodata && Sh2-210 & \nodata                  & \nodata \\
WK120 &  8.2 $\pm$ 2.6  & \nodata           & \nodata && \nodata & \nodata                  & \nodata \\
WK121 & (5.9 $\pm$ 2.0) &  7.39 $\pm$ 0.44  & \nodata && Sh2-211 &   1.664                  & O9V; O9.5Ib; one B-type \\
WK123 & (3.8 $\pm$ 1.8) &  4.37 $\pm$ 0.87  & \nodata && Sh2-217 &   0.725                  & O9.5V \\
WQ014 & \nodata         &  4.19 $\pm$ 0.27\tablenotemark{b} & \nodata && BFS 44 & \nodata   & \nodata \\
WK126 & \nodata         &  4.2 $\pm$ 1.3\tablenotemark{d} & \nodata && Sh2-226 & \nodata    & \nodata \\
WK132 & \nodata         &  1.6 $\pm$ 0.3\tablenotemark{d} & \nodata && Sh2-233 & \nodata    & \nodata \\
WK139 & \nodata         & (3.76 $\pm$ 0.28) & 2.07 $\pm$ 0.06\tablenotemark{a} && Sh2-237 & 0.723 & B0V; B0.5V; B4V \\
WK141 & \nodata         &  4.84 $\pm$ 0.97  & \nodata && Sh2-241 &  0.621                   & O9V \\
WQ022 &  0.48\tablenotemark{c} & \nodata    & \nodata && BFS 50  & \nodata                  & \nodata \\
WK144 & \nodata         &  1.8 $\pm$ 0.3\tablenotemark{d} & \nodata && LBN 840; BFS 51 & \nodata & \nodata \\
WK153 & \nodata         & (2.27 $\pm$ 0.45) & 1.6 $\pm$ 0.1 && Sh2-255 & 1.180              & B0V \\
WK154 & \nodata         &  3.03 $\pm$ 0.61  & \nodata && Sh2-258 &   1.398                  & B3V \\
WK158 & \nodata         &  3.89 $\pm$ 0.28  & \nodata && Sh2-267 &   1.123                  & O9V; one B-type \\
WK162 & \nodata         & (3.90 $\pm$ 0.47) & 3.25 $\pm$ 0.12\tablenotemark{a} && Sh2-271 & 0.980 & B0V \\
WK170 & \nodata         &  7.39 $\pm$ 1.21\tablenotemark{b} & \nodata && Sh2-283 (Star 5)\tablenotemark{b} & 0.994\tablenotemark{b} & B1V\tablenotemark{b} \\
WK171 & \nodata         &  1.17 $\pm$ 0.29\tablenotemark{b} & \nodata && BFS 54 & \nodata  & \nodata \\
WK172 & (6.4 $\pm$ 1.8) &  7.89 $\pm$ 0.27\tablenotemark{b} & \nodata && Sh2-284 & \nodata & \nodata \\
WC066 & \nodata         &  7.89 $\pm$ 0.27\tablenotemark{b} & \nodata && Sh2-284 & \nodata & \nodata \\
WK173 &  9.8 $\pm$ 2.6  & \nodata           & \nodata && \nodata & \nodata                  & \nodata \\
WG051 & \nodata         &  7.89 $\pm$ 0.27\tablenotemark{b} & \nodata && Sh2-284 & \nodata & \nodata \\
\enddata
\tablenotetext{a}{From \citet{2022MNRAS.510.4436M}.}
\tablenotetext{b}{From \citet{2007AA...470..161R}.}
\tablenotetext{c}{From \citet{2014AA...569A.125H}.}
\tablenotetext{d}{From \citet{2003AA...397..133R}.}
\tablecomments{Kinematic and parallax distances are obtained from version 2.2 of the {\it WISE} \ion{H}{2} region catalog (\url{http://astro.phys.wvu.edu/wise/}), while spectrophotometric distances primarily come from Table 1 of \citet{2015AJ....150..147F}, except where noted in the footnotes. For sources with multiple distance values, the distances with the highest distance-to-uncertainty ratios are listed without parentheses and were applied to Figure \ref{fig:cecorr}(d). Information regarding corresponding \ion{H}{2} regions is also derived from Table 1 of \citet{2015AJ....150..147F}, except as noted in the footnotes. For six Pa$\alpha$ sources, some ionizing stars mentioned in the references were excluded because they lie outside the angular boundaries of the Pa$\alpha$ sources. The selected ionizing stars are listed in parentheses in the ``Name'' column.}
\end{deluxetable*}
\end{longrotatetable}

\end{document}